\documentclass[a4paper,11pt]{article}

\usepackage[breaklinks,bookmarks]{hyperref}
\usepackage{fullpage,amsmath,amssymb,amsthm,epsfig,color,bbold}
\usepackage[noadjust]{cite}

\makeatletter
\renewcommand\section{\@startsection{section}%
  {1}%
  {\z@}%
  {-3.5ex \@plus -1ex \@minus -.2ex}%
  {2.3ex \@plus.2ex}%
  {\normalfont\large\bfseries}}
\renewcommand\subsection{\@startsection{subsection}%
  {2}%
  {\z@}%
  {-3.25ex\@plus -1ex \@minus -.2ex}%
  {1.5ex \@plus .2ex}%
  {\normalfont\normalsize\bfseries}}
\renewcommand{\@seccntformat}[1]{{\csname the#1\endcsname}.\ \ }

\makeatother

\newcommand{\IC}{\mathbb{C}}
\newcommand{\IN}{\mathbb{N}}
\newcommand{\IR}{\mathbb{R}}
\newcommand{\ZZ}{\mathbb{Z}}
\newcommand{\B}{B} 
\newcommand{\C}{\mathbb{C}}
\renewcommand{\H}{\mathcal{H}}
\newcommand{\M}{\mathcal{M}}
\renewcommand{\P}{\mathcal{P}}
\newcommand{\Q}{\mathbb{Q}}
\newcommand{\floor}[1]{\left\lfloor #1\right\rfloor}
\newcommand{\ceil}[1]{\left\lceil #1\right\rceil}
\renewcommand{\ldots}{\mathellipsis} 
\newcommand{\shortldots}{\mathinner{\ldotp \ldotp }}
\renewcommand{\epsilon}{\varepsilon}
\renewcommand{\phi}{\varphi}

\newcommand{\poly}{\operatorname{poly}}
\renewcommand{\k}[1]{\mathopen|#1\mathclose\rangle}
\renewcommand{\b}[1]{\mathopen\langle#1\mathclose|}
\newcommand{\bk}[2]{\mathopen\langle#1\,|\,#2\mathclose\rangle}
\newcommand{\lbl}{\operatorname{\rm label}}
\newcommand{\var}{\operatorname{\rm var}}
\renewcommand{\sp}{\operatorname{\rm span}}
\newcommand{\Conf}{{\cal C}}

\newcommand{\adv}{\operatorname{\rm adv}}
\newcommand{\PERM}{\mbox{PERM}}
\newcommand{\IND}{\mbox{$\mathrm{IND}$}}
\newcommand{\ISA}{\mbox{$\mathrm{ISA}$}}
\newcommand{\asn}{\operatorname{\rm asn}}
\newcommand{\tr}{\operatorname{\rm tr}}
\newcommand{\wt}{\widetilde}
\newcommand{\SU}{\mbox{\rm SU}}
\newcommand{\pre}{\text{pre}}
\newcommand{\bdot}{\mathbin{\text{\boldmath$\cdot$}}}
\newcommand{\bigk}[1]{\mathopen{\bigl|}#1\mathclose{\bigr\rangle}}

\newcommand{\bigbk}[2]{\mathopen{\bigl\langle}#1\,|\,#2\mathclose{\bigr\rangle}}
\newcommand{\DISJ}{\mbox{$\mathrm{DISJ}$}}
\newcommand{\IP}{\mbox{$\mathrm{IP}$}}
\newcommand{\wbar}[1]{\overline{#1}}

\newenvironment{shortindent}[1]
  {\begin{list}{}%
   {\settowidth{\labelwidth}{#1}\leftmargin\labelwidth
    \advance\leftmargin by\labelsep\parsep0pt\topsep0pt}
    \sloppy\clubpenalty4000\widowpenalty4000
    }%
  {\end{list}}

\newcommand{\listnewline}{\leavevmode\newline}
\newtheoremstyle{defaultthm}
  {\topsep}
  {\topsep}
  {\let\\=\listnewline\itemsep0pt}
  {}
  {\bfseries}
  {:}
  {0.5em}
  {}
\newtheoremstyle{defaultit}%
  {\topsep}%
  {\topsep}%
  {\itshape\let\\=\listnewline\itemsep0pt}%
  {}%
  {\bfseries}%
  {:}%
  {0.5em}%
  {}%
\newtheoremstyle{clmstyle}%
  {\topsep}%
  {\topsep}%
  {\itshape\itemsep0pt}%
  {}%
  {\itshape}%
  {.}%
  {0.5em}%
  {}%
\renewenvironment{proof}[1][\proofname]{\par
  \normalfont
  \topsep6pt plus6pt \trivlist
  \item[\hskip\labelsep\itshape#1.]\ignorespaces
}{%
  \hspace*{\fill}$\Box$\endtrivlist
}
\theoremstyle{defaultit}
\newtheorem{theorem}{Theorem}[section]
\newtheorem*{theorem*}{Theorem}
\newtheorem{corollary}[theorem]{Corollary}
\newtheorem{proposition}[theorem]{Proposition}
\newtheorem{lemma}[theorem]{Lemma}
\newtheorem*{lemma*}{Lemma}
\newtheorem*{remark*}{Remark}

\newtheorem*{fact*}{Fact}
\theoremstyle{clmstyle}

\newtheorem*{claim*}{Claim}
\newtheorem*{observation*}{Observation}
\theoremstyle{defaultthm}
\newtheorem{definition}[theorem]{Definition}

\begin{document}

\title{\vspace*{-\baselineskip}\Large\bfseries Quantum Branching Programs and\\
Space-Bounded Nonuniform Quantum Complexity}
\author{\normalsize Martin Sauerhoff\footnotemark[1]~~ and Detlef
  Sieling\footnotemark[7]\\
\normalsize FB Informatik, Univ. Dortmund, 44221 Dortmund, Germany\\
\normalsize Email: {\{sauerhoff$\mid$ds01\}@ls2.cs.uni-dortmund.de}}
\date{}
\maketitle

\renewcommand{\thefootnote}{\fnsymbol{footnote}}
\footnotetext[1]{Supported by DFG grant Sa 1053/1-1.}
\footnotetext[7]{Supported by DFG grant Si 577/1-1.}

\vspace*{-\baselineskip}
\begin{quote}
\textbf{Abstract.}\ \ 
In this paper, the space complexity of nonuniform quantum algorithms
is investigated using the model of quantum branching programs (QBPs).
In order to clarify the relationship between QBPs and nonuniform quantum Turing machines, 
simulations between these two models are presented
which allow to transfer upper and lower bound results. 
Exploiting additional insights about the connection between the running time
and the precision of amplitudes, it is shown that nonuniform quantum Turing machines with
algebraic amplitudes and QBPs with a suitable analogous set of amplitudes
are equivalent in computational power if both models
work with bounded or unbounded error.
Furthermore, quantum ordered binary decision diagrams (QOBDDs) are 
considered, which are restricted QBPs that can be regarded as a 
nonuniform analog of one-way quantum finite automata.
Upper and lower bounds are proved that allow a classification of the computational power of 
QOBDDs in comparison to usual deterministic and randomized variants of the model.
Finally, an extension of QBPs is proposed where the performed unitary operation may depend 
on the result of a previous measurement. A simulation of randomized BPs by this generalized
QBP model as well as exponential lower bounds for its ordered variant are presented.
\end{quote}

\section{Introduction}

The intriguing open question behind the research on quantum
computing is whether there are problems that can be solved more
efficiently by quantum computers than by classical ones.
Shor's famous quantum algorithm for factoring 
integers in polynomial time~\cite{Sho97} provides the
most conclusive evidence so far in favor of an affirmative answer of
this question.
The notion of a quantum algorithm is made precise by models of computation
such as quantum Turing machines (QTMs), quantum circuits, quantum finite automata
(QFAs), and quantum communication protocols. For an introduction to
these models, we refer to the textbooks of
Gruska~\cite{Gru99}, Kitaev, Shen, and Vyalyi~\cite{Kit02}, and
Nielsen and Chuang~\cite{Nie00}. 

Apart from the obviously important computation time, different other complexity 
measures for quantum algorithms have been investigated.
Space is a crucial resource due to inherent technical constraints in the current 
physical realizations of quantum computers. As pointed out by Ambainis 
and Freivalds~\cite{Amb98}, the goal of obtaining systems with a small
quantum mechanical part was one of the motivations for considering
quantum finite automata.  
In his seminal paper~\cite{Wat99} and its later extensions~\cite{Wat99a,Wat03}, Watrous investigated
the space complexity of quantum algorithms in the more general model
of quantum Turing machines. 
The quantum Turing machines considered by Watrous may
have algebraic transition amplitudes and are 
unidirectional, i.\,e., the direction of the head movements is
a function of the state entered in a computation step.
Among other results, he has shown for this scenario that 
space~$O(s)$ probabilistic Turing machines with unbounded error and
quantum Turing machines with unbounded error are
equivalent in computational power, where~$s$ is a space-constructible
function.
It is open whether similar statements hold for other types of error, e.\,g.,
bounded error. It is also not known whether the requirement of
algebraic transition amplitudes is crucial for space-restricted
quantum Turing machines, despite the results of 
Adleman, DeMarrais, and Huang~\cite{Adl97} that allow us to restrict
the set of amplitudes to $\{0,\pm 3/5,\pm 4/5,\pm 1\}$ for
polynomial time, bounded error quantum Turing machines. 
Finally, even the standard assumption of unidirectionality remains
to be justified for QTMs with sublinear space-bounds,
since the known simulations for the time-bounded case 
due Bernstein and Vazirani~\cite{Ber97} and Yao~\cite{Yao93} or  
Nishimura and Ozawa~\cite{Nis02a} can not be applied in an
obvious way.

Already classical Turing machines have turned out to be a quite
cumbersome device for proving upper and lower bounds.  Branching
programs are a graphic representation of boolean functions and as
such are more amenable to combinatorial arguments than Turing
machines. Furthermore, it is well-known that the logarithm of the size
of branching programs is asymptotically equal to the space complexity
for the nonuniform (advice taking) variant of Turing machines
(Cobham~\cite{Cob66}, Pudl\'{a}k and \v{Z}\'{a}k~\cite{Pud83}).
Recently obtained lower bound results for branching
programs~\cite{Bea01b,Ajt99a,Ajt99b,Bea00b,Bea02a}, which imply
time-space tradeoffs for sequential computations, underline the
significance of branching programs in the investigation of space
complexity.

\begin{figure}
\centerline{\input{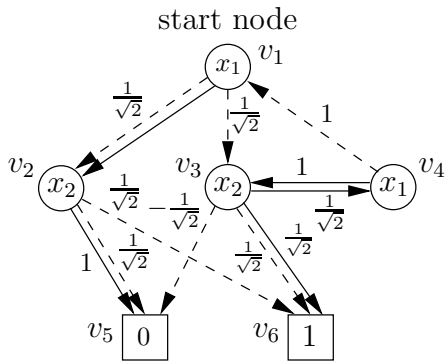}}
\caption{An example of a QBP.}
\label{FigBspQBP}
\end{figure}

In this paper we deal with a quantum variant of branching programs.
In order to give a feeling of how quantum branching programs (QBPs) 
work, we consider the example in
Figure~\ref{FigBspQBP}. For the formal definition and the technical
details we refer to Definitions~\ref{def:qbp} and~\ref{def:comp_qbps}.
The QBP in the figure represents a boolean function depending on the
variables $x_1$ and $x_2$.  Each node $v\in V = \{v_1,\ldots,v_6\}$ 
of the QBP is associated with a vector $\k{v}$ of an orthonormal basis 
of the Hilbert space $\H=\IC^{|V|}$. Each
intermediate state of the computation of the QBP is a vector in $\H$.
The initial state of the QBP is $\k{v_1}$, where $v_1$ is the start node
of the QBP. Each computation step consists of a first phase, where a
projective measurement is used to decide whether the computation continues or
whether it stops with the result $0$ or $1$, and a second phase, where
a unitary transformation described by the edge labels is applied to
the state.  If $x_i=0$ ($x_i=1$), only the dashed (solid) edges
leaving each $x_i$-node contribute to this transformation.  In our
example the projections describing the measurement are 
$E_{\rm cont}=\k{v_1}\b{v_1}+\cdots+\k{v_4}\b{v_4}$, $E_0=\k{v_5}\b{v_5}$,
and $E_1=\k{v_6}\b{v_6}$, i.\,e., the projections on the subspaces
spanned by the vectors corresponding to interior nodes and sinks
labeled by $0$ and $1$, resp. Assume that $x_1=x_2=0$. The initial state
is $\k{v_1}$. The projective measurement yields that the computation is 
continued with probability~$1$. The dashed edges leaving $v_1$ are
labeled by $1/\sqrt{2}$, hence, the next state is
$(1/\sqrt{2})(\k{v_2}+\k{v_3})$. In the second step the
computation again continues with probability $1$ and according to the
labels of the edges leaving~$v_2$ and~$v_3$ the next state is~$\k{v_6}$. 
Hence, in the third step the computation stops 
with probability~$1$ and the result is~$1$.

The most important complexity measures for QBPs are the size of the QBP, i.\,e.,
its number of nodes, and the (expected or worst-case) computation time.
QBPs may be cyclic or acyclic. For acyclic
QBPs one can furthermore consider the width of the QBP, i.\,e., the
maximum number of nodes with the same distance from the start node.
Before we present our results on the relationship between the
complexity measures for QBPs and other complexity measures for boolean
functions, in particular the space complexity of quantum Turing
machines, we discuss previous work on QBPs.

\goodbreak
Ablayev, Gainutdinova, and Karpinski~\cite{Abl01a} and
Nakanishi, Hamaguchi, and Kashiwabara~\cite{Nak00} have introduced 
quantum OBDDs (quantum ordered binary decision diagrams), i.\,e., 
acyclic QBPs where the input variables may only be read once in
a fixed order during each computation. Ablayev, Gainutdinova, and
Karpinski have presented a function that requires linear width in the
input length for deterministic OBDDs, but only logarithmic width for
quantum OBDDs. Nakanishi, Hamaguchi, and Kashiwabara have obtained a
similar gap, but their lower bound even holds for randomized OBDDs.
More recently, Ablayev, Moore, and Pollett~\cite{Abl02a} have proved
that the class of functions that can be exactly computed by oblivious
width-$2$ QBPs of polynomial size coincides with
the class~$\text{NC}^1$, while width~$5$ is necessary classically
unless $\text{NC}^1 = \text{ACC}$.
Finally, \v{S}palek~\cite{Spa02a} has studied a general model of QBPs 
and has independently come up with a definition
similar to that used here. Furthermore, he has also presented exact
simulations between QBPs whose transition
function is composed of unitary matrices from a finite basis and
quantum Turing machines defined analogously.
In the following, we describe the contributions of our paper.
For the sake of a clearer presentation, we group the results into
three parts.

\medskip
\emph{First Part: Simulations (Sections~2--5).}\ \ 
In Sections~2 and~3 we define quantum branching programs and extend the definition of
quantum Turing machines (QTMs) to the nonuniform case. Following
Watrous~\mbox{\cite{Wat99,Wat99a,Wat03}}, we include unidirectionality
as a part of our definition of QBPs and we usually consider unidirectional
nonuniform QTMs. Simulations between QBPs and unidirectional nonuniform QTMs are 
presented in Section~4.
Our first result shows that unidirectional nonuniform QTMs using space $O(\log S)$ 
can be simulated by QBPs of size~$\poly(S)$ taking the same number of computation 
steps as the simulated machine. In the opposite direction, we obtain an
approximate simulation of QBPs of size~$S$ by unidirectional nonuniform QTMs 
that carry out $T$ simulation steps with approximation error~$\epsilon$ 
in space $\poly(S+\log\log(T/\epsilon))$ and time $\poly(S,T,\log(1/\epsilon))$.
These results are for QBPs and QTMs whose amplitudes are
arbitrary complex numbers. 

As remarked above, the standard set of transition amplitudes for 
QTMs in the space-bounded scenario are algebraic numbers. 
As an analogous standard set for QBPs we propose short amplitudes, i.\,e.,
amplitudes that can be represented in polynomial bit length in the size of the QBP
as rational polynomials on finitely many algebraic numbers.
Using our general simulation results and additional insights about the
connection between running time and the precision of amplitudes, we show that in 
the case of bounded and unbounded error, QBPs with short amplitudes and 
size~$\poly(S)$ and unidirectional nonuniform QTMs with algebraic amplitudes 
using space~$O(\log S)$ are of the same computational power.

In Section~5, we justify our standard assumption of unidirectionality for
the considered models. We provide a space-efficient approximate simulation of
(general) nonuniform QTMs by unidirectional ones. In particular, this result
yields that $O(\log S)$ space nonuniform QTMs, $O(\log S)$ space unidirectional nonuniform QTMs, and 
$\poly(S)$ size QBPs are of the same computational power if these models work with algebraic
and short amplitudes, resp., and with bounded or unbounded error. Altogether, these arguments show that
 QBPs are a suitable model for exploring space-bounded nonuniform
quantum complexity.

\medskip
\emph{Second Part: QOBDDs (Section~6).}\ \ 
We explore the relationship between
the size of quantum OBDDs (QOBDDs) and classical OBDDs.  First, we design
polynomial size QOBDDs for a function that classical
deterministic OBDDs can only represent in exponential size, as well as
for a partially defined function for which even randomized
OBDDs require exponential size.  On the other hand, even very simple
functions can be hard for QOBDDs.  We show that for the
disjointness function $(\wbar{x}_1\vee \wbar{x}_2)\wedge
(\wbar{x}_3\vee\wbar{x}_4)\wedge\cdots\wedge
(\wbar{x}_{n-1}\vee\wbar{x}_n)$ as well
as the inner product function $x_1x_2\oplus x_3x_4\oplus\cdots\oplus
x_{n-1}x_n$, QOBDDs require exponential size, while deterministic OBDDs can
represent these functions in linear size.  Finally, we prove that zero
error QOBDDs of polynomial size are no more powerful than polynomial
size reversible OBDDs.

\medskip
\emph{Third Part: QBPs with Generalized Measurements (Section~7).}\ \ 
For quantum OBDDs as well as for quantum finite automata, the
unitarity requirement of quantum algorithms is a serious restriction.
Intuitively, the problem is that it is difficult in these models to
forget input already read. 
In Section~7 we study the question of whether it may help to allow measurements 
to choose the unitary transformation for the next computation step (apart from
checking whether the computation has stopped). For quantum circuits
this question has already been considered by Aharonov, Kitaev and
Nisan~\cite{Aha98}, who have proposed to describe the states and the
computations of quantum circuits by mixed states and
superoperators, resp.  We define natural variants of QBPs and QOBDDs
with generalized measurements and investigate some of their
properties. QBPs and QOBDDs with generalized
measurements can simulate their randomized counterpart without
increase in size.  On the other hand, we prove an exponential lower
bound on the size of QOBDDs with generalized measurements for all
so-called $k$-stable functions. This class includes, e.\,g., the function
checking for the presence of a clique in a graph and the determinant
of a boolean matrix.

\section{Quantum Branching Programs}\label{sec:qbps}

In this section, we define classical and quantum variants of branching
programs and discuss basic properties of the quantum variant.  An
extensive survey of results for classical branching programs is given
in the monograph of Wegener~\cite{Weg00}.

\begin{definition}\label{def:bps}
A \emph{(deterministic) branching program (BP)} on the variable set 
$X = \{x_1,\ldots,x_n\}$ is a directed acyclic graph with a designated 
\emph{start node} and two sinks. The sinks are labeled by the constants~$0$ and~$1$, resp. 
Each interior node is labeled by a variable from~$X$ and has two outgoing edges
carrying labels~$0$ and~$1$, resp. 
This graph computes a boolean function $f$ defined on~$X$ as
follows. To compute $f(a)$ for some input $a=(a_1,\ldots,a_n)\in\{0,1\}^n$,
start at the start node. For an interior node labeled by $x_i$, 
follow the edge labeled by~$a_i$ (this is called \emph{testing} the variable).
Iterate this until a sink is reached, whose label gives the value $f(a)$.
For a fixed input~$a$, the sequence of nodes 
visited in this way is called the 
\emph{computation path for $a$}.
The \emph{size}~$|G|$ of a branching program is the number of its 
nodes. Its  \emph{width} is the maximum number
of nodes with the same distance from the start node.
The \emph{branching program size} of a function~$f$ is the minimum
size of a branching program that computes it.
\end{definition}

{\sloppy\hbadness=2300
BPs are a nonuniform model of computation, so we usually consider a
sequence $(G_n)_{n\in\IN}$ of BPs representing a sequence of boolean
functions $(f_n)_{n\in\IN}$, where $G_n$ represents the function
${f_n\colon\{0,1\}^n\to\{0,1\}}$. We will encounter the following
variants of BPs.

}

\begin{definition}\label{def:restr_BPs}\item[]
\begin{shortindent}{--\ }
\item[--] A BP is called \emph{read-once} if, for each
variable~$x_i$, each of the paths in the BP contains at most one node
labeled by~$x_i$.
\item[--] A BP is called \emph{leveled} if the set of its nodes can be 
partitioned into disjoint sets $V_1,\ldots,V_{\ell}$, where $V_i$ is called the
\emph{$i$th level}, such that for $1\le i\le\ell-1$, each edge leaving
a node in~$V_i$ reaches a node in $V_{i+1}$.
\item[--] An {\em OBDD} (ordered binary decision diagram) is a
  read-once BP where on each computation path the variables are tested
  according to the same order. For the variable order $\pi$ it
  is also called $\pi$-OBDD.
\end{shortindent}
\end{definition}

\begin{definition}
A \emph{randomized BP} is defined as a deterministic BP, 
but may additionally contain unlabeled \emph{randomized nodes}
with two unlabeled outgoing edges, may contain cycles, and may have
sinks labeled by $0$, $1$, or ``?''. The computation for 
an input~$a$ is carried out by starting at the start node, following the outgoing edge 
labeled by~$a_i$ for an $x_i$-node as for deterministic BPs, and taking one 
of the outgoing edges with probability~$1/2$ for randomized nodes until a sink 
is reached, where different randomized decisions are independent 
of each other. The probability that the randomized BP computes the 
output~$r\in\{0,1,{\rm ?}\}$ for the input~$a$ is the probability that
the computation for~$a$ reaches a sink labeled by~$r$.
\end{definition}

Different modes of acceptance with unbounded, bounded (two-sided), one-sided,
and zero error are defined as usual (see, e.\,g., \cite{Sak96,Weg00}). 
Randomized variants of the restricted models of BPs from
Definition~\ref{def:restr_BPs} are obtained by applying the respective
restriction to the nodes labeled by variables.

Next, we define a quantum variant of BPs.
This definition contains the alternative definitions in the
literature as special cases.

\begin{definition}\label{def:qbp}\sloppy
A \emph{quantum branching program (QBP) over the variable set
$X = \{ x_1,\ldots,x_n\}$} is a directed multigraph $G = (V,E)$
with a \emph{start node} $s\in V$, a set $F\subseteq V$ of sinks,
and \emph{(transition) amplitudes} $\delta\colon V\times V\times\{0,1\}\to\C$.
Each node $v\in V-F$ is labeled by a variable $x_i\in X$ and
we define $\var(v) = i$. Each node $v\in F$ carries a 
label from $\{0,1,{\rm ?}\}$, denoted by $\lbl(v)$.
Each edge $(v,w)\in E$ is labeled by a boolean constant $b\in\{0,1\}$ and 
the amplitude~$\delta(v,w,b)$.
An edge with boolean label~$b$ is called \emph{$b$-edge} for short.
We assume that there is at most one edge carrying the same boolean label
between a pair of nodes and set $\delta(v,w,b) = 0$ 
for all $(v,w)\not\in E$ and $b\in\{0,1\}$. 

The graph~$G$ is required to satisfy the following two constraints.
First, it has to be \emph{well-formed}, meaning that for each pair of 
nodes $u,v\in V-F$ and all assignments $a=(a_1,\ldots,a_n)$ to the variables in~$X$,
\begin{equation}\tag{W}\label{wellformed}
  \sum_{w\in V} \delta^*(u,w,a_{\var(u)})\delta(v,w,a_{\var(v)}) 
  \ =\  \left\{\begin{array}{ll}
    1, & \mbox{if $u = v$; and}\\
    0, & \mbox{otherwise.}
    \end{array}\right.
\end{equation}
Second, $G$ has to be \emph{unidirectional}, which means that for each $w\in V$, 
all nodes $v\in V$ such that $\delta(v,w,b)\neq 0$ for some $b\in\{0,1\}$ are labeled by
the same variable.
\end{definition}

The well-formedness constraint implies that 
the QBP has a unitary time evolution operator (see below) 
and is, therefore, motivated by the
laws of quantum theory. Unidirectionality is a property that makes
understanding and manipulating models of quantum computation much
easier. We discuss this issue in more detail in Section~\ref{sec:qtms}.
Since unidirectionality is crucial for our simulations, we include
this requirement in the definitions of QBPs.
Next, we define the semantics of QBPs.

\begin{definition}[Computation of a QBP]\label{def:comp_qbps}
Let $G = (V,E)$ be a QBP on $n$ variables with start node~$s\in V$,
sinks~$F\subseteq V$, and transition amplitudes~$\delta$.
Let $\H = \C^{|V|}$ and let $(\k{v})_{v\in V}$ be an orthonormal basis of~$\H$. 
Let $a = (a_1,\ldots,a_n)$ be an assignment to the variables of~$G$. Let $L(a)$ be the linear 
transformation from the subspace spanned by all $\k{v}$, \mbox{$v\in V-F$}, into~$\H$ 
such that for $v\in V-F$,
\[
  L(a)\k{v} \ =\  \sum_{w\in V} \delta(v,w,a_{\var(v)})\k{w}.
\]
Due to the well-formedness constraint~(\ref{wellformed}), 
$L(a)$ can be extended to a unitary transformation~$U(a)$ on $\H$.
Call $U(a)$ a \emph{time evolution operator} of the QBP
for input~$a$. 
Define projection operators on~$\H$ by setting
\[
  E_{\rm cont} \ = \sum_{v\in V-F} \k{v}\b{v},\ \ 
  E_{\rm stop} \ =\  \sum_{v\in F} \k{v}\b{v},\ \ \mbox{and}\ \
  E_r \ = \!\!\!\!\!\sum_{v\in V,\,\lbl(v) = r}\!\!\!\!\!
  \k{v}\b{v},\;\;\text{for $r\in\{0,1,{\rm ?}\}$}.
\]
For $T\in\IN_0$ and $r\in\{0,1,{\rm ?}\}$ define
\[
  p_{G,\,r}(a,T) \ =\  \sum_{t=0}^T \bigl\Vert E_r (U(a)E_{\rm cont})^t \k{s}\bigr\Vert^2\quad\text{and}\quad
  p_{G,\,r}(a)   \ =\  p_{G,\,r}(a,\infty),
\]
the \emph{probability that $G$ outputs~$r$ for input~$a$ during the first $T$ time steps} and 
the \emph{(absolute) probability that $G$ outputs~$r$ for input~$a$}, resp.

{\sloppy
QBPs computing a function~$f\colon\{0,1\}^n\to\{0,1\}$ with \emph{unbounded error}, 
\emph{bounded (two-sided) error}, and \emph{one-sided error} are defined in the straightforward way.
We say that $G$ computes~$f$ with \emph{zero error and failure probability~$\epsilon$}, 
$0\le\epsilon < 1$, if $p_{G,\,\neg f(a)}(a) = 0$ and $p_{G,\,{\rm ?}}(a) \le \epsilon$ 
for all $a\in\{0,1\}^n$. We say that $G$ {\em computes~$f$ exactly} if it computes~$f$
with zero error and failure probability~$0$.

}

Let the \emph{(worst-case) running time of $G$ on $a$} be
\[
  T_G(a) \ =\  \min\{ T \mid T\in\IN_0\cup\{\infty\},\, p_{G,\,0}(a,T)+p_{G,\,1}(a,T)+p_{G,\,{\rm ?}}(a,T) = 1 \}.
\]
The running time can be in $\IN_0$, infinite, or undefined. 
The {\em expected running time of $G$ on $a$} is
defined by
\[
  \wbar{T}_G(a) \ =\  \sum_{t=0}^\infty t\cdot \bigl\Vert E_{\rm
  stop}(U(a) E_{\rm cont})^t \k{s}\bigr\Vert^2.
\]
We say that $G$ \emph{runs in time $T$} if $T_G(a)\le T$ for all
$a\in\{0,1\}^n$. Furthermore,  $G$ \emph{runs in expected time~$T$} 
if $\wbar{T}_G(a)\le T$ for all
$a\in\{0,1\}^n$.

\end{definition}

Since the QBP does not have edges leaving the sinks, the time
evolution operator is merely an extension of the mapping $L(a)$ and,
therefore, not necessarily uniquely determined.  

In the remainder of this section we discuss the relationship between (classical) BPs and QBPs,
and some variants of the definition of QBPs.
Because of the well-formedness and the unidirectionality requirements of QBPs
it is not obvious whether functions with small size BPs also have
small size QBPs. In order to prove such a statement, we introduce the
notion of reversibility.

\begin{definition}
A BP is {\em reversible} if each node is reachable
from at most one node $v$ by a $0$-edge and from at most one node
$w$ by a $1$-edge and $v$ and $w$ are labeled by the same variable.
\end{definition}

Reversible BPs are obviously special QBPs. Furthermore,
as proved by \v{S}palek~\cite{Spa02a} using a similar construction of Lange, McKenzie, and Tapp~\cite{Lan97}
for Turing machines, any (possibly non-reversible) BP of size~$s(n) = \Omega(n)$ 
can efficiently be simulated by a reversible one of size~$\poly(s(n))$. This implies:

\begin{proposition}[\cite{Spa02a}]
  \label{Prop_BPvsQBP}
  If the sequence of functions $(f_n)_{n\in\IN}$ has BPs $(G_n)_{n\in\IN}$ of size $s(n)=\Omega(n)$, it also has
  QBPs $(G'_n)_{n\in\IN}$ of size $\poly(s(n))$.
\end{proposition}

Adleman, DeMarrais, and Huang~\cite{Adl97} have shown that uniform QTMs with
arbitrary complex amplitudes can decide certain languages
of arbitrarily high Turing degree in polynomial time
and are thus too powerful to be realistic.
For randomized classical as well as quantum models of computation,
practical considerations (depending on the details of the physical
implementation of the model) lead to restrictions on the set of 
allowed amplitudes. 
However it is not obvious what a natural restriction in the nonuniform,
space-bounded scenario is. 
The following definition is motivated by the goal of finding the least 
restrictive definition that still allows the resulting QBPs to be simulated
efficiently by the corresponding standard QTM model. Recall that 
an \emph{algebraic number (over~$\Q$)} is an $x\in\C$ such that there
is a rational polynomial with root~$x$.

\begin{definition}\sloppy
\label{def:shortamplitudes}
A sequence $(G_n)_{n\in\IN}$ of QBPs has \emph{short amplitudes} if for some
number~$k$ independent from the input length there are algebraic 
numbers $\alpha_1,\ldots,\alpha_k$,  such that each amplitude
of each $G_n$ can be written as $p(\alpha_1,\ldots,\alpha_k)$ for some $k$-variate
rational polynomial~$p$ of degree $\poly(|G_n|)$ whose coefficients
are fractions with numerator and denominator each of bit length at
most $\poly(|G_n|)$.
\end{definition}

The requirements of this definition are obviously satisfied in the
special case that the sequence of QBPs uses only amplitudes from
a fixed, finite set of algebraic numbers. This is the situation investigated
for uniform, space-restricted QTMs by Watrous~\cite{Wat99a,Wat03}. 
Among other results, we show in Section~\ref{sec:sim} that
unidirectional nonuniform QTMs with algebraic amplitudes and QBPs 
with short amplitudes are equivalent in computational
power under space restrictions, which serves as a motivation for the
above definition. 

We conclude the discussion on reasonable restrictions for the
amplitudes with some simple observations. First, QBPs with complex
amplitudes can be transformed into equivalent QBPs with real
amplitudes, where the number of nodes increases by a factor of at most
$2$ (cf.~Proposition~5.3 in~\cite{Wat03}). The main idea is to replace each
node $v$ with two nodes $v_{\rm r}$ and $v_{\rm i}$ such that the
corresponding vectors $\k{v_{\rm r}}$ and $\k{v_{\rm i}}$  carry the
real and imaginary part of the amplitude of $\k{v}$, resp. Second, in
Definition~\ref{def:shortamplitudes} the number $k$ of algebraic
numbers can be replaced with $1$, since by the primitive element
theorem from algebra, the algebraic numbers
$\alpha_1,\ldots,\alpha_k$ can be represented as polynomials in a
single algebraic number $\alpha$. Since $k$ as well as $\alpha_1,\ldots,\alpha_k$ are
independent from the input size, these polynomials have a constant
number of constant
coefficients such that the resulting QBP still has short amplitudes.
Finally, since the bit lengths of the denominators of all coefficients
are bounded by ${\poly(|G_n|)}$ and the numbers of edges and,
therefore, the number of denominators is bounded by $2|G_n|^2$, all
the coefficients have a common denominator $m$ of bit length
${\poly(|G_n|)}$. We obtain the following result.

\begin{proposition}
\label{prop:simpleshortamps}
Each sequence $(G_n)_{n\in\IN}$ of QBPs with short amplitudes can be simulated
by a sequence $(G'_n)_{n\in\IN}$ of QBPs with $|G'_n|\leq 2|G_n|$ such that
there is a single algebraic number $\alpha$ and a number
$m=2^{\poly(|G'_n|)}$ such that each amplitude of $G_n'$ can be written
as $p(\alpha)/m$ for an integer polynomial $p$ with a degree bounded
by $\poly(|G'_n|)$ and coefficients bounded above in absolute value by
$2^{\poly(|G'_n|)}$. 
\end{proposition}

As for classical BPs, it is possible to simplify the structure
of QBPs without increasing their size too much.
The following has been observed by \v{S}palek~\cite{Spa02a}.

\begin{proposition}[\cite{Spa02a}]\label{Prop_leveled_QBP}
  Let $G$ be a QBP and let $t\in\IN_0$. 
  Then there is a leveled~QBP $G'$ with~$t+1$ levels that for each input~$a$
  computes an output~$r\in\{0,1,{\rm ?}\}$ with probability~$p_{G,r}(a,t)$
  after carrying out exactly~$t$ computation steps and that does not stop 
  before. The size of $G'$ is bounded above by $(t+1)^2|G|$.
\end{proposition}

For the construction of QBPs, it is convenient to allow {\em unlabeled
nodes} with an arbitrary number of outgoing edges carrying only
amplitude labels. An unlabeled node $v$ can be understood as an
abbreviation for a node that is labeled by some input
variable, where the value of this variable does not influence the
computation. This means that each edge leading from the unlabeled node
$v$ to $w$ has to
be replaced with a $0$-edge and a $1$-edge from $v$ to $w$ which both
have the same amplitude label as the original edge from $v$ to $w$. 
When using unlabeled nodes we have to make sure that the QBP resulting from this
transformation is unidirectional and well-formed.

\section{Definitions and Tools for Quantum Turing Machines}\label{sec:qtms}

{\sloppy
We first introduce a nonuniform variant of quantum Turing machines (QTMs).
The definition is similar to those of Bernstein and Vazirani~\cite{Ber97} and
Nishimura and Ozawa~\cite{Nis02a} for the uniform setting. 
Afterwards, we collect tools for approximately performing arbitrary unitary
transformations  by QTMs.

}

\begin{definition}
  \sloppy
  A \emph{nonuniform (or advice-taking) quantum Turing machine} is a
  QTM $M = (Q,\Sigma,\delta)$ together with an
  advice function $\adv\colon\IN\to\Sigma^*$, 
  where $Q$ is a finite set containing $q_0,q_f$ and  
  $\Sigma = \Sigma_1\times\cdots\times\Sigma_k$ with finite sets 
  $\Sigma_1,\ldots,\Sigma_k$ each containing $\{0,1,\mbox{?},\B\}$.
  The QTM~$M$ has the initial state~$q_0$ and the unique final 
  state~$q_f$, and ``$\B$'' is used as the blank symbol. The machine is equipped
  with three tapes, a read-only input tape, a read-only advice tape,
  and the work tape. All tapes are two-way infinite and indexed
  by~$\ZZ$ and each is split into $k$ separate tracks that may contain symbols from $\Sigma_1,\ldots,\Sigma_k$.  
  We have $\delta\colon (Q\times\Sigma^3)\times(Q\times\Sigma\times\{-1,0,1\}^3)\to\C$, and
  $\delta\bigl((q,\sigma_{\rm i},\sigma_{\rm a},\sigma_{\rm w}),
  (q',\sigma_{\rm w}',d_{\rm i},d_{\rm a},d_{\rm w})\bigr)$ is the amplitude for a
  transition from state~$q$, with symbols $\sigma_{\rm i},\sigma_{\rm a},\sigma_{\rm w}$ 
  on the input, advice, and work tape, resp., to
  state~$q'$, writing~$\sigma_{\rm w}'$ on the work tape and moving the heads
  on the three tapes according to $d_{\rm i},d_{\rm a},d_{\rm w}$.  
  Upon start of the machine, the input tape is loaded with the input
  string~$x\in\{0,1\}^*$ at positions~$0,\ldots,|x|-1$ of the first track. 
  The advice tape is loaded with the advice string $\adv(|x|)\in\Sigma^*$ at
  positions~$0,\ldots,|\adv(|x|)|-1$.  All other tape positions contain
  blanks, all heads are at position~$0$ and the finite control of $M$
  is in its initial state.  
  A \emph{configuration} of~$M$ is a tuple
  $(q,w,i,j,k)$, with the current state of the finite control $q\in
  Q$, the contents $w\in\Sigma^*$ of the work tape, and the positions
  $i,j,k\in\ZZ$ of the heads on the input, advice, and work tape,
  resp. Let $\Conf_n(M)$ be the set of all configurations of $M$ for
  inputs of length~$n$.
  Let $\H = \C^{|\Conf_n(M)|}$ be the Hilbert space spanned by all
  configurations from $\Conf_n(M)$, which we identify with vectors
  from an orthonormal basis. The {\em time evolution operator} $U(a)$
  describes the application of the transition function~$\delta$ to a
  superposition of configurations, where the input is $a$. The {\em
    well-formedness constraint} requires $U(a)$ to be unitary for all
  inputs $a$.
\end{definition}

\begin{definition}[Computation of a nonuniform QTM]
  Let $M = (Q,\Sigma,\delta)$ be as in the above definition.
  A QTM indicates stopping by entering~$q_f$ and signals its output 
  by an entry at position~$0$, called the \emph{output cell}, of a designated track of the 
  work tape, called the \emph{output track}.
  Define $E_{\rm state}(A)$ as the projection operator over $\H$
  onto the subspace spanned by all configurations with state in~$A\subseteq Q$.
  Then the projections $E_{\rm state}(\{q_f\})$, $E_{\rm state}(Q-\{q_f\})$ describe
  the measurement checking whether the current state is
  equal to~$q_f$. This measurement is performed before each
  computation step. If the QTM does not stop, $U(a)$ is applied to the
  state after the measurement. Let $E_{\rm result}(r)$, $r\in\{0,1,\mbox{?}\}$, be
  the projection onto the subspace spanned by the configurations with result~$r$ in
  the output cell. If stopping of the QTM has been detected, the
  measurement described by these latter projections is carried out in
  order to determine the result of the computation.
For $T\in\IN_0$ and $r\in\{0,1,\mbox{?}\}$, let
\begin{align*}
  p_{M,\,r}(a,T) &\ =\  
    \sum_{t=0}^T \bigl\Vert E_{\rm result}(r) E_{\rm state}(\{q_f\}) 
      (U(a)E_{\rm state}(Q-\{q_f\}))^t \k{s}\bigr\Vert^2
\end{align*}
be the \emph{probability that $M$ outputs~$r$ on input~$a$ during the first $T$ computation steps}.
Based on these probabilities, acceptance of the QTM with different types of error is 
defined as usual. The \emph{(expected) running time of $M$ on $a$}, 
denoted by $T_M(a)$ ($\wbar{T}_M(a)$), is defined analogously to QBPs (Definition~\ref{def:comp_qbps}).
The \emph{space used by~$M$ on input~$a\in\{0,1\}^*$}
is the maximum number of cells on the work tape between the leftmost and rightmost
non-blank symbol taken over all configurations which are reached with
nonzero amplitude during the computation on input~$a$ and in which the
machine has not yet halted.  The \emph{(total) space $s_M(a)$ used by~$M$ 
on input~$a\in\{0,1\}^*$} is defined as the sum of the space
on the work tape and $\lceil{\log|\adv(|a|)|}\rceil$.  Finally, we say
that \emph{$M$ runs in space $s\colon\IN\to\IN_0$} if for all
$a\in\{0,1\}^n$, $s_M(a) \le s(n)$.
\end{definition}

\begin{definition}
A {\em reversible Turing machine (RTM)} is a deterministic TM where
each configuration has at most one predecessor. 
A TM or QTM $M$ is called {\em unidirectional} if each state can be entered
from only one direction on each tape, i.\,e., if there are functions
$D_{\rm i}, D_{\rm a}, D_{\rm w}:Q\to\{-1,0,1\}$ such that 
$\delta\bigl((q,\sigma_{\rm i},\sigma_{\rm a},\sigma_{\rm w}),
  (q',\sigma'_{\rm w},d_{\rm i},d_{\rm a},d_{\rm w})\bigr)\neq 0$ only
  if $D_{\rm i}(q')=d_{\rm i}$,
  $D_{\rm a}(q')=d_{\rm a}$ and $D_{\rm w}(q')=d_{\rm w}$.
\end{definition}

Unidirectionality is a crucial property of QTMs that makes working with them much easier.
The property has first been investigated by Bernstein and Vazirani~\cite{Ber97}
for single-tape QTMs that are additionally \emph{two-way}, i.\,e., are required
to move their head in each computation step. Their results include that single-tape RTMs 
(even with stationary tape heads allowed) are automatically unidirectional and, furthermore, 
that single-tape two-way QTMs can be simulated time and space efficiently by unidirectional ones.
Furthermore, it is well known that also QTMs with stationary tape heads allowed
can be time efficiently simulated by unidirectional ones
using the simulations 
of QTMs by quantum circuits and vice versa
due to Yao~\cite{Yao93} and Nishimura and Ozawa~\cite{Nis02a}.
These results cannot be applied in an obvious way in
the space-bounded scenario.
Already for TMs with only one additional input tape, reversibility does no longer
imply unidirectionality, as simple examples show. In Section~\ref{sec:unidir} we show that 
general nonuniform QTMs with sublinear space can be space efficiently simulated by 
unidirectional ones.

For constructing unidirectional nonuniform QTMs, we need the usual
toolbox of programming primitives that allows us to work with multiple
tracks, combine TMs, construct looping TMs and so on. 
We use appropriate versions of lemmas for these tasks due to 
Bernstein and Vazirani~\cite{Ber97}. We only remark
that, by going through their proofs, it is straightforward to extend these
lemmas to unidirectional RTMs and unidirectional QTMs, resp., with an arbitrary number
of read-only input tapes. This includes nonuniform machines as a special case.

In simulations of other models of quantum computation by QTMs, we face the problem of
carrying out an arbitrary given unitary transformation over a finite-dimensional Hilbert space
using only a finite program for the QTM. For doing this,
we use a result due to Harrow, Recht, and Chuang~\cite{Har01} that
allows us to approximate
any unitary operator over a finite-di\-men\-sion\-al Hilbert space by a product of ``few'' 
elements from a finite collection of ``simple'' unitary transformations. The approximation
is with respect to the \emph{operator norm}, defined for an operator~$A$ over a Hilbert space~$\H$
by $\|A\| = \sup\{ \|Ax\| \mid x\in\H,\, \|x\|\le 1 \}$. We say that $A'$ is an 
\emph{$\epsilon$-approximation of~$A$} or \emph{approximates~$A$ with error~$\epsilon$} 
if $\Vert A'-A\Vert \le \epsilon$.

{\sloppy
Define the unitary matrices
{\def\I{\sqrt{\!\text{\small$-$}1\!}}
\begin{align*}
  V_1 = \frac{1}{\sqrt{5}}\begin{pmatrix} 1 & 2\I\\ 2\I & 1 \end{pmatrix},\;
  V_2 = \frac{1}{\sqrt{5}}\begin{pmatrix} 1 & 2\\ -2 & 1 \end{pmatrix},\;\text{and}\;
  V_3 = \frac{1}{\sqrt{5}}\begin{pmatrix} 1+2\I & 0\\ 0 & 1-2\I \end{pmatrix}.
\end{align*}}%
For $i\in\{1,2,3\}$ let $V_{i+3}=V_i^{-1}$.
Let 
${\cal G}_2 = \{ V_1,\ldots,V_6 \}$.
For $i\in\{1,\ldots,6\}$ and ${j\in\{1,\ldots,d-1\}}$ define 
the unitary $d\times d$-matrix $W_{i,j}$ by setting
\[
  W_{i,j}\k{k} \ =\ 
  \begin{cases}
    (V_i)_{1,1}\k{j} + (V_i)_{2,1}\k{j+1}, & \text{if $k=j$;}\\
    (V_i)_{1,2}\k{j} + (V_i)_{2,2}\k{j+1}, & \text{if $k=j+1$;}\\
    \k{k},                                 & \text{otherwise.}
  \end{cases}
\]}%
Let ${\cal G}_d$ be the set of all $W_{i,j}$ with
$i\in\{1,\ldots,6\}$ and $j\in\{1,\ldots,d-1\}$. Recall that $\SU(d)$
denotes the set of all unitary $d\times d$-matrices.
Harrow, Recht, and Chuang~\cite{Har01} have proved the following
lemma, where we have 
added the estimate of the  bound for $k$ depending on $d$, while
in~\cite{Har01} the dimension is regarded as a constant.

\begin{lemma}[\cite{Har01}]\label{lem:hrc}\label{lem:myhrc}
\sloppy
There is a constant $c > 0$ such that for all $\epsilon > 0$, $U\in\SU(d)$, and
${k = \lceil cd^2\log(d/\epsilon)\rceil}$, there are $U_1,\ldots,U_k\in{\cal G}_d$
such that ${\Vert U-U_1\cdots U_k\Vert \le \epsilon}$.
\end{lemma}

Call the matrices $W_{i,j}$ with $i\in\{1,\ldots,6\}$ and
$j\in\{1,\ldots,d-1\}$ \emph{elementary}.  Let $d=2^m$ and let
$\k{\psi}\in\C^d$ be encoded in $m = \log d$ qubits on the work tape
of a QTM.  Given~$i,j$ as additional inputs, we would like to
compute $W_{i,j}\k{\psi}$, as required for the application of Lemma~\ref{lem:hrc}.  
Bernstein and Vazirani~\cite{Ber97} have shown how to implement this for a different set of 
two-dimensional transformations. 
By an easy adaptation of their construction and an application of the 
simulation of single-tape two-way QTMs by unidirectional ones also from their paper, 
we obtain:

\begin{lemma}[\cite{Ber97}]\label{lem:machine_elem}\sloppy\hbadness=2000
There is a unidirectional single-tape QTM~$M_{\rm elem}$ with multiple tracks that works as follows.
Let $d = 2^m$ and let $\k{\psi}\in\C^d$ be a superposition of $m$~qubits.
Let $c(i,j)$ consist of the binary codes of $i\in\{1,\ldots,6\}$ and  $j\in\{1,\ldots,d-1\}$.
Started with $\k{\psi}$ in tape cells $0,\ldots,m-1$ of the first track and $\k{c(i,j)}$ in the 
tape cells $0,\ldots,|c(i,j)|-1$ of the second
track, $M_{\rm elem}$ computes the output $W_{i,j}\k{\psi}$ on the first track, replacing~$\k{\psi}$,
in time and space $O(m)$. Furthermore, the running time of $M_{\rm elem}$ only depends on $m$, the length of
the contents on the first track.
\end{lemma}

Combining Lemmas~\ref{lem:hrc} and~\ref{lem:machine_elem},
we can use a QTM to compute a good approximation of any desired 
finite-dimensional unitary transformation. We still have to make sure that measuring the 
state after applying the approximate transformation gives a result
that agrees with that after applying the original transformation
with high probability. 
This can be shown using the following statements. The first one is due
to Bernstein and Vazirani~\cite{Ber97}, the proof of second one is
analogous to that of a similar statement in~\cite{Nie00}, page~195. 

\begin{proposition}\label{prop:product_approx}\sloppy\hbadness=6000
Let $U$, $U_1,\ldots,U_n$, and $V_1,\ldots,V_n$ be operators over a 
Hilbert space~$\H$ with 
\mbox{$\|U_i\|,\,\|V_i\|\le 1$} and $\|U_i-V_i\| \le \epsilon_i$ for $i=1,\ldots,n$. 
Then $\Vert U_1\cdots U_n - V_1\cdots V_n \Vert \le \epsilon_1+\cdots+\epsilon_n$.
\end{proposition}

\begin{lemma}\label{lem:approx_sim}
Let $\epsilon > 0$ and $t\in\IN$.
Let $U$ and $U'$ be unitary operators over a Hilbert space~$\H$
with $\|U-U'\|\le\epsilon$. 
Let $P,Q$ be projections over~$\H$. Let $\k{v}\in\H$ with $\|\k{v}\|=1$. 
Define $p = \|Q(UP)^t\k{v}\|^2$ and $p' = \|Q(U'P)^t\k{v}\|^2$.
Then $|p-p'| \le 2t\epsilon$.
\end{lemma}


\section{\!\!Equivalence of QBPs and Space-Bounded\,Unidirectional\,Nonuniform\,QTMs}\label{sec:sim}

We prove our simulation results for QBPs and unidirectional nonuniform QTMs.
We first provide a basic theorem that allows a step-by-step simulation of unidirectional nonuniform
QTMs by QBPs and vice versa. Each step of a QBP can only be done
approximately by a unidirectional nonuniform QTM. 
In order to control the total error, 
we have to specify the number of simulation steps in advance.
This raises the problem of bounding the computation time of space-bounded
algorithms that is studied afterwards. We first define a suitable
notion of simulations. 

\begin{definition}\label{def:sim}
Let $M_1,M_2$ be nonuniform QTMs or QBPs. As defined in
Sections~\ref{sec:qbps} and~\ref{sec:qtms}, let
$p_{M_i,r}(a,T)$ be the probability that $M_i$ computes the output $r$
on the input $a$ during the first $T$ computation steps. We say that
{\em $M_1$ simulates $T$ steps of $M_2$ in $T'$ steps with accuracy~$\epsilon\ge 0$},
if for all $a\in\{0,1\}^*$ and $r\in\{0,1,?\}$:
$\mbox{$|p_{M_1,\,r}(a,T') - p_{M_2,\,r}(a,T)| \le \epsilon$}$. We say
that {\em $M_1$ simulates $M_2$} if $M_1$ simulates $T$ steps of $M_2$
in the same number of steps with accuracy $\epsilon=0$ for arbitrary~$T$.
\end{definition}

\subsection{Basic Step-by-Step Simulations}

{\samepage
\begin{theorem}\label{the:sim}\item[]\sloppy
\begin{shortindent}{(ii)}
\item[(i)]
  Let $M$ be a unidirectional nonuniform QTM that runs in space $S(n) = \Omega(\log n)$. 
  Then there is a sequence of QBPs $(G_n)_{n\in\IN}$ with
  $|G_n|=2^{O(S(n))}$ that simulate~$M$. 
\item[(ii)]
  Let $(G_n)_{n\in\IN}$ be a sequence of QBPs with $|G_n| = \Omega(n)$.
  Let $\epsilon\colon\IN\to(0,1)$ and
  ${T\colon\IN\to\IN_0}$. Then there is a unidirectional nonuniform QTM
  that for each $n\in\IN$ simulates $T(n)$
  steps of $G_n$ in $\poly(|G_n|,T(n),\log(1/\epsilon(n)))$ steps with
  accuracy~$\epsilon(n)$ and runs in space \mbox{$O(\log|G_n|+\log\log (T(n)/\epsilon(n)))$}.
\end{shortindent}
\end{theorem}

}

We discuss the consequences of this 
theorem for the motivation of our QBP model and the relationship between QBPs and QTMs 
in detail in Section~\ref{subsec:prec_and_rt}.

\begin{proof}[Proof of Theorem~\ref{the:sim}, Part~(i)]
This follows by an easy adaptation of the proof of
the analogous result for classical BPs and TMs.
Let $M = (Q,\Sigma,\delta)$ be a unidirectional nonuniform QTM 
with advice function ${\adv\colon\IN\to\Sigma^*}$ that
runs in space $S(n) = \Omega(\log n)$.
We ensure that the heads on the input and advice tape stay in the area 
consisting of the non-blank cells (see~\cite{Wat98b} 
for details). Then $M$ has at most $2^{O(S(n))}$ configurations. 

We construct the QBP $G$
over the variable set $X = \{ x_0,\ldots,x_{n-1} \}$
with $\Conf_n(M)$ as its node set.
For a configuration $c\in \Conf_n(M)$ of $M$ where the
head on the input tape is at position~$i\in\{0,\ldots,n-1\}$,
define $\var(c) = i$ in $G$ (recall that $\var(c)$ denotes the index
of the variable with which a QBP node is labeled). 
For an input with bit $b\in\{0,1\}$ at position~$i$ on
the input tape of $M$, let the application of the transition 
function $\delta$ of $M$ to $\k{c}$ yield the superposition
\[
  \sum_{c'\in\Conf_n(M)} \alpha(c,c',b) \k{c'},\quad \alpha(c,c',b)\in\C.
\]
For each $\alpha(c,c',b) \neq 0$, we add a $b$-edge from~$c$ to~$c'$ 
in $G$ and use $\alpha(c,c',b)$ as the amplitude label of this edge.
We define the start node of~$G$ as the initial configuration of~$M$
and identify the set of final nodes~$F$ with the set of final 
configurations of~$M$.

The graph $G$ defined above fulfills the well-formedness requirement
of QBPs since the time evolution operator of the QTM~$M$ is unitary.
In order to prove that $G$ is unidirectional assume for a contradiction that the node $v$
has predecessors $v_1$ and $v_2$ labeled by different variables. Then
during the transitions of $M$ that correspond to the transition of
$v_1$ to $v$ and $v_2$ to $v$ the head on the input tape makes
different moves in contradiction to the unidirectionality of $M$.
Since $|\Conf_n(M)| = 2^{O(S(n))}$, the
branching program is of the required size. It is easy to verify that
$G$ simulates $M$ because of the similarity of the definitions of the
semantics for the two models. 
\end{proof}

\begin{proof}[Proof of Theorem~\ref{the:sim}, Part~(ii)] 
Let $G$ be the QBP to be simulated and let $X = \{ x_0,\ldots,x_{n-1}\}$ be the
variable set of~$G$. In a first step, we show how to transform $G$
into an equivalent QBP $G'$ which has the additional property that
all nodes that are reachable from the start node by a path of length
$t$ are labeled by $x_{t\bmod n}$. This allows us to decompose the
time evolution operator into~$n$ factors where each factor only depends on
the value of one of the variables.  In a second step we construct a
nonuniform QTM and its advice string from the decomposed time
evolution operator of~$G'$ and prove the claims on the resources
required by this QTM.

Let $G = (V,E)$ and let $s$ and $F$ denote the start node and the set of sinks of $G$, resp.
Due to the unidirectionality of~$G$, all predecessors of a node $v\in V$ are labeled
by the same variable, whose index is  denoted by $\pre(v)$. If the start node does not have any predecessor, let $\pre(s) = n-1$.
Furthermore, we set $\var(v) = 0$ for $v\in F$. 

{\sloppy\hbadness=3700
We construct the QBP $G'=(V',E')$ from~$G$ by adding dummy nodes. Let
$V' = \bigl{\{ (v,i) \bigm| v\in V, i\in\{ {\pre(v)+1},\ldots,{n-1},0,\ldots,\var(v)\} \bigr\}}$.
Let $s' = (s,0)$ be the start node of~$G'$ and let $F' = \{ (v,0) \mid v\in F\}$ be its set of sinks.
Define ${\var(v,i) = i}$ for all $v\in V$ and $\lbl(v,0) = \lbl(v)$ for all $v\in F$.
For each $(v,w)\in E$, add an edge $((v,\var(v)),(w,(\pre(w)+1)\bmod n))$ to~$E'$ that inherits all labels
of the edge~$(v,w)$. Furthermore, for each $w\in V$, ${i\in\{\pre(w)+1,\ldots,n-1,0,\ldots,\var(w)-1\}}$, and $b\in\{0,1\}$,
add an edge $((w,i),(w,(i+1)\bmod n))$ to~$E'$ with boolean label~$b$ and amplitude~$1$. 
Let $\delta'$ be the transition amplitudes of~$G'$ defined in this way.
It is easy to see that $G'$ is a QBP. Well-formedness and unidirectionality of~$G'$ follow from the respective 
properties of~$G$ for the subgraph induced by the nodes in $\{
(v,\var(v)), (v,(\pre(v)+1)\bmod n) \mid v\in V\}$ 
and are obvious for the rest of the graph. 
It is easy to see that ${|G'| = O(n|G|)}$.

}

\begin{claim*}
$G'$ simulates $T$ steps of $G$ in $nT$ steps with accuracy $0$. Furthermore,
there are unitary operators $U_i(b)$ with $0\leq i\leq n-1$ and
$b\in\{0,1\}$ such that for any time evolution operator $U'(a)$ of
$G'$ with $a\in\{0,1\}^n$, the projection $E'_{\rm cont}$ to the space
spanned by the non-sink nodes of $G'$, the start node $s'$ of $G'$,
and any $T\in\IN_0$, $(U'(a)E'_{\rm
  cont})^{nT}\k{s'}={\bigl((U_{n-1}(a_{n-1})\bdot\cdots\bdot
  U_0(a_0))E_{\rm cont}'\bigr)^T\k{s'}}$. 
\end{claim*}

\begin{proof}[Proof of the claim]
 For the proof that $G'$ simulates $T$ steps of~$G$
with $nT$ of its own steps, let~$\phi$ be the linear embedding of the superpositions
of~$G$ into those of~$G'$ induced by setting $\phi(\k{v}) = \k{(v,0)}$ for $v\in V$.
Let $U(a)$ and $U'(a)$ be time evolution operators of~$G$ and~$G'$, resp.,
for the input~$a\in\{0,1\}^n$. Let $E_{\rm cont}$, $E_r$ and $E_{\rm cont}'$, $E_r'$
be the projections to the spaces spanned by the non-sink nodes and nodes with output label~$r$, resp., 
for the graphs~$G$ and~$G'$, resp. An easy induction shows that for each $T\in\IN_0$,
$(U'(a) E_{\rm cont}')^{nT}\k{s'} = \phi\bigl((U(a)E_{\rm cont})^T\k{s}\bigr)$.
Furthermore, $E_r'\phi(\k{v}) = \phi(E_r\k{v})$ for all $v\in V$. Hence,
$p_{G',r}(a,nT) = p_{G,r}(a,T)$ for all~$T\in\IN_0$ and $G'$ simulates $T$ steps of~$G$
with~$nT$ steps.

Furthermore, it is also easy to prove by induction that for any~$T\in\IN_0$, $i = T\bmod n$, 
and any $v\in V'-F'$ with $\var(v)\neq i$, 
$\b{v}E_{\rm cont}'(U'(a)E_{\rm cont}')^T\k{s'} = \b{v}(U'(a)E_{\rm cont}')^T\k{s'} = 0$.
Hence, instead of applying~$U'(a)$ in the $(T+1)$-st computation step, we may
apply a unitary extension $U_i(a_i)$ of the mapping defined by
$\k{v}\mapsto \sum_{w\in V'}\delta'(v,w,a_i)\k{w}$ for $v\in V'$ with $\var(v)=i$, 
without changing the computed superposition. 
Finally, for all $v\in F'$ and ${T\bmod n\neq 0}$, we have
$\bk{v}{(U'(a)E'_{\rm cont})^T\,|\,s'}=0$.
By induction, it follows that for any~${T\in\IN_0}$,
$(U'(a)E_{\rm cont}')^{nT}\k{s'} = \bigl((U_{n-1}(a_{n-1})\bdot\cdots\bdot U_0(a_0))E_{\rm cont}'\bigr)^T\k{s'}$,
as claimed.
\end{proof}

Now we describe the second step of the proof, the construction of the
QTM from $G'$.
Let $s=O(n|G|)$ be the number of nodes of $G'$. Let
$m=\ceil{\log s}$.  It is convenient to assume that the node
numbers have the length $m+2$, where the numbers of interior nodes begin
with $00$ and the numbers of $0$- and $1$-sinks with $01$ and $11$,
resp.  Furthermore, we assume that the start node has the number $0$.

\medskip
{\em Construction of the advice string.}
First, we define approximate representations for each matrix $U_i(b)$, $0\le i\le n-1$ and $b\in\{0,1\}$,
as a list of elementary matrices using Lemma~\ref{lem:hrc}.
Choosing $\epsilon' ={\epsilon}/(2nT^2)$ as the error bound and $s$ as the dimension of the Hilbert space,
Lemma~\ref{lem:hrc} yields $s\times s$-matrices $U_{i,0}(b),\ldots,U_{i,k-1}(b)$ whose product 
is an $\epsilon'$-approximation of~$U_i(b)$, where $k = O(s^2\log(s/\epsilon'))=O(s^2\log (nsT/\epsilon))$
is the number of matrices obtained from the lemma.
Observe that the number of elementary matrices in the representation of~$U_i(b)$
is the same for all $i$ and $b$. 
Elementary matrices are encoded such that the corresponding unitary transformations 
can be applied using the QTM provided in Lemma~\ref{lem:machine_elem}. The code for an elementary matrix
$W_{j,j'}$ consists of the binary codes of $j\in\{1,\ldots,6\}$ and 
 $j'\in\{1,\ldots,s-1\}$. 

On the advice tape, we store the codes of the elementary matrices $U_{i,\ell}(b)$ for $0\le i\le n-1$,
$b\in\{0,1\}$, and $\ell\in\{0,\ldots,k-1\}$, as well as some additional administrative information.
The information is organized using four tracks, where the non-blank part of each track starts at position~$0$:
\begin{shortindent}{Track 5: }\itemsep=0.25\baselineskip
\item[{\bf Track 1:}] Binary code of the input length~$n$.
\item[{\bf Track 2:}] Binary code of $k$.
\item[{\bf Track 3:}] Binary code of the length of the code for an elementary matrix.
\item[{\bf Track 4:}] List of codes for all $U_{i,\ell}(b)$.
\end{shortindent}

{\sloppy
The length of the code of each elementary matrix is $O(\log s)$. Each of the $2n$ matrices $U_i(0)$ and $U_i(1)$
is encoded using $O(k\log s)$ bits. We have $k = O(s^2\log (nsT/\epsilon))$. 
Hence, the length of the information on track~$4$ is bounded by
$O(2n\cdot k\log s) = \poly(s,\log(T/\epsilon))$,
which is also a bound on the overall length of the advice string.
The logarithm of this, ${O(\log s+\log\log(T/\epsilon))} = {O(\log|G|+\log\log(T/\epsilon))}$, is
the contribution of the advice tape to the space.

}

\medskip
{\em Construction of the QTM.}
The QTM uses the following tracks on the work tape:
\begin{shortindent}{Track 6: }\itemsep=0.25\baselineskip
\item[{\bf Track 1:}] Output track. The output of the QTM is in cell~$0$ of this track upon termination.
\item[{\bf Track 2:}] Node register consisting of $m+2$ cells that contains the current superposition of node numbers of $G'$.
\item[{\bf Track 3:}] Buffer for the code of $U_{i,\ell}(x_i)$.
\item[{\bf Track 4:}] Counter~$i$ with values in $\{0,\ldots,n-1\}$.
\item[{\bf Track 5:}] Counter~$\ell$ with values in $\{0,\ldots,k-1\}$.
\item[{\bf Track 6:}] Buffer for the value of the current input bit.
\item[{\bf Track 7:}] Buffer for the position of the currently applied $U_{i,\ell}(x_i)$ on the advice tape.
\end{shortindent}

Initially, the work tape only contains blanks. By choosing an
appropriate encoding of binary numbers 
(see, e.\,g.,~\cite{Wat99}), we ensure that a string of blanks
represents the number~$0$. Hence, the counters
on track~4 and track~5 are initialized with~$0$. Since the start node has the number $0$, the blanks from the initialization of the node register encode the start node.

\begin{figure}\normalsize\fboxsep=10pt
{\centering\fbox{\parbox{0.99\textwidth}{\normalsize
\vspace*{-\baselineskip}
\begin{tabbing}
10.\=\quad\=\kill
1. \> Forever do\\[2pt]
2. \> \>
\begin{minipage}[t]{13.7cm}
{\em Termination check.} Swap the contents of cell $0$ of
  the node register (signaling the
  output if the current node is a sink)
  and the output cell. If cell $1$ of the node
  register contains a $1$ (signaling a sink), enter $q_f$. Otherwise,
  swap again the contents of cell $0$ of the node register and the
  output cell.\vspace*{0.25\baselineskip}
  \end{minipage}\\[2pt]
3. \> \> For \=$i:=0$ to $n-1$ do\\[2pt]
4. \> \>     \> XOR track 6 with the value of $x_i$.\\[2pt]
5. \> \>     \> For \=$\ell:=0$ to $k-1$ do\\[2pt]
6. \> \>     \>    \> XOR track 7 with the position of the code of~$U_{i,\ell}(x_i)$ on the advice tape.\\[2pt]
7. \> \>     \>    \> XOR track 3 with the code of $U_{i,\ell}(x_i)$ from the advice tape.\\[2pt]
8. \> \>     \>    \> Apply $U_{i,\ell}(x_i)$ to the node register.\\[2pt]
9. \> \>     \>    \> Repeat step 7; this erases track 3.\\[2pt]
10.\> \>     \>    \> Repeat step 6; this erases track 7.\\[2pt]
11.\> \>     \> Repeat step~4; this erases track 6.
\end{tabbing}
\vspace*{-\baselineskip}
}}}
\hfuzz5pt 
\caption{Algorithm for the nonuniform QTM simulating $G'$.}\label{fig:sim_qtm}
\end{figure}

The algorithm performed by the QTM is shown in Figure~\ref{fig:sim_qtm}. 
The algorithm consists of an infinite loop whose body, steps~2--11, simulates
one computation step of the QBP~$G'$. The loop is left and the algorithm terminates in step~2
if a sink has been reached. We only bother to simulate the first~$nT$ computation steps of~$G'$
and thus the first~$T$ computation steps of~$G$ with sufficient accuracy.
In the following, we describe how this algorithm is implemented.

We construct unidirectional RTMs for steps 2, 4, 6, and~7 with the following additional 
properties. We ensure that these machines only use the space already allotted on the work tape, 
that the time can be bounded by~$O(1)$ and $O(n)$ for step~2 and~4, resp., and by a polynomial in  
the length of the advice tape, i.\,e., $\poly(s,\log(T/\epsilon))$, for steps~6 and~7.
For step~2, we additionally take care that the running time only
depends on the length of the node register, but not on the actual contents of the node register.
It is not hard to construct these machines from scratch. 
Furthermore, Lemma~\ref{lem:machine_elem} yields a unidirectional QTM for step~8 that has
space and running time bounded by the length of the node register, i.\,e., $O(\log s)$ and 
whose running time is independent of the actual contents of the node register.

For constructing the final QTM from these basic RTMs, we apply appropriate versions of the lemmas 
of Bernstein and Vazirani~\cite{Ber97} for dealing with unidirectional nonuniform RTMs
and unidirectional nonuniform QTMs. The finite loops are realized as described by Watrous~\cite{Wat99}.
At the beginning of a loop, we check a starting/stopping condition for the loop
and switch the state of being outside or inside the loop, resp., when this condition 
is met. For the loops beginning in step~3 and~5, we use counters modulo~$n$ and~$k$, resp.,
and check as the starting/stopping condition whether the counter is equal to zero. 

Using these tools, we first combine the machines for the steps~4 and 6--11, implementing the loops in step~3 and~5 
as described above, to get a QTM $M_{\text{3--11}}$ for steps 3--11. The outermost, endless loop is then realized 
by modifying the RTM  for step~2. We use a simple unidirectional RTM constructed from scratch that 
carries out the described termination check, enters two special states as placeholders depending
on the value of cell~1 of the node register, and then restarts its computation.
We insert $M_{\text{3--11}}$ into the state for the value~$0$ of cell~$1$ (non-sink) and replace the state 
for the value~$1$ (sink) with the final state~$q_f$ of the whole QTM. This yields
the desired QTM for simulating~$G'$ and thus~$G$.

We note that a space-bounded RTM performing an infinite loop cannot carry out initialization 
steps before the loop. By our choice of the encoding of the contents of the tracks, we do not 
need such an initialization. Furthermore, we have ensured that the running time for the body
of the outermost loop is the same for all possible classical inscriptions in the node register. 
Hence, even if the simulated QBP is in a superposition, step~2 is always reached simultaneously
for all nodes in the superposition.

{\sloppy\hbadness=1600
{\em Space and time requirements}. The space on tracks~1--6 of the work tape
is obviously bounded  by $O(1)$, $O(\log s)$, $O(\log s)$, $O(\log n)$, 
$O(\log k)=O(\log s+\log \log (T/\epsilon))$, and $O(1)$, resp. 
The space on track~7 is bounded above by the logarithm of the length of the advice string,
which is $ {O(\log s+\log\log(T/\epsilon))}$ as computed above.
Since this is also the contribution of the advice string to the space, 
the overall space complexity is of the same order.
We can estimate the running time for simulating one computation step of~$G'$ (steps~2--11
of the algorithm) as follows. The running time of steps~4 and~11 is $O(n)$. 
The running time of steps~6, 7, 9, and~10 is dominated by the length of 
the advice tape, which is of order $\poly(s,\log(T/\epsilon))$. 
Step~8 can be performed in time proportional to the length of
the node register, i.\,e., $O(\log s)$. Hence, also the overall time for one
computation step is of order $\poly(s,\log(T/\epsilon)) = \poly(|G|,\log(T/\epsilon))$.

}

{\em Correctness.}  Let us assume for a moment that the product
$U_{i,k-1}(x_i)\cdots U_{i,0}(x_i)$ equals~$U_i(x_i)$. 
Then it is easy to see that steps 4--10 exactly apply $U_i(x_i)$ and that
steps 3--11 exactly apply $U_{n-1}(x_{n-1})\cdots U_0(x_0)$ to the node register. 
Together with the termination check in step~2 which realizes 
the projection $E_{\rm cont}'$ to the non-sink nodes of~$G'$,
steps~2--11 exactly apply $U_{n-1}(x_{n-1})\cdots U_0(x_0) E_{\rm cont}'$
to the node register if the QTM does not stop. Due to the above claim, we know that
this simulates $n$ successive computation steps of~$G'$ and thus one 
computation step of the original QBP~$G$. 

\sloppy
However, the product
$U_{i,k-1}(x_i)\cdots U_{i,0}(x_i)$ is merely an
$\epsilon'$-approximation of $U_i(x_i)$, where $\epsilon' ={\epsilon}/(2nT^2)$.
By Proposition~\ref{prop:product_approx} we may estimate the error in the
application of $U_0(x_0),\ldots,U_{n-1}(x_{n-1})$ by $n\epsilon'$. 
 Let $\hat{p}_{G,r}(a,t)$ be the probability that $G$ halts after
{\em exactly}~$t$ steps at a sink labeled by $r\in\{0,1,\mbox{?}\}$.  Let
$\hat{p}_{M,r}(a,t)$ be the probability that~$M$ halts after {\em exactly}~$t$
iterations of steps 2--11 and outputs~$r$.  As remarked
above, the error of one iteration of the outer loop is bounded by
$n\epsilon'$. By Lemma~\ref{lem:approx_sim}, $|\hat{p}_{G,r}(a,t) -
\hat{p}_{M,r}(a,t)| \le 2t\epsilon'n \le \epsilon/T$ for all
$t=0,\ldots,T$. Hence,
\[
  \Bigl|\,\sum_{t=0}^T \hat{p}_{G,r}(a,t) - \sum_{t=0}^T \hat{p}_{M,r}(a,t)\,\Bigr|  
  \ \le\  \sum_{t=0}^T |\hat{p}_{G,r}(a,t) - \hat{p}_{M,r}(a,t)|
  \ \le\  \epsilon.
\]
Altogether, we have proved that $M$ simulates $T$ steps of  $G$ in
$\poly(|G|,T,\log(1/\epsilon))$ steps with accuracy~$\epsilon$.
\end{proof}

\subsection{High-Level Simulation Theorems}
\label{subsec:prec_and_rt}

Here we use the basic, technical simulations from the last subsection for proving
that the logarithm of the size of QBPs and the space complexity of QTMs
asymptotically agree for the standard models of QBPs and QTMs.
On the way, we investigate the relationship between precision and running time for QBPs.
All proofs are given in Section~\ref{subsec:proofs_prec}.
We assume throughout this subsection that the logarithm of the size of the 
considered QBPs and the space complexity of the QTMs are at least logarithmic in
the input length.

We begin with a simple corollary from the basic simulations.
If we want to apply the approximate simulation of QBPs by QTMs,
we have to specify a bound~$\epsilon$ on the simulation error and 
a bound~$T$ on the number of simulation steps in advance. These 
parameters turn up in a term of $O(\log\log(T/\epsilon))$ in the space 
complexity of the simulating machine. If we restrict ourselves to
bounded error computation and to exponential running time, 
Theorem~\ref{the:sim}  immediately yields:

\begin{corollary}\label{cor:equiv_i}
The logarithm of the size of QBPs and the space complexity of 
unidirectional nonuniform QTMs are asymptotically equal if 
both models are restricted to bounded error and exponential 
running time in the worst case.
Furthermore, the classes of functions computable by sequences of
QBPs with polynomial size and by unidirectional nonuniform QTMs 
with logarithmic space are the same if both models are restricted 
to bounded error and polynomial running time. 
\end{corollary}

It is obviously practically motivated to work with bounded running time, 
but it is not clear what kind of bounds can be chosen without 
restricting the computational power of the space-bounded models
considered here. In~\cite{Wat98b} and implicitly also in~\cite{Wat99}, 
Watrous has investigated this question for
unidirectional uniform QTMs and has obtained answers analogous to the 
situation for probabilistic TMs.
He has shown that unidirectional uniform QTMs with rational amplitudes and running 
in space $S(n) = \Omega(\log n)$ have an expected running time that is at most 
doubly exponential in~$S(n)$. 
This result can be extended to unidirectional uniform QTMs with algebraic amplitudes 
using the ideas from his later papers~\cite{Wat99a,Wat03}.

These considerations provide the motivation to look at the relationship between the precision 
allowed for the amplitudes and the running time also for the nonuniform model of QBPs.
In turns out that short amplitudes take over a role analogous to 
algebraic amplitudes for QTMs.

\begin{theorem}\label{th:equiv_1}\item[]
\begin{shortindent}{(ii)}
\item[(i)]  Sequences of QBPs $(G_n)_{n\in\IN}$ with bounded error and short amplitudes and
  sequences of QBPs $(G_n')_{n\in\IN}$  with bounded error and expected running time
  $2^{\poly(|G_n'|)}$ have polynomially related size complexities.
\item[(ii)]  Sequences of QBPs $(G_n)_{n\in\IN}$ with unbounded error and short
  amplitudes can be simulated by sequences of QBPs $(G_n')_{n\in\IN}$ of size
  $\poly(|G_n|)$ and with expected running time $2^{\poly(|G_n'|)}$.
\end{shortindent}
\end{theorem}

Our final and main result of this subsection provides a justification to
regard QBPs with short amplitudes as the natural standard variant of the model analogous
to QTMs with algebraic amplitudes.

\begin{theorem}\label{th:equiv_3}
  The logarithm of the size of QBPs with bounded or unbounded error and short
  amplitudes and the space complexity of unidirectional nonuniform QTMs with algebraic
  amplitudes and the same type of error are asymptotically equal.
\end{theorem}

\subsection{Proofs of Theorems~\ref{th:equiv_1} and~\ref{th:equiv_3}}
\label{subsec:proofs_prec}

For the proofs of the theorems we need a couple of technical lemmas,
which are concerned with the analysis of a matrix series that
describes the acceptance probability of a QBP.  Using these lemmas we
provide two results on QBPs with short amplitudes, which are the basic
tools for proving Theorems~\ref{th:equiv_1} and~\ref{th:equiv_3}.  First,
even in the case of unbounded error there is some gap between the
error probability and $1/2$. Second, in QBPs with short amplitudes
a probabilistic clock can be added by which computations
lasting too long are aborted.

For the following, consider an arbitrary QBP~$G$ with $s$ nodes. 
For any fixed input~$a$ 
for~$G$ let $U = U(a)$ be a unitary time evolution matrix of
$G$. Recall that $E_{\rm cont}$ is the
projection operator in the measurement of the output label which belongs
to the result ``no label.'' Let $D = UE_{\rm cont}$ and $M = \wbar{D}\otimes D$,
where $\wbar{D}$ denotes the matrix obtained from $D$ by taking
the complex conjugate of each of its entries. Let $N = s^2$ denote
the dimension of~$M$ and let $\k{1},\ldots,\k{N}$ be the standard basis
of~$\C^N$. For $v\in\{1,\ldots,s\}$, define $i_v = v+s(v-1)\in\{1,\ldots,N\}$. 
Then, for any $v,w\in\{1,\ldots,s\}$, 
$M_{i_w,i_v} = (\b{w}\otimes\b{w})M(\k{v}\otimes\k{v})$.

\goodbreak
{\samepage
\begin{lemma}\item[]
\begin{shortindent}{(ii)}
\item[(i)]
  The probability that the node~$w$ is reached after
  exactly $k$ computation steps in~$G$ when starting at the node~$v$ is 
  equal to~$(M^k)_{i_w,i_v}$.
\item[(ii)]
  The absolute value of each eigenvalue of~$M$ is bounded above by~$1$.
\end{shortindent}
\end{lemma}

}

\begin{proof}
Part~(i) follows from $
  (M^k)_{i_w,i_v} = (\b{w}\otimes\b{w}) (\wbar{D}^k \otimes D^k) (\k{v}\otimes\k{v}) 
  = (\wbar{D}^k)_{w,v}\cdot (D^k)_{w,v} = |((UE_{\rm cont})^k)_{w,v}|^2$,
which is obviously the desired probability. 

\sloppy\hbadness=1500
For part~(ii) it suffices to prove that $\Vert M\Vert \le 1$, since $\Vert M\Vert$ provides
an upper bound on the absolute value of the eigenvalues of~$M$
(see, e.\,g., \cite{Hou65}, page~45). 
We have $M^{\dag} M = {(\wbar{D}\otimes D)^{\dag}(\wbar{D}\otimes D)} 
= {((\wbar{D})^{\dag}\wbar{D})\otimes (D^{\dag} D)}$.
Furthermore, $D^{\dag} D = (UE_{\rm cont})^{\dag} (UE_{\rm cont}) = E_{\rm cont}^{\dag} E_{\rm cont} = E_{\rm cont}$. The eigenvalues
of $D^{\dag} D$ are thus from $\{0,1\}$,
and the same holds for $(\wbar{D})^{\dag}\wbar{D}$.
Since the eigenvalues of $M^{\dag} M$ are obtained as products of the
eigenvalues of $(\wbar{D})^{\dag}\wbar{D}$ and $D^{\dag} D$, it
follows that ${\Vert M\Vert \le 1}$.
\end{proof}

The above lemma yields that, for each pair of nodes $(v,w)$ in $G$, 
$\lim_{k\to\infty} \bigl(\sum_{\ell=0}^k M^{\ell}\bigr)_{i_w,i_v}$
is the probability of reaching node~$w$ from node~$v$ in~$G$. 
In particular, the acceptance probability of~$G$ can expressed as the sum of all
such terms where~$v$ is the start node and~$w$ is a $1$-sink.

We use the technique of Watrous~\cite{Wat99,Wat99a,Wat03} to 
analyze the series $\bigl(\sum_{\ell=0}^{\infty} M^{\ell}\bigr)_{i_w,i_v}$.
Since the matrix series $\sum_{\ell=0}^{\infty} M^{\ell}$ does not
converge in general, we look at the series $\sum_{\ell=0}^{\infty} (zM)^{\ell}$ for
some $z\in[0,1)$ instead and let $z$ tend to~$1$ afterwards. Using the restrictions
on the involved numbers, we then show two facts:
First, $\lim_{z\uparrow 1}\bigl(\sum_{\ell=0}^{\infty} (zM)^{\ell}\bigr)_{i_w,i_v}$
can be approximated with sufficient precision
by choosing $z = 1 - 2^{-\poly(N)}$. Second, if the limit 
$\bigl(\sum_{\ell=0}^{\infty} M^{\ell}\bigr)_{i_w,i_v}$ is not exactly~$1/2$,
then it can be bounded away from~$1/2$ by a gap of size at least $2^{-\poly(N)}$.

For a multivariate polynomial~$f$, the \emph{height of~$f$}, denoted by $\|f\|$,
is the maximum absolute value of any of its coefficients and 
$\deg(f)$ is the maximum degree of~$f$ with respect to any of its variables.
Using the form of the entries of~$U = U(a)$ obtained by Proposition~\ref{prop:simpleshortamps},
it is easy to see that there is a real algebraic number~$\alpha$ not depending on~$N$ and 
a number $m = 2^{\poly(N)}$ such that each entry of~$M = \overline{UE_{\rm cont}}\otimes UE_{\rm cont}$ 
can be written as $p(\alpha)/m$ for an integer polynomial~$p$ with $\deg(p) = \poly(N)$ and $\|p\| = 2^{\poly(N)}$. 
The following three technical lemmas yield properties of general 
matrices of this form (not necessarily derived from QBPs). 
The first two lemmas are extracted from~\cite{Wat03} 
(Lemma~4.6 and its proof and the beginning of the proof of Lemma~4.2, resp.).

\begin{lemma}[\cite{Wat03}]\label{lem:wat_gaps_low}\sloppy
Let $\alpha$ be any real algebraic number.
\begin{shortindent}{(ii)}
\item[(i)]
  If $f$ is a univariate polynomial with $\|f\|\leq 2^d$, $\deg(f)\leq d$ and $f(\alpha)\neq 0$, then ${|f(\alpha)|\ge 2^{-O(d^2)}}$.
\item[(ii)]
  Let  $f$, $g$ be bivariate integer polynomials with  $\|f\|,\|g\| \le 2^d$,
  $\deg(f),\deg(g)\le d$ and $g(\alpha,1)\neq 0$. 
  Then there is a constant~$c > 0$ such that for any $\delta$ with $0 < \delta < 2^{-c d^2}$ 
  and~$d$ sufficiently large,
  \[
    \left|\frac{f(\alpha,1)}{g(\alpha,1)} - 
      \frac{f(\alpha,1-\delta)}{g(\alpha,1-\delta)}\right| \ \le\  \delta\, 2^{c d^2}.
  \]
\end{shortindent}
\end{lemma}

\begin{lemma}[\cite{Wat03}]\label{lem:wat_matrix}\sloppy
Let $\alpha$ be any real algebraic number and let $m\in\IR$.
Let $M$ be an $N\times N$-matrix such that for each entry~$x$ there is an
integer polynomial~$p$ with $x = p(\alpha)/m$ and $\deg(p) = \poly(N)$, $\|p\| = 2^{\poly(N)}$.
Further suppose that the eigenvalues of $M$ are bounded above in absolute
value by~$1$. Let $1\leq i,j\leq N$ and
let $S = \bigl(\sum_{\ell=0}^{\infty} M^{\ell}\bigr)_{i,j}$ be convergent.
For $z\in[0,1)$, define
$\widetilde{S}(z) = \bigl(\sum_{\ell=0}^{\infty} (z M)^{\ell}\bigr)_{i,j}$.
Then there are bivariate integer polynomials $f,g$ such that $\|f\|,\|g\| \le m^N 2^{\poly(N)}$, 
$\deg(f),\deg(g) = \poly(N)$, $g(\alpha,1)\neq 0$, and 
\begin{align*}
  &f(\alpha,z)/g(\alpha,z) \ =\  \widetilde{S}(z),\ \text{for $z\in[0,1)$, and}\\
  &f(\alpha,1)/g(\alpha,1) \ =\  S.
\end{align*}
\end{lemma}

\begin{lemma}\label{lem:wat_gaps}\sloppy
Let $m = 2^{\poly(N)}$.
Let $M$ be an $N\times N$-matrix as in the previous lemma.
Let $S_{i,j} = \bigl(\sum_{\ell=0}^{\infty} M^{\ell}\bigr)_{i,j}$ for $1\le i,j\le N$.
\begin{shortindent}{(ii)}
\item[(i)]
  Suppose that $S_{i,j}$ converges. For $z\in[0,1)$, let
  $\widetilde{S}_{i,j}(z) = \bigl(\sum_{\ell=0}^{\infty} (z M)^{\ell}\bigr)_{i,j}$.
  Then there is a polynomial~$p$ such that for any $z = 1-\delta$ with $0<\delta< 2^{-p(N)}$,
  ${|S_{i,j} - \widetilde{S}_{i,j}(z)| \le \delta 2^{p(N)}}$.
\item[(ii)] Let $I\subseteq\{1,\ldots,N\}^2$ and suppose that for each $(i,j)\in I$, $S_{i,j}$
  converges. Let $S = \sum_{(i,j)\in I} S_{i,j}$. 
  Then there is a polynomial~$p$ such that 
  $S \neq 1/2$ implies ${|S - 1/2| \ge 2^{-p(N)}}$.
\end{shortindent}
\end{lemma}

\begin{proof}
\emph{Part~(i):}
Use Lemma~\ref{lem:wat_matrix} to get bivariate integer polynomials $f_{i,j},g_{i,j}$ such that
\begin{align*}
  &f_{i,j}(\alpha,z)/g_{i,j}(\alpha,z) \ =\  \widetilde{S}_{i,j}(z),\ \text{for $z\in[0,1)$, and}\\
  &f_{i,j}(\alpha,1)/g_{i,j}(\alpha,1) \ =\  S_{i,j}.
\end{align*}
By the lemma and the fact $m = 2^{\poly(N)}$, there is a polynomial~$q$
such that $\|f_{i,j}\|,\|g_{i,j}\|\le 2^{q(N)}$ and $\deg(f_{i,j}),\deg(g_{i,j})\le q(N)$ and, furthermore,
$g_{i,j}(\alpha,1)\neq 0$. By Lemma~\ref{lem:wat_gaps_low}(ii) applied to~$f_{i,j}$ and~$g_{i,j}$
with $d = q(N)$, it follows that there is a constant~$c > 0$ such that for all
$0 < \delta < 2^{-c q(N)^2}$ and $N$ sufficiently large,
\[
  |S_{i,j}-\widetilde{S}_{i,j}(1-\delta)|
  \ = \ \left|\frac{f_{i,j}(\alpha,1)}{g_{i,j}(\alpha,1)} - 
    \frac{f_{i,j}(\alpha,1-\delta)}{g_{i,j}(\alpha,1-\delta)}\right| 
    \ \le\  \delta\, 2^{c q(N)^2}.
\]
Choosing $p(N) = c q(N)^2$ yields the desired
bound for any $z = 1-\delta$ with $0 < \delta < 2^{-p(N)}$.

\emph{Part~(ii):} \sloppy\hbadness=3500
By Lemma~\ref{lem:wat_matrix}, it follows that for each $(i,j)\in I$,
\[
  S_{i,j} \ =\  \Bigl(\,\sum_{\ell=0}^{\infty} M^{\ell}\Bigr)_{i,j}
          \ =\  \frac{f_{i,j}(\alpha,1)}{g_{i,j}(\alpha,1)},
\]
where $f_{i,j}$ and~$g_{i,j}$ are bivariate integer polynomials with
$\|f_{i,j}\|,\|g_{i,j}\| \leq 2^{q(N)}$ and $\deg(f_{i,j}), \deg(g_{i,j})\le q(N)$
for some polynomial~$q$, and $g_{i,j}(\alpha,1)\neq 0$ for all~$i,j\in I$. Then
\begin{align*}
  S  = \sum_{(i,j)\in I} \frac{f_{i,j}(\alpha,1)}{g_{i,j}(\alpha,1)} \neq  1/2 ~~\Rightarrow~~
  2\sum_{(i,j)\in I} f_{i,j}(\alpha,1) \!\!\!\prod_{(i',j')\neq (i,j)}\!\!\! g_{i',j'}(\alpha,1)\;
    -\! \prod_{(i,j)\in I} g_{i,j}(\alpha,1) \neq 0.
\end{align*}
The left hand side of the last inequality is a polynomial in~$\alpha$ with height at most $2^{O(|I|\cdot q(N))} = 2^{\poly(N)}$
and degree at most $|I|\cdot q(N) = \poly(N)$, since $|I| \le N^2$.
Lemma~\ref{lem:wat_gaps_low}(i) implies that the absolute value of this expression is lower bounded
by $2^{-q'(N)}$ for a suitable polynomial~$q'$ and~$N$ large enough. Hence, 
\[
  \left|\sum_{(i,j)\in I} \frac{f_{i,j}(\alpha,1)}{g_{i,j}(\alpha,1)} - \frac{1}{2}\right|  
  \ \ge\  \frac{2^{-q'(N)-1}}{\prod_{(i,j)\in I} |g_{i,j}(\alpha,1)|}.
\]
We have $\|g_{i,j}\| \le 2^{q(N)}$, $\deg(g_{i,j})\le q(N)$, and 
$\alpha,\alpha^2,\ldots,\alpha^{q(N)} = 2^{\poly(N)}$ since $\alpha$ is a constant.
This implies that $|g_{i,j}(\alpha,1)| \le 2^{q''(N)}$ for a polynomial~$q''$ and $N$ sufficiently large.
Thus,
\[
  |S-1/2| \ \ge\  \frac{2^{-q'(N)-1}}{2^{|I|\cdot q''(N)}} \ \ge\  2^{-p(N)}
\]
for $p(N) = q'(N) + N^2 q''(N) + 1$, which proves the claim.
\end{proof}

Now we can state and prove our first main lemma that allows us to bound the error
probability of QBPs away from~$1/2$.

\begin{lemma}\label{lem:err_bound}
For each QBP $G$ with short amplitudes there
exists a polynomial~$q$ such that for each input~${a\in\{0,1\}^n}$,
$p_{G,1}(a) > 1/2$ implies $p_{G,1}(a) \ge 1/2+2^{-q(|G|)}$
and $p_{G,1}(a) < 1/2$ implies $p_{G,1}(a) \le 1/2-2^{-q(|G|)}$.
\end{lemma}

\begin{proof}
Let $G$ be a QBP with short amplitudes on $n$ variables.  By
Proposition~\ref{prop:simpleshortamps} we may assume that the amplitudes in $G$ are of
the form $p(\alpha)/m$, where $p$ is an integer polynomial with 
$\deg(p) = \poly(|G|)$ and $\|p\| = 2^{\poly(|G|)}$ and 
where $\alpha$ is the same algebraic number and $m = 2^{\poly(|G|)}$ is 
the same natural number for all amplitudes.
Let $v$ be the start node of~$G$ and let $F_1 = \{ w \mid \text{$w$ is a $1$-sink of~$G$ } \}$.
Let $N = |G|^2$ and let the $N\times N$-matrix~$M$ describing the 
computation of $G$ on an input~$a\in\{0,1\}^n$ as well as the indices $i_v\in\{1,\ldots,N\}$ corresponding
to nodes~$v\in\{1,\ldots,|G|\}$ be defined as above.
Then the probability of $G$ accepting~$a$ in the $k$th computation step 
is given by $\sum_{w\in F_1} (M^k)_{i_w,i_v}$, 
and the total probability of accepting~$a$ is
$p_{G,1}(a) = \sum_{w\in F_1} \bigl(\sum_{k=0}^{\infty} M^k\bigr)_{i_w,i_v}$.
Since $G$ only contains labels of the form $p(\alpha)/m$, the entries of~$M$ are
of the form $p'(\alpha)/m'$, where $p'$ is a polynomial with
$\deg(p') = \poly(|G|)$ and $\|p'\| = 2^{\poly(|G|)}$ and $m' = m^2 = 2^{\poly(|G|)}$.
Hence, part~(ii) of Lemma~\ref{lem:wat_gaps} yields the claimed result.
\end{proof}

The other main argument in our proofs is the construction of a
probabilistic clock, which works in the case of bounded as well as
unbounded error.  

\begin{lemma}\label{lem:comp_time}
  For each sequence of QBPs $(G_n)_{n\in\IN}$ with bounded or
  unbounded error and 
  short amplitudes, there is a sequence of QBPs $(G_n')_{n\in\IN}$ for the same
  function with short amplitudes, the same type of error, size
  $\poly(|G_n|)$, and expected running time $2^{\poly(|G'_n|)}$.
\end{lemma}

\begin{proof}
  The main idea is similar to that of Simon~\cite{Sim81} for limiting
  the running time of probabilistic Turing machines.  We simulate $G$
  step-by-step. Before each simulation step, we stop and reject the
  input with fixed, small probability.  
  A similar construction for unidirectional QTMs has been  given in Lemma~4.6 of
  Watrous~\cite{Wat99}. 

Let $G$ be a QBP on $n$ variables of size~$s$. By
Proposition~\ref{prop:simpleshortamps} we may assume that the amplitudes of $G$ are the
fraction of some integer polynomial in an algebraic number and a common
denominator $m=2^{\poly(s)}$. 
 Let $q$ be some polynomial. We
construct a QBP $G'$ with size polynomial in $s$, expected running
time $2^{\poly(s)}$, and such that for all $a\in\{0,1\}^n$, 
${p_{G,1}(a) - 2^{-q(s)}} \le p_{G',1}(a) \le p_{G,1}(a)$. 
Together with Lemma~\ref{lem:err_bound}, this implies the claim.

\begin{figure}
\centerline{\input{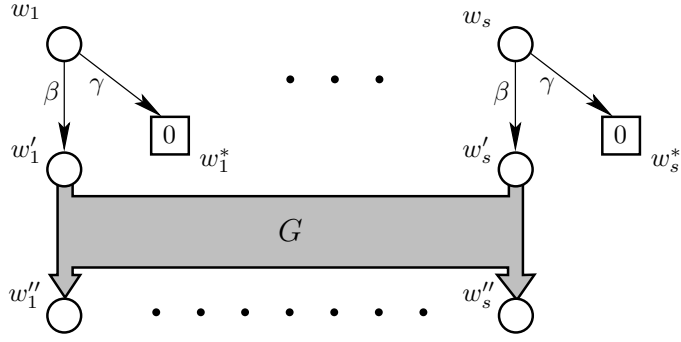}}
\caption{The QBP $G_0'$ used in the proof of Lemma~\ref{lem:comp_time}.}
\label{fig:probclock}
\end{figure}

Let $t=t(s)=q(s)+p(s)+\log s$, where $p(s)$ is a polynomial defined later on.
Let $v_1,\ldots,v_s$ be the nodes of $G$.  The new QBP~$G'$ is obtained
from the QBP $G_0'$ shown schematically in
Figure~\ref{fig:probclock}.  We use unlabeled nodes introduced in
Section~\ref{sec:qbps} to simplify the presentation.  
The start node of $G_0'$ is
$w_{1}$.  The edges in the upper part of the figure represent the
transformation $\k{w_i}\mapsto \beta \k{w_i'}+\gamma \k{w_i^*}$, where 
\[
  \beta \ =\  \frac{2^{2t+1}+2^{t+1}}{2^{2t+1}+2^{t+1}+1}\quad\text{and}\quad
    \gamma \ =\  \frac{2^{t+1}+1}{2^{2t+1}+2^{t+1}+1}. 
\]
Then $\beta^2+\gamma^2=1$, which is used to prove that the QBP is
well-formed.
Each node $w_{i}'$,
${i\in\{1,\ldots,s\}}$, is a copy of the node~$v_i$ in $G$ and is labeled by
the same variable as $v_i$.  For each edge $(v_i,v_j)$ in $G$, an edge
$(w_{i}',w_j'')$ is inserted in $G_0'$ that carries the same labels.
The shaded part in the figure represents these edges. The node
$w_i''$ is a sink if the corresponding node~$v_i$ in $G$ is, and each
non-sink node $w_i''$ is unlabeled and has an outgoing edge with amplitude~1 to
node~$w_{i}$ (not shown in the figure). The only nodes labeled by
variables are $w_{1}',\ldots,w_{s}'$, all other nodes are unlabeled.
We remove all unlabeled nodes from~$G_0'$ to obtain the desired
QBP~$G'$.  It is easy to see that~$G'$ constructed in this way is
well-formed and unidirectional. The only numbers added as amplitudes here,
$1$, $\beta$, and $\gamma$, are rational and have representations of polynomial 
length. Hence, $G'$ also has short amplitudes. 

We observe that the probability of $G'$ terminating during a traversal of the upper part
is $\delta = |\gamma|^2\leq 2^{-2t}$. Hence, its expected running time
is bounded by $2^{O(t)} = 2^{\poly(s)}$.  Furthermore, for all inputs
$a\in\{0,1\}^n$, $p_{G',1}(a) \le p_{G,1}(a)$.  It remains to show
that for all inputs~$a$, $p_{G',1}(a) \ge p_{G,1}(a) - 2^{-q(s)}$.

{\sloppy
Fix any input $a\in\{0,1\}^n$. Let $N = s^2$, let the $N\times
N$-matrix~$M$ describing the computation of~the original QBP $G$ on
$a$, and let the mapping of nodes~$v\in\{1,\ldots,s\}$ to
indices~$i_v\in\{1,\ldots,N\}$ be defined as above.  Let $v$ be the
start node of $G$ and let  $F_1 = {\{ w \mid \text{$w$ is a $1$-sink of~$G$ } \}}$.  
As in the proof of Lemma~\ref{lem:err_bound}, the total
probability of~$G$ accepting~$a$ is $p_{G,1}(a) \ =\ \sum_{w\in F_1}
\bigl(\sum_{k=0}^{\infty} M^k\bigr)_{i_w,i_v}$.  Now recall that $G'$
performs the same computation as $G$ with the only exception that it
terminates the computation with the probability $\delta$ before each
step of $G$.
Hence, the probability of
$G'$ accepting~$a$ in the $k$th simulation step of~$G$ after not
rejecting $k$ times in the first phase of the computation is
$\sum_{w\in F_1} \bigl((1-\delta)^k M^k\bigr)_{i_w,i_v}$.  We obtain
\[
  p_{G',1}(a) 
  \ =\  \sum_{w\in F_1} \Bigl(\,\sum_{k=0}^{\infty} (1-\delta)^k M^k\Bigr)_{i_w,i_v}.
\]}%
Now choose~$p$ as the polynomial obtained when Lemma~\ref{lem:wat_gaps}(i) is
applied with $z = 1-\delta$, ${S_{i,j} = p_{G,1}(a)}$, and $\wt{S}_{i,j}(z) = p_{G',1}(a)$. 
The lemma implies that 
\[
  |p_{G',1}(a) - p_{G,1}(a)|
  \ \le\ \sum_{w\in F_1} \left| \Bigl(\,\sum_{k=0}^{\infty} (1-\delta)^k M^k\Bigr)_{i_w,i_v}
           - \Bigl(\,\sum_{k=0}^{\infty} M^k\Bigr)_{i_w,i_v} \right| 
  \ \le\ |F_1|\cdot\delta\cdot 2^{p(s)},
\]
provided that $0 < \delta < 2^{-p(s)}$.
The restriction on~$\delta$ is easily seen to be satisfied 
since $\delta \le 2^{-2t}$ and $t = t(s) = p(s) + q(s) + \log s$.
Using that $|F_1|\le s$, we obtain
\[
  |F_1|\cdot\delta\cdot 2^{p(s)} 
  \ \le\  |F_1|\cdot 2^{-2(q(s)+p(s)+\log s)}\,\cdot 2^{p(s)} 
  \ \le\  2^{-q(s)}
\]
and thus $|p_{G',1}(a)-p_{G,1}(a)| \le 2^{-q(s)}$. Hence, $G'$ has
all required properties.
\end{proof}

Now we have collected all tools for the proofs of
Theorems~\ref{th:equiv_1} and~\ref{th:equiv_3}. For the convenience
of the reader, we restate the theorems here. We begin with the
proof of Theorem~\ref{th:equiv_3}.

\setcounter{theorem}{4}
\begin{theorem}[restatement]
  The logarithm of the size of QBPs with bounded or unbounded error and short
  amplitudes and the space complexity of unidirectional nonuniform QTMs with algebraic
  amplitudes and the same type of error are asymptotically equal.
\end{theorem}

\begin{proof}
  A simulation of unidirectional nonuniform QTMs by QBPs 
  is already provided in Theorem~\ref{the:sim}. It is easy to see that
  the resulting QBP has short amplitudes if the amplitudes of the QTM
  are algebraic numbers.
  
  Now let a sequence $(G_n)_{n\in\IN}$ of QBPs with short amplitudes be given.
  By Lemma~\ref{lem:comp_time} we can simulate $G_n$ by a QBP $G'_n$
  with size $\poly(|G_n|)$, the same type of error, short amplitudes
  and expected running time $T(n)=2^{\poly(|G_n'|)}$. In the case of
  bounded error, let $\epsilon$ be the error bound of $G'_n$. In the case of
  unbounded error, by Lemma~\ref{lem:err_bound}, there is some
  polynomial $q(n)$ such that the acceptance and rejection
  probabilities of $G'_n$ are strictly larger than $1/2+2^{-q(|G_n'|)}$
  or strictly smaller than $1/2-2^{-q(|G_n'|)}$, resp.  In this case let
  $\epsilon=\epsilon(n)=1/2-2^{-q(|G_n'|)}$ be the error bound of $G'_n$.  We
  choose $\epsilon'=(1/2-\epsilon)/3$ and
  $T'(n)=T(n)/\epsilon'=2^{\poly(|G_n'|)}$. Then we apply the simulation of
  QBPs by QTMs from Theorem~\ref{the:sim} for the accuracy $\epsilon'$ and
  the running time $T'(n)$. The space complexity of the QTM is $O(\log
  |G_n'|+\log\log (T'(n)/\epsilon'))=O(\log |G_n|)$.  By Markov's
  inequality, the probability that the running time of $G'_n$ and thus
  the number of performed simulation steps exceeds
  $T'(n)=T(n)/\epsilon'$ is bounded by $\epsilon'$. Hence, the probability
  of an error caused by running more than~$T'(n)$ simulation steps is bounded
  by $\epsilon'$ and the overall error probability is bounded by
  $\epsilon+2\epsilon'= 1/2-\epsilon'$.
\end{proof}

\setcounter{theorem}{3}
{\samepage
\begin{theorem}[restatement]\item[]
\begin{shortindent}{(ii)}
\item[(i)]  Sequences of QBPs $(G_n)_{n\in\IN}$ with bounded error and short amplitudes and
  sequences of QBPs $(G_n')_{n\in\IN}$  with bounded error and expected running time
  $2^{\poly(|G_n'|)}$ have polynomially related size complexities.
\item[(ii)]  Sequences of QBPs $(G_n)_{n\in\IN}$ with unbounded error and short
  amplitudes can be simulated by sequences of QBPs $(G_n')_{n\in\IN}$ of size
  $\poly(|G_n|)$ and with expected running time $2^{\poly(|G_n'|)}$.
\end{shortindent}
\end{theorem}

}

\begin{proof}
  A simulation of QBPs $(G_n)_{n\in\IN}$ with short amplitudes by QBPs
  $(G_n')_{n\in\IN}$ with expected running time $2^{\poly(|G_n'|)}$ for
  bounded and unbounded error is contained in
  Lemma~\ref{lem:comp_time}. 
  This proves one direction of part~(i) as
  well as part~(ii). It remains to prove the missing direction of part~(i),
  i.\,e., to provide a simulation of QBPs with bounded error and
  an expected exponential running time by QBPs with bounded error and short amplitudes. 
  Let $(G_n)_{n\in\IN}$ be a sequence of QBPs with
  expected running time $2^{\poly(|G_n|)}$ and error probability
  $\epsilon\in [0,1/2)$. As in the proof of Theorem~\ref{th:equiv_3}, we
  choose $\epsilon'=(1/2-\epsilon)/3$ and
  $T'(n)=T(n)/\epsilon'=2^{\poly(s(n))}$ and apply the simulation of QBPs
  by QTMs of Theorem~\ref{the:sim} for the accuracy $\epsilon'$ and the
  running time $T'(n)$. By the same arguments as in the proof of
  Theorem~\ref{th:equiv_3}, we obtain a unidirectional nonuniform QTM
  simulating the given QBP with bounded error, expected running time
  $T(n)$, and space complexity $O(\log |G_n|)$. The transition
  function of the QTM only contains a constant number of algebraic
  numbers.

  In a second step we apply the simulation of unidirectional nonuniform QTMs by QBPs
  from Theorem~\ref{the:sim}. The resulting QBP has 
  an error probability of at most~$\epsilon'$. Its size is bounded above by 
  $2^{O(\log |G_n|)} = \poly(|G_n|)$. The amplitudes occurring in the QBP 
  are the amplitudes of the transition function of the QTM and thus
  are short.  
\end{proof}

\section{Simulation of Nonuniform QTMs by Unidirectional Nonuniform QTMs}
\label{sec:unidir}

In this section, we consider nonuniform RTMs and QTMs that are, different from the
previous sections, not necessarily unidirectional.  We show that they
can be simulated space-efficiently by their unidirectional
counterparts. 
We discuss some consequences of the simulation
result at the end of this section.

Our simulation result uses
the construction of the universal QTM due to Yao~\cite{Yao93}
and Nishimura and Ozawa~\cite{Nis02a} based on a simulation of QTMs by quantum circuits 
and vice versa as intermediate steps. The original simulations cannot be applied
since they use markers on the work tape of the simulating machine to store the
positions of the simulated tape heads and (which is more serious) 
generate a quantum circuit for the simulated machine online on the work tape.
Both of this is too costly in terms of space.
These obstacles are overcome here by using a space-efficient encoding of the 
positions of the input tape heads and by storing a representation of the required
quantum circuit on the advice tape.

As a preparation for the proof of our simulation result, we state a simple necessary property of 
the transition function of QTMs with two read-only input tapes
which is extracted from the proof of Theorem~4.5 in~\cite{Oza00a}. 
In the following the expression $[A=B]$ has the value $1$, if $A=B$, and $0$ otherwise.

\begin{lemma}[\cite{Oza00a}]\label{lem:qtm_char}
\sloppy
Let $M = (Q,\Sigma,\delta)$ be a QTM with two 
read-only input tapes. Let ${p,p'\in Q}$, $\Delta=(\Delta_1,\Delta_2)\in\ZZ^2$ and 
$a_1,a_2,a_1',a_2',{v,w,v',w'}\in\Sigma$.
\begin{shortindent}{(ii)\ } 
\item[(i)] 
  $\displaystyle
  0 \ \ \ =   \hspace*{-0.2cm}
  \sum_{\substack{q\in Q,\,d''\in\{0,1\},\\d,d'\in\{-1,0,1\}^2}}\hspace*{-0.2cm}
     \hspace*{0cm}\parbox[t]{11cm}{$\displaystyle\delta(p,(a_1,a_2,v),q,w,(d,d''-1))^*\\
     \hspace*{1cm}\cdot\;\,\delta(p',(a_1',a_2',v'),q,w',(d',d''))\cdot\bigl[d'-d = \Delta\bigr]$.}
  $
\item[(ii)]
  $\displaystyle
  0 \ \ \ =  \hspace*{-0.2cm}
  \sum_{\substack{q\in Q,\\d,d'\in\{-1,0,1\}^2}}\hspace*{-0.2cm}
     \hspace*{0cm}\parbox[t]{11cm}{$\displaystyle\delta(p,(a_1,a_2,v),q,w,(d,-1))^*\\
     \hspace*{1cm}\cdot\;\,\delta(p',(a_1',a_2',v'),q,w',(d',1))\cdot\bigl[d'-d = \Delta\bigr]$.}
  $
\end{shortindent}
\end{lemma}

Now we can state and prove our result. 

{\samepage
\begin{theorem}\item[]\vspace*{-2pt}\label{the:unidir_sim}
\begin{shortindent}{(ii)}\nopagebreak
\item[(i)] 
  Each nonuniform RTM that runs in space~$S$ 
  at least logarithmic in the input length and
  time~$T$ can be simulated by a unidirectional nonuniform RTM 
  running in time $\poly(S,T)$ and space~$O(S)$.
\item[(ii)]  
  Let $\epsilon > 0$ and~${T\colon\IN\to\IN_0}$.
  For each nonuniform QTM~$M$ running in space~$S$ at least logarithmic in the input length,
  there is a unidirectional nonuniform QTM that simulates~$M$ for~$T$ steps in 
  $\poly(2^{O(S)}, T, \log(1/\epsilon))$ steps with accuracy~$\epsilon$ using 
  space ${O(S + \log\log(T/\epsilon))}$.
\end{shortindent}
\end{theorem}
}

\begin{proof}
In the main part of the proof, we deal with part~(ii). 
We handle necessary changes for part~(i) and RTMs at the end. 
We first describe how we encode the information about the simulated machine 
on the work tape of the simulating machine. 
Then we present a high-level algorithm carrying out a whole simulation step
and define a unitary transformation realizing a single transition of the
simulated machine. Afterwards, this unitary transformation is 
implemented approximately by the simulating unidirectional nonuniform QTM. 

\medskip
\emph{Storage layout on the work tape.}\ \ 
Let $M = (Q,\Sigma,\delta)$ be a
nonuniform QTM  that is to be simulated unidirectionally. 
We regard the advice tape simply as an additional read-only input tape.
We assume that for input length~$n$ and space bound~$S\geq\log n$
the heads on the input tapes~$i\in\{1,2\}$ of $M$ only reach the
positions $0,\ldots,n_i-1$, where $n_1 = n+2$, $n_2 = \poly(n)$,
and that the work tape head only reaches the positions $0,\ldots,n_3-1$ with $n_3 = S+2$ 
(this may be achieved using end markers). 
We assume that $\{0,1,2\}\subseteq\Sigma$.

Let $\ell = \ell_1 + 6\ell_2+1$ with $\ell_1 = \ceil{\log|Q|}$ and
$\ell_2 = \max\{\ceil{\log n_i}\mid i\in\{1,2,3\}\} = O(S)$, 
and assume w.\,l.\,o.\,g. that $\ell \ge 3$. 
The information about the simulated machine is stored on two tracks of the work tape
of the simulating machine as shown below.

\bigskip
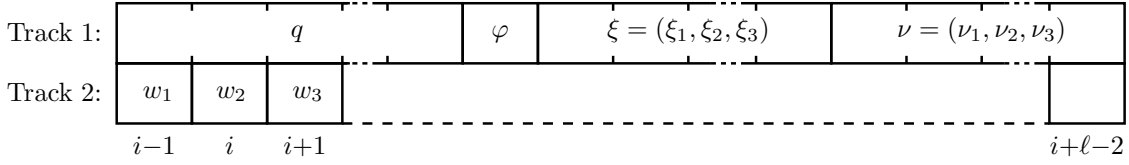
\centerline{\hspace*{-0.6cm}\hbox{\begin{picture}(0,0)%
\includegraphics{infoblock.pstex}%
\end{picture}%
\setlength{\unitlength}{4144sp}%
\begingroup\makeatletter\ifx\SetFigFont\undefined%
\gdef\SetFigFont#1#2#3#4#5{%
  \reset@font\fontsize{#1}{#2pt}%
  \fontfamily{#3}\fontseries{#4}\fontshape{#5}%
  \selectfont}%
\fi\endgroup%
\begin{picture}(6881,971)(297,-2675)
\put(1036,-2311){\makebox(0,0)[rb]{\smash{{\SetFigFont{10}{12.0}{\familydefault}{\mddefault}{\updefault}{\color[rgb]{0,0,0}Track 2:}%
}}}}
\put(1036,-1951){\makebox(0,0)[rb]{\smash{{\SetFigFont{10}{12.0}{\familydefault}{\mddefault}{\updefault}{\color[rgb]{0,0,0}Track 1:}%
}}}}
\put(1375,-2290){\makebox(0,0)[b]{\smash{{\SetFigFont{10}{12.0}{\familydefault}{\mddefault}{\updefault}{\color[rgb]{0,0,0}$w_1$}%
}}}}
\put(1813,-2290){\makebox(0,0)[b]{\smash{{\SetFigFont{10}{12.0}{\familydefault}{\mddefault}{\updefault}{\color[rgb]{0,0,0}$w_2$}%
}}}}
\put(2270,-2290){\makebox(0,0)[b]{\smash{{\SetFigFont{10}{12.0}{\familydefault}{\mddefault}{\updefault}{\color[rgb]{0,0,0}$w_3$}%
}}}}
\put(1351,-2626){\makebox(0,0)[b]{\smash{{\SetFigFont{10}{12.0}{\familydefault}{\mddefault}{\updefault}{\color[rgb]{0,0,0}$i{-}1$}%
}}}}
\put(1801,-2626){\makebox(0,0)[b]{\smash{{\SetFigFont{10}{12.0}{\familydefault}{\mddefault}{\updefault}{\color[rgb]{0,0,0}$i$}%
}}}}
\put(2251,-2626){\makebox(0,0)[b]{\smash{{\SetFigFont{10}{12.0}{\familydefault}{\mddefault}{\updefault}{\color[rgb]{0,0,0}$i{+}1$}%
}}}}
\put(6931,-2626){\makebox(0,0)[b]{\smash{{\SetFigFont{10}{12.0}{\familydefault}{\mddefault}{\updefault}{\color[rgb]{0,0,0}$i{+}\ell{-}2$}%
}}}}
\put(6301,-1926){\makebox(0,0)[b]{\smash{{\SetFigFont{10}{12.0}{\familydefault}{\mddefault}{\updefault}{\color[rgb]{0,0,0}$\nu = (\nu_1,\nu_2,\nu_3)$}%
}}}}
\put(4546,-1926){\makebox(0,0)[b]{\smash{{\SetFigFont{10}{12.0}{\familydefault}{\mddefault}{\updefault}{\color[rgb]{0,0,0}$\xi = (\xi_1,\xi_2,\xi_3)$}%
}}}}
\put(2206,-1926){\makebox(0,0)[b]{\smash{{\SetFigFont{10}{12.0}{\familydefault}{\mddefault}{\updefault}{\color[rgb]{0,0,0}$q$}%
}}}}
\put(3421,-1926){\makebox(0,0)[b]{\smash{{\SetFigFont{10}{12.0}{\familydefault}{\mddefault}{\updefault}{\color[rgb]{0,0,0}$\phi$}%
}}}}
\end{picture}%
}}

\smallskip
{\sloppy
Track~2 contains the work tape of the simulated machine.
In~$\ell$ consecutive cells on track~1, which are called the \emph{info block},
we encode all administrative information for the simulation.
The position of the info block is used to indicate the position
of the head on the work tape in a classical configuration.
If the cells of the info block are located at positions
${i-1,i,i+1,\ldots,i+\ell-2}$ on the work tape as shown
in the figure, we say that the info block is at position~$i$.
In this situation, the inscription in the info block together with the 
symbols $w_1,w_2,w_3\in\Sigma$ in cells $i-1,i,i+1$ on track~2 
are called the \emph{info window} induced by the info block.

}

The information stored in the info block consists of 
the local state~$q\in Q$ of the simulated machine encoded in binary, 
a flag $\phi\in\{0,1\}$ showing whether the actual transition step has already
been carried out, and vectors $\xi=(\xi_1,\xi_2,\xi_3),\nu=(\nu_1,\nu_2,\nu_3)$
in $\{0,\ldots,n_1-1\}\times\cdots\times\{0,\ldots,n_3-1\}$
encoded in binary. The coordinates of $\xi$ are the positions of the
tape heads of the simulated machine. Similarly, $\nu_1$ and $\nu_2$ are the
positions of the heads on the input tapes of the simulating machine. Finally,
$\nu_3$ is the position of the info block. 
We write the contents of the info window shown above as
$(q,\phi,\xi,\nu,w)$, where $w = (w_1,w_2,w_3)$.

\medskip
\emph{Carrying out a simulation step.}\ \ 
We first give an outline of our approach.
For the simulation of a single step of~$M$, we let the input tape heads of the simulating
machine as well as the info block on the work tape
successively move to all combinations of positions in
$\{0,\ldots,n_1-1\}\times\cdots\times\{0,\ldots,n_3-1\}$
on the tapes that may be accessed. 
If during this sweep the machine reaches a configuration where the positions of the  
heads of the input tapes as well as the position of the info block, which are encoded in~$\nu$, all agree with the stored positions of those of the 
simulated machine and $\phi = 0$, then a local transition of the
simulated machine is applied, for which we update the contents of the
info window and set $\phi = 1$. After the sweep through all positions is complete,
the flag $\phi$ is negated.

\begin{figure}\normalsize\fboxsep=10pt
{\centering\fbox{\parbox{0.95\textwidth}{\normalsize
Loop with starting/stopping condition $\nu = (0,0,0)$:
\vspace*{2mm}
\begin{shortindent}{3.}
\item[1.]
  Move the real input tape heads and the info block on the work tape
  to the positions in~$\nu$.
\item[2.]
  Transition: Let $(p,\phi,\xi,\nu,(w_1,w_2,w_3))$ 
    be the contents of the current info window and let 
    $a_1,a_2\in\Sigma$ be the symbols under the input tape heads.
  \begin{shortindent}{2.2.}
  \item[2.1.] If $\xi = \nu$ and $\phi = 0$, 
    replace the contents of the info window with the superposition
    {\abovedisplayskip=0.5\abovedisplayskip\belowdisplayskip\abovedisplayskip
    \[
      \sum_{\substack{q\in Q,b\in\Sigma,\\d\in\{-1,0,1\}^3}}
        \!\!\!\delta\bigl(p,(a_1,a_2,w_2),q,b,d\bigr)\k{q, 1, \xi+d, \xi, w_1\,b\,w_3}.
    \]}%
  \item[2.2.] For all inscriptions of the info window that do not satisfy the condition
    of step~2.1 and can actually arise during the computation, do nothing.
  \end{shortindent}
\item[3.]
  Move real input tape heads and the info block on the work tape
  to positions~$(0,0,0)$.
\item[4.] Update~$\nu$ to a new vector~$\nu'$ such that
  $|\nu'| \equiv (|\nu|+1)\bmod n_1\cdot n_2\cdot n_3$.
\end{shortindent}
\vspace*{2mm}
Set $\phi = 1-\phi$. End of simulation step.
}}}
\hfuzz5pt 
\caption{High-level description of the simulation step.}\label{fig:single_step_algo}
\end{figure}

{\sloppy
In Figure~\ref{fig:single_step_algo} this is described in more detail as a high-level algorithm.
We use the following notation.
For $x=(x_1,x_2,x_3)\in\{0,\ldots,n_1-1\}\times\cdots\times\{0,\ldots,n_3-1\}$, let
$|x| = {x_3 n_2 n_1 + x_2 n_1 + x_1}$. Furthermore, let
$\k{q,\phi,\xi,\nu,w_1\,w_2\,w_3}$ denote
an ON-basis indexed by the different possible classical inscriptions of the info window.

}

\medskip
\emph{Realizing a Transition Unitarily.}\ \ 
Next we show that step~2 of the high-level algorithm
can be described by a unitary transformation.
For this, let the heads on the input tapes of the simulating machine
as well as the info block on the work tape  be at fixed positions. 
Let $a_1,a_2\in\Sigma$ be the symbols under the input tape heads.
Our goal is to specify a unitary transformation $U_{\rm trans} = U_{\rm trans}(a_1,a_2)$
that changes the contents of the info window according to the high-level algorithm.
Using an idea due to Yao~\cite{Yao93}, we only carry out the identity in step~2.2
for those inscriptions of the info window that can actually arise during the computation
at this point. This is required to allow the transformations of steps~2.1
and~2.2 to be combined to a unitary one.

For a precise definition of $U_{\rm trans}$, 
we introduce the collections of vectors in Figure~\ref{fig:unidir_sim_vectors}.
For these definitions, 
let $p\in Q$, $\xi=(\xi_1,\xi_2,\xi_3),\nu=(\nu_1,\nu_2,\nu_3)\in\{0,\ldots,n_1-1\}\times\cdots\times\{0,\ldots,n_3-1\}$, 
and $w,b,w_1,w_2,w_3\in\Sigma$.
The summations are over all $q\in Q$, $b\in\Sigma$, and $d=(d_1,d_2,d_3)\in\{-1,0,1\}^3$ if
not indicated otherwise.
Let $V_i$ be the set of vectors with upper index~$i\in\{1,\ldots,5\}$.

\begin{figure}
{\centering\def\arraystretch{1.75}\def\smallsub#1{\text{\small$\scriptstyle#1$}}
\fbox{\parbox{0.99\textwidth}{%
$
\begin{array}{@{}l@{}l@{}}
  \bigk{v^{(1)}_{\smallsub{p,\xi,w_1,w_2,w_3}}} &\>=\> \k{p, 0, \xi, \xi, w_1\,w_2\,w_3}\\
  \bigk{v^{(2)}_{\smallsub{p,\xi,w_1,w_2,w_3}}} &\>=\>
    \displaystyle\sum_{q,b,d}
      \delta\bigl(p,(a_1,a_2,w_2),q,b,d\bigr)\k{q, 1, \xi+d, \xi, w_1\,b\,w_3}\\
  \bigk{v^{(3)}_{\smallsub{p,\xi,\nu,w_1,w_2,w_3}}} &\>=\>
     \k{p, \phi, \xi, \nu, w_1\,w_2\,w_3}\ \ \text{with $\phi=0\land\nu\neq\xi$ or $\phi=1\land\nu_3\ge\xi_3+2$}\\
  \bigk{v^{(4)}_{\smallsub{p,\xi,\nu_1,\nu_2,w,w_2,w_3}}} &\>=\>
    \displaystyle\hspace*{-6pt}\sum_{\substack{q,b,d {\rm ~with} \\d_3\in\{0,1\}}}
      \hspace*{-6pt}\delta\bigl(p,(a_1,a_2,w),q,b,d)\bigr)
        \bigk{q, 1, \xi{+}d, (\nu_1,\nu_2,\xi_3{+}1), b\,w_2\,w_3}\\
  \bigk{v^{(5)}_{\smallsub{p,\xi,\nu_1,\nu_2,w,b,w_1,w_2,w_3}}} &\>=\>
    \displaystyle\sum_{\substack{q,d {\rm ~with}\\d_3 = 1}}
      \delta\bigl(p,(a_1,a_2,w),q,b,d\bigr)\bigk{q, 1, \xi{+}d, (\nu_1,\nu_2,\xi_3{+}2), w_1\,w_2\,w_3}
\end{array}
$
}}}
\hfuzz5pt 
\caption{Vectors for the definition of $U_{\rm trans}$.}
\label{fig:unidir_sim_vectors}
\end{figure}

We require that the transformation~$U_{\rm trans}$ satisfies
\[
  U_{\rm trans}\bigk{v^{(1)}_{p,\xi,w_1,w_2,w_3}} \ =\  
    \bigk{v^{(2)}_{p,\xi,w_1,w_2,w_3}}
\]
for all $p$, $\xi$, and $w_1,w_2,w_3$
and that $U_{\rm trans}\k{v} = \k{v}$ for all $\k{v} \in V_3\cup V_4\cup V_5$.
The following claim implies
that the above requirements can be satisfied by a unitary operator~$U_{\rm trans}$,
completing this part of the proof.

\begin{claim*}
The sets $V_1$, $V_2$, and $V_3\cup V_4\cup V_5$
are mutually orthogonal and the vectors in~$V_2$ form an ON-basis.
\end{claim*}

\begin{proof}[Proof of the claim]
The claim follows from the fact that~$M$ 
is a legal QTM and thus has a unitary time evolution operator. 
We use the notion ``superposition of~$M$'' to describe a unit vector from the
Hilbert space spanned by the classical configurations of~$M$ as an ON-basis.

\emph{The vectors in $V_2$\/ form an ON-basis}: 
We regard the vectors in $V_1$ and $V_2$ as
unique descriptions of superpositions of~$M$. This is possible since the contents 
of the work tape of~$M$ that is outside the three symbols in the info window is fixed.
Each vector $\bigk{v^{(2)}_{p,\xi,w_1,w_2,w_3}}$ uniquely describes the image of the 
classical configuration described by $\bigk{v^{(1)}_{p,\xi,w_1,w_2,w_3}}$ under the 
time evolution operator of~$M$. Since this time evolution operator is 
unitary and the vectors in~$V_1$ obviously form an ON-basis, the vectors 
from~$V_2$ also form an ON-basis.

\smallskip
\emph{The vectors in $V_1$, $V_2$, $V_3\cup V_4\cup V_5$ are mutually orthogonal}:
We write $M_1\bot M_2$ for two sets of vectors $M_1$ and $M_2$ if
$\bk{v}{w} = 0$ for all $v\in M_1$ and $w\in M_2$ and prove the statement
by considering all possible pairs of sets in the list.

\emph{$V_1 \bot V_2$, $V_1 \bot V_3\cup V_4\cup V_5$, $V_2\bot V_3$}: 
This follows immediately, since either the component for the flag~$\phi$ or
that for the position vector~$\nu$ distinguishes vectors from the
considered sets. 

\emph{$V_2 \bot V_4$}: We consider any pair of vectors
$\bigk{v^{(2)}_{p,\xi,w_1,w_2,w_3}}$ and $\bigk{v^{(4)}_{p',\xi',\nu_1',\nu_2',w',w_2',w_3'}}$. We may assume that
$w_3' = w_3$, $\nu_i' = \xi_i$ for $i\in\{1,2\}$ and $\xi_3' = \xi_3 - 1$
since otherwise the inner product of these vectors is obviously zero.
By keeping only the summands in the inner product for which the basis vectors
meet, 
we get
\begin{align*}
  &\bigbk{v^{(2)}_{p,\xi,w_1,w_2,w_3}}{v^{(4)}_{p',(\xi_1',\xi_2',\xi_3-1),\xi_1,\xi_2,w',w_2',w_3}}\ =\ \\
  &\hspace*{1cm}\sum_{\substack{q\in Q,d,d'\in\{-1,0,1\}^3,\\ {\rm
  with~} d_3'\in\{0,1\}}}
      \parbox[t]{11cm}{$\displaystyle\delta\bigl(p,(a_1,a_2,w_2),q,w_2',d\bigr)^*\\
      \hspace*{1cm}\bdot\;\delta\bigl(p',(a_1',a_2',w'),q,w_1,d'\bigr)
      \cdot\bigl[d'-d = \xi-\xi'\bigr]$.}
\end{align*}
For the $d,d'$ over which the summation is done, it is required that $d_3'-d_3 = \xi_3-\xi_3' = 1$,
i.\,e., $d_3 = d_3'-1$. The sum may thus be rewritten as
\[
  \sum_{\substack{q\in Q,\,d''\in\{0,1\},\\d,d'\in\{-1,0,1\}^2}}
     \hspace*{-6pt}\parbox[t]{13cm}{$\displaystyle\delta\bigl(p,(a_1,a_2,w_2),q,w_2',(d,d''-1)\bigr)^*\\
     \hspace*{1cm}\bdot\;\delta\bigl(p',(a_1',a_2',w'),q,w_1,(d',d'')\bigr)
     \cdot\bigl[d'-d = (\xi_1,\xi_2)-(\xi_1',\xi_2')\bigr]$.}
\]
For $\Delta=(\xi_1,\xi_2)-(\xi_1',\xi_2')$,  Lemma~\ref{lem:qtm_char}(i) implies that the sum takes the value $0$.
Thus the considered vectors are orthogonal.

\emph{$V_2 \bot V_5$}: This case is handled similarly to the latter one now using part~(ii) of 
Lemma~\ref{lem:qtm_char}.
\end{proof}

\emph{Constructing the Simulating QTM.}\ \ 
We now describe how the QTM simulating the given QTM~$M$ unidirectionally is constructed.
This simulating QTM carries out an endless loop executing single simulation steps
until the simulated machine terminates, similar to the machine constructed
for part~(ii) of Theorem~\ref{the:sim}. It is initialized as follows.
\begin{shortindent}{--}
\item[--] The info block belonging to the initial configuration of~$M$
  is located at
  position~$0$ of track~1 of the work tape. The complete contents of the respective
  info window is then $(q_0,0,\xi,\nu,w)$, where $q_0$ is the initial state of~$M$,
  $\xi = \nu = (0,0,0)$, and $w$ only consists of blanks.
\item[--] All input tape heads of the simulating machine are at position~$0$.
\end{shortindent}
As in the last section, this initialization is realized by choosing the 
encoding for the information on the work tape such that the blank tape is
consistent with the above requirements.

We realize the high-level algorithm by first constructing a unidirectional RTM for
everything except for step~2, for which the RTM has a special state as
a placeholder. This is easy by putting together machines for basic tasks using
appropriate versions of the lemmas of Bernstein and Vazirani~\cite{Ber97},
as in the last section.
Afterwards, we insert a QTM for carrying out step~2 which has still
to be constructed. We ensure that the running time of this QTM is independent
of the inscriptions of the info window. Then the complete QTM for the high-level algorithm 
obtained by the insertion has a running time independent of the contents of the different tapes.

The transformation~$U_{\rm trans}$ operates on a Hilbert space of
dimension~$O(\ell) = O(S)$. The number of iterations of the loop is $n_1n_2n_3=\poly(n)S$.
Reusing the calculations in the proof of Theorem~\ref{the:sim}(ii),
it follows that a description of~$U_{\rm trans}$  with accuracy~$\epsilon' = \epsilon/(2n_1n_2n_3T^2)$ by  elementary 
matrices adds~$O(S + \log\log(T/\epsilon))$
to the total space complexity if it is stored on the advice-tape.
This is within the required bound for part~(ii) of the theorem.
The chosen accuracy~$\epsilon'$ is sufficient to carry 
out the~$T$ simulation steps with accuracy~$\epsilon$. This
corresponds to  $n_1n_2n_3T$ executions of
$U_{\rm trans}$. 
The transformation $U_{\rm trans}$ is  realized by carrying out the respective
elementary transformations as described in the last section, using Lemma~\ref{lem:machine_elem}.

\medskip
{\sloppy
\emph{Resources.}\ \ 
The running time for carrying out $U_{\rm trans}$ is dominated by the  length of its description
on the advice tape and can be estimated by 
$2^{O(S)}\log(T/\epsilon)$. The number of iterations of the loop is $\poly(n)S$.
Thus the total time required for one simulation step can be estimated by
$O(\poly(n) 2^{O(S)}\log(T/\epsilon)) = \poly(2^{O(S)},\log(T/\epsilon))$.

}

\medskip
\emph{Correctness.}\ \ 
We show that each single computation step is performed correctly. 
We first consider step~2.1 of the high-level algorithm
and the case that the condition in this step is met.
We assume that the current configuration of the simulating machine is consistent with
our described invariants, that track~2 and the info block contain classical
inscriptions, and that the latter is at a fixed position.
Then it is easy to see that
$U_{\rm trans}$ correctly realizes a single transition of~$M$.

It remains to check that step~2.2 does not change anything.
We observe that before the transition of~$M$ has been carried out in step~2.1, $U_{\rm trans}$ performs
the identity in step~2.2, since all encountered info window inscriptions correspond to
vectors from~$V_3$. Immediately after the transition, the info window operated upon contains 
a vector $\k{v}\in V_2$. If after one or two shifts of the info window to the right 
on the work tape we adapt~$\k{v}$ by inserting the new~$\nu$, this
yields a vector from~$V_4$ or~$V_5$, resp., on which~$U_{\rm trans}$ also
performs the identity. If the window is shifted further to the right, the distance of
the info window from the stored position of the work tape head 
in each classical inscription contained in the current superposition is at least two.
Then the vector obtained by adapting~$\k{v}$ as described belongs to~$V_3$ and $U_{\rm trans}$
also performs the identity. Hence, $U_{\rm trans}$ behaves as desired.
Altogether, we have completed the proof of part~(ii).

\medskip
\emph{Simulation of RTMs.}
We can use the same construction as above, but replace the implementation of $U_{\rm trans}$.
In this case, $U_{\rm trans}$ is just a permutation of inscriptions of the
info window. This permutation can be computed exactly by a reversible
circuit of size~$\poly(\ell)$ consisting only of Toffoli gates. 
The description of this circuit on the advice tape adds an 
amount of $O(\log\ell) = O(\log S)$ to the space complexity  
and its simulation takes time $\poly(\ell) = \poly(S)$, which yields
an overall bound on the time of~$\poly(S,T)$. Hence, also part~(i) follows.
\end{proof}

Since the simulation of QTMs in Theorem~\ref{the:unidir_sim} is done
only approximately and the space ${O(S+\log\log (T/\epsilon))}$ needed for
the simulation increases with the running time we again obtain the
question in which cases we can bound the running time without
restricting the computational power of the model. Here we need a
statement for bounding the error probability away from $1/2$ in the case of
unbounded error and a construction of a probabilistic clock for QTMs.

\begin{lemma}\label{lem:err_boundQTM}
For each nonuniform QTM $M$ with algebraic amplitudes and running in space $S(n)$ there
exists a polynomial~$q$ such that for each input~${a\in\{0,1\}^n}$,
$p_{M,1}(a) > 1/2$ implies $p_{M,1}(a) \ge     1/2+2^{-q(2^{S(n)})}$
and $p_{M,1}(a) < 1/2$ implies $p_{M,1}(a) \le 1/2-2^{-q(2^{S(n)})}$.
\end{lemma}

\begin{lemma}\label{lem:comp_timeQTM}
  For each nonuniform QTM $M$ with bounded or unbounded error,
  algebraic amplitudes and running in space  $S(n)$, there is a QTM for the same
  function with algebraic amplitudes, the same type of error, the space bound $O(S(n))$ 
  and expected running time $2^{2^{O(S(n))}}$.
\end{lemma}

Lemma~\ref{lem:err_boundQTM} is proved in the same way as Lemma~\ref{lem:err_bound} since the matrix describing
the transition probabilities
in the proof in the same way describes transition probabilities of nonuniform QTMs. For the proof of
Lemma~\ref{lem:comp_timeQTM} we modify the given QTM $M$ in a way similar to the construction of
the QBP in the proof of Lemma~\ref{lem:comp_time}. Using the proof of Lemma~4.6 in Watrous~\cite{Wat99} 
it is easy to construct a QTM $M_t$ that for an appropriate $t=2^{O(S)}$ stops with probability $2^{-\Theta(t)}$ and continues
with probability $1-2^{-\Theta(t)}$. Using suitable versions of the lemmas of Bernstein and 
Vazirani~\cite{Ber97} for the construction of QTMs 
we modify $M$ in such a way that, before each computation step, it additionally performs $M_t$. 
By a reasoning similar to the proof of Lemma~\ref{lem:comp_time} we obtain a QTM with the behavior
claimed in Lemma~\ref{lem:comp_timeQTM}.
Using these results we easily obtain the following.

\begin{theorem}\label{the:nonunidir}\sloppy
The space complexity of nonuniform QTMs with algebraic amplitudes and bounded or unbounded
error is asymptotically equal to the space complexity of unidirectional nonuniform QTMs 
with the same kind of amplitudes and the same type of error, provided that these
space complexities are at least logarithmic in the input length.
\end{theorem}

\begin{proof}
Applying Lemmas~\ref{lem:err_boundQTM} and~\ref{lem:comp_timeQTM} to a nonuniform QTM that according to 
the hypothesis runs in space~$S$, we obtain a nonuniform QTM of the same kind running in expected time
$2^{2^{O(S)}}$. Analogously to the proofs in the last section, using Markov's inequality
to estimate the error of computations that take longer than time $2^{2^{O(S)}}$,
 Theorem~\ref{the:unidir_sim} yields a unidirectional nonuniform QTM of the desired kind running in space $O(S)$.
\end{proof}

 \section{Quantum OBDDs}\label{sec:qobdds}

{\sloppy
 Since for unrestricted branching programs no powerful lower bound
 methods are known, restricted variants of branching programs have been
 investigated in order to develop lower bound methods and to compare
 different modes of nondeterminism and randomization. 
 A simple variant of branching programs closely related to the uniform model
 of DFAs and to one-way communication complexity are ordered binary 
 decision diagrams (OBDDs). OBDDs are also used as a data structure for 
 the representation and manipulation of boolean functions, see, e.\,g.,
 Wegener~\cite{Weg00}. Hence, it is
 natural to investigate also the quantum variant of OBDDs.

}

 \begin{definition}\label{def_qobdd}
 A {\em quantum OBDD (QOBDD)} is a read-once QBP where on each path
 the variables are tested according to the same order. 
 \end{definition}
 
 Below, we prove upper and lower bound results for QOBDDs.  Before we
 do that, we discuss the definition of QOBDDs and their relationship
 to quantum finite automata. Furthermore, we define complexity classes in
 terms of the size of QOBDDs and compare them with the corresponding
 complexity classes for OBDDs.

Since on each path from the start node to a sink each variable is
tested at most once, QOBDDs are always acyclic.  Because of the definition of
QBPs, also QOBDDs are unidirectional. Different from
Definition~\ref{def_qobdd}, Ablayev, Gainutdinova, and
Karpinski~\cite{Abl01a} require QOBDDs to be leveled such that there
are edges only between adjacent levels. 
Proposition~\ref{Prop_leveled_QBP} shows that this
restriction is not crucial, because QOBDDs according to
Definition~\ref{def_qobdd} can be transformed into leveled QOBDDs
where the size increases by a factor of at most $(n+1)^2$.

Despite their superficial similarity,
there are some important differences between QOBDDs and 
($1$-way) quantum finite automata (QFAs). At the definition level, observe 
that, unlike QFAs, QOBDDs may read their input in an order different from $x_1,\ldots,x_n$.
Furthermore, they are a nonuniform model while QFAs are uniform.
This implies two less obvious  differences between QOBDDs and QFAs.
In general, measuring whether the computation has stopped and, if yes,
with which result, is allowed also during the computation of a QOBDD. 
The more restrictive
definition that allows end nodes to be reached only after exactly~$n$ computation steps
have been performed
is equivalent to our definition because of Proposition~\ref{Prop_leveled_QBP}. On the other hand, it is known 
that QFAs with and without such intermediate measurements are of different
power (Kondacs and Watrous~\cite{Kon97}). Furthermore, one
can decrease the error probability of a QOBDD with bounded 
error by \emph{probability amplification} below
any given constant, as for randomized OBDDs (see~\cite{Weg00} for the
randomized case). 
Again, this does not work for QFAs: Ambainis and Freivalds~\cite{Amb98} have 
shown that the language $\{a\}^*\{b\}^*$ can be recognized by QFAs with 
two-sided error~$0.318$, but not with error smaller than~$2/9$.

For QOBDDs, we distinguish the same types of error as for general QBPs
(see Definition~\ref{def:comp_qbps}). For characterizing the relative
power of the resulting different types of QOBDDs, it is useful to
define complexity classes with a naming convention analogous to that
used for QTMs. The class of functions that can be computed exactly by
polynomial size QOBDDs is called \mbox{EQP-OBDD}, and the class of functions
with polynomial size zero error (bounded-error) QOBDDs is called
ZQP-OBDD (BQP-OBDD). Similarly, the classes P-OBDD and BPP-OBDD of
functions with polynomial size deterministic OBDDs and polynomial size
randomized OBDDs with bounded error are defined. 
Furthermore, let Rev-OBDD
denote the class of functions with polynomial size reversible
OBDDs. The inclusions
$\mbox{Rev-OBDD}\subseteq\mbox{EQP-OBDD}\subseteq\mbox{ZQP-OBDD}\subseteq\mbox{BQP-OBDD}$
and $\mbox{Rev-OBDD}\subseteq\mbox{P-OBDD}\subseteq\mbox{BPP-OBDD}$
immediately follow from the definitions.

{\sloppy
In this section we present simple, concrete example functions
in order to prove that QOBDDs with bounded error and classical,
deterministic OBDDs are incomparable in power, i.\,e.,
$\mbox{P-OBDD}\not\subseteq\mbox{BQP-OBDD}$ and
$\mbox{BQP-OBDD}\not\subseteq\mbox{P-OBDD}$. We also present a partially
defined function in order to show a similar result  for QOBDDs and classical,
randomized OBDDs for partial functions.  Finally, we study the power of zero error and exact
quantum computation for OBDDs.   We 
prove that $\mbox{ZQP-OBDD}\subseteq\mbox{Rev-OBDD}$, i.\,e., zero error
QOBDDs can \emph{at best} be as good as reversible OBDDs. This
implies that  the three classes $\mbox{Rev-OBDD}$,
$\mbox{EQP-OBDD}$, and $\mbox{ZQP-OBDD}$ coincide and are strictly
contained in $\mbox{P-OBDD}$.

}

\subsection{A Function with Small QOBDDs that Requires Large Deterministic OBDDs}
\label{subsec_perm}

The \emph{permutation matrix test function} $\PERM_n$ is defined
on $n^2$ boolean variables that are arranged in a quadratic
matrix. The function takes the value $1$ iff each row and each column
contains exactly one entry $1$. It is well-known that $\PERM = (\PERM_n)_{n\in\IN}$ 
does not have polynomial size read-once branching programs (Krause, Meinel and
Waack~\cite{Kra91c}) and, therefore, no polynomial size OBDDs
either. In~\cite{Sa98a} (see also~\cite{Weg00}), 
a polynomial size randomized OBDD with one-sided error for $\PERM$ 
has been designed using the so-called fingerprinting
technique. We show here how this construction can be modified
to work for QOBDDs.

Let $X$ denote the input matrix and let
$x_j=(x_{j,0},\ldots,x_{j,n-1})$ denote the $j$th row of $X$. Let
$|x_j|=\sum_k x_{j,k}2^k$ denote the value of the $j$th row
interpreted as a binary number. The crucial observation is that
\[
 \PERM_n(X) \ =\  1\;\;\Leftrightarrow\;\; \sum_{j=0}^{n-1}|x_j|-(2^n-1)\ =\ 0\; \wedge\;\mbox{each
 $x_j$ contains exactly one entry $1$}.  
\]
The exact evaluation of the sum $S = \sum_{j=0}^{n-1} |x_j|-(2^n-1)$ requires OBDDs of exponential size.
Hence, $S$ is evaluated modulo a randomly chosen prime number
$p$. It is straightforward to construct a reversible OBDD $G^{(p)}$
that evaluates  $S\bmod p$ and 
simultaneously checks that each $x_j$ contains exactly one entry $1$.
In $G^{(p)}$ the variables are tested in a rowwise order. For each row it
has to be stored whether an entry $1$ has already been found. If a
second $1$ is found in some row, a $0$-sink is reached. Furthermore,
in each level the OBDD stores the partial sum of the terms
corresponding to the bits already read.  Since the partial sums are only stored
modulo~$p$, this increases the width merely by a factor of
$p$. Altogether, each level contains at most $2p$ interior nodes. 
Hence, the size of $G^{(p)}$ is $O(pn^2)$. It only accepts if
$S\bmod p$ is equal to $0$.

Now we construct a QOBDD $G$ for $\PERM_n$. Let $m=2n^2$ and let
$p_1,\ldots,p_m$ denote the $m$ smallest primes. By the prime number
theorem, $p_m=O(m\log m)=O(n^2\log n)$. We construct
$G^{(1)},\ldots,G^{(p_m)}$ and combine these reversible OBDDs by a
node labeled by the first variable with $m$ outgoing $c$-edges with amplitudes
$1/\sqrt{m}$ leading to the $c$-successors of the start nodes of $G^{(1)},\ldots,G^{(p_m)}$. This
 realizes a random choice between
$G^{(p_1)},\ldots,G^{(p_m)}$. The size of $G$ is bounded by $O(n^6\log n)$.

We estimate the error probability. The sum $S$ is
bounded above by $n2^n$. Hence, if $S$ is different from~$0$, it
has at most $n+\log n$ prime factors. Thus the probability of randomly
choosing a prime dividing $S$ is bounded above by $(n+\log n)/(2n^2)\leq 1/n$. 
This is also an upper bound on the error
probability of $G$. The error is one-sided, i.\,e., if
${\PERM_n(X)=1}$, then the QOBDD~$G$ always computes $1$, while it may
err if $\PERM_n(X)=0$. 
The probability can even be made smaller than $1/p(n)$ for any polynomial~$p$
by increasing the number of primes, which only increases the size of~$G$
polynomially. We have proved:

\begin{theorem}\label{the:PERM}
 There are QOBDDs for $\neg\PERM_n$ with one-sided error $1/n$
 and size $O(n^6\log n)$.  
\end{theorem}

\begin{corollary} $\mbox{BQP-OBDD}\not\subseteq\mbox{P-OBDD}$.
\end{corollary}

\subsection{Functions with Small Deterministic OBDDs that Require Large QOBDDs} 
\label{subsec_no}
 
The disjointness function and the inner product function are defined by
$\DISJ_n(x_1,\ldots,x_n)=(\wbar{x}_1\vee \wbar{x}_2)\wedge
(\wbar{x}_3\vee\wbar{x}_4)\wedge\cdots\wedge
(\wbar{x}_{n-1}\vee\wbar{x}_n)$ and
$\IP(x_1,\ldots,x_n)=x_1x_2\oplus\cdots\oplus x_{n-1}x_n$, where $n$
is an even number. Both functions are extensively investigated in
communication complexity, see, e.\,g., \cite{Kus97}. For the variable
order $x_1,\ldots,x_n$ they have OBDD size $O(n)$, since it suffices
to store at most two bits at each level of the OBDD, namely, the value
of the variable read in the last step and the value that the function
takes on the variables up to the last variable with an even index.
However, both functions are difficult for QOBDDs and, therefore, also for
reversible OBDDs, since these OBDD models have difficulties in
``forgetting'' variables read.

The lower bound proof uses some ideas due to Nayak~\cite{Nay99a} based on quantum
information theory. We
briefly introduce the required notions and facts. For a proper introduction
to quantum information theory we refer to~\cite{Nie00}. 
Recall that a \emph{mixed state} of a quantum system is a probability distribution 
of pure quantum states. A mixed state is usually described by its 
\emph{density matrix}, which is a positive matrix with unit trace. The density 
matrix for the probability distribution $(p_i,\k{\phi_i})_i$ is $\sigma=\sum_i
p_i\k{\phi_i}\b{\phi_i}$. A state resulting
from the application of the unitary transformation $U$ to the state
described by the density matrix $\sigma$ is described by the density
matrix $U\sigma U^\dagger$. Now assume that $(\k{\psi_i})_i$ is an
orthonormal basis of eigenvectors of $\sigma$ and that $\lambda_i$ is
the eigenvalue belonging to~$\k{\psi_i}$. Then the {\em von Neumann entropy} of
$\sigma$ is defined as $S(\sigma)=-\sum_i \lambda_i\log\lambda_i$.  
The von Neumann entropy is invariant under unitary
transformations~$U$, i.\,e., $S(U\sigma U^\dagger)=S(\sigma)$.
Furthermore, if $\sigma$ is a density matrix over a 
(finite-dimensional) Hilbert space~$\H$, then $S(\sigma) \le \log(\dim(\H))$.
Finally, we formally introduce the kind of measurements that are relevant here.

\begin{definition}
Let $J$ be a finite index set and let $\M = (P_i)_{i\in J}$ be a family of projection operators 
over the finite-dimensional Hilbert space~$\H$ with $\sum_{i\in J} P_i = I$. 
Then call $\M$ a \emph{projective measurement over $\H$ with results in $J$}.
For any density matrix~$\sigma$ over~$\H$, 
define the \emph{probability of measuring result~$i\in J$ in the state
described by~$\sigma$} by $\Pr\{ \M(\sigma) = i \} = \tr(\sigma P_i)$.
\end{definition}

The following lemma is due to
Nayak. In the lemma, $H(p)$ denotes the binary entropy function
defined by $H(p) = -p\log p - (1-p)\log(1-p)$.

\begin{lemma}[\cite{Nay99a}]\label{lem:nayak}
 Let $\sigma_0$ and $\sigma_1$ be density matrices over the finite-dimensional 
 Hilbert space~$\H$ and let $\sigma=1/2\cdot (\sigma_0+\sigma_1)$. Suppose there is a
 projective measurement $\M = (P_0,P_1)$ over~$\H$ with results in~$\{0,1\}$
 such that for $b\in\{0,1\}$, $\Pr\{ \M(\sigma_b) = b \} \ge p \ge 1/2$.
 Then $S(\sigma)\geq(S(\sigma_0)+S(\sigma_1))/2+(1-H(p))$.
\end{lemma}

Now we are ready to prove the main result of this section, which is
stated in the following theorem. The corollary directly follows from
the upper bound on the OBDD size mentioned above.

\begin{theorem}\label{the:NO}
  The size of each QOBDD with bounded error for $\DISJ_n$ or $\IP_n$
  is  $2^{\Omega(n)}$.
 \end{theorem}

 \begin{corollary} $\mbox{P-OBDD}\not\subseteq\mbox{BQP-OBDD}$.
 \end{corollary}

\begin{proof}[Proof of Theorem~\ref{the:NO}]
  We only prove the statement for disjointness, the claim for the
  inner product follows in the same way.  Let a QOBDD $G$ with some
  variable order $\pi$ for $\DISJ_n$ be given. W.l.o.g.\ let $G$ be
  leveled. Due to the symmetry of the OR-function, we may assume
  w.l.o.g.\ that for each $i\in\{1,\ldots,n/2\}$ the variable
  $x_{2i-1}$ is tested before $x_{2i}$ in $\pi$.
Let $p=1/2+\varepsilon$ be a lower bound on the success probability of $G$. We generate
random inputs $x$ for $\DISJ_n$ in the following way. Each variable
  with an odd index  gets one of the values $0$ and $1$ with a probability
of $1/2$ each. All variables with an even index get the value $0$. 
Let $\sigma(k)$
denote the density matrix describing the state of the QOBDD after
reading the $k$th variable with an odd index. By induction we prove
$S(\sigma(k))\geq (1-H(p))k$.  Since the state of the QOBDD before
reading the first randomly chosen variable is a pure state, we have
$S(\sigma(0))=0$. Now let $k\geq 1$. By induction hypothesis
$S(\sigma(k-1))\geq (1-H(p))(k-1)$.  
Let $x_i$ be the $k$th variable with an odd index.
Let $U_0$ and $U_1$ be the unitary
transformations performed by the QOBDD while reading all the variables after
the \mbox{$(k-1)$-st} variable with an odd index and up to $x_i$
inclusively, where the latter gets the value~$0$ or~$1$, resp. Since $x_i$ is
chosen to be~$0$ or~$1$ at random,
\[
 \sigma(k) \ =\  \frac{1}{2}\left(U_0\sigma(k-1) U_0^\dagger +
 U_1\sigma(k-1) U_1^\dagger\right).  
\] 
Let $U$ denote the composition of the unitary transformations
performed by the QOBDD if the partner $x_{i+1}$ of $x_i$ gets the value $1$ and
all other variables read after $x_i$ get the value~$0$. Then the
function $\DISJ_n$ attains the value $c\in\{0,1\}$ if $x_i = \wbar{c}$.  Let
$\sigma=U\sigma(k)U^\dagger$. Since the QOBDD computes the function
$\DISJ_n$, the measurement of the QOBDD on $\sigma$ yields the result
${c}$ with a probability of at least $p$ if $x_i$ has the value $\wbar{c}$. By
Lemma~\ref{lem:nayak} and the invariance of the von Neumann entropy
under unitary transformations, 
\begin{align*} 
  S(\sigma(k)) &\ =\  S(\sigma) \ \geq\  \frac{1}{2}\left( S(UU_0 \sigma(k-1) U_0^\dagger U^\dagger) +S(UU_1 \sigma(k-1) U_1^\dagger U^\dagger) \right) +1-H(p)\\ 
               &\ \geq\ \frac{1}{2}\left(S(\sigma(k-1))+S(\sigma(k-1))\right)+1-H(p).  
\end{align*} 
\sloppy\hbadness=3000
Then the claim follows by the induction hypothesis. We obtain the
lower bound $(1-H(p))\cdot n/2$ on the von Neumann entropy of
the density matrix describing the state of $G$ after reading all
variables with odd indices. By the above remark,
this implies the lower bound $2^{(1-H(p))\cdot n/2}$ on the 
dimension of the state space of~$G$ and, therefore, also on the 
size of~$G$.
\end{proof}

\subsection{A Partial Function with Small QOBDDs that Requires Large 
Randomized OBDDs}
\label{subsec_raz}

An OBDD or QOBDD for a partially defined function has to
compute the correct value of the function only on the domain of the
function, while it may compute an arbitrary result on inputs outside
the domain. We present a partially defined function with polynomial
size QOBDDs but only exponential size randomized OBDDs. The idea behind the
construction of the function is based on a result of Raz~\cite{Raz99}
for communication protocols.

The function we consider gets unitary matrices as inputs. In order to
obtain a finitely representable function, we redundantly encode
sufficiently precise approximations of the desired matrices by 
boolean variables. The redundancy in the encoding will allow us to
prove a lower bound for arbitrary variable orders.

For the following, fix an even $n\in\IN$ and let $\epsilon>0$.
Let $b = 6(n-1)$ and let $W_0,\ldots,W_{b-1}$ be some fixed enumeration of
the matrices in ${\cal G}_n$ from Lemma~\ref{lem:myhrc}.
Let $k=k(n,\epsilon)=O(n^2\log(n/\epsilon))$ be the number from this lemma. 
For $\ell\ge k$ and $m\ge b-1$ 
the {\em universal $(\epsilon,\ell,m)$-code} of $n\times n$-matrices consists of the
$\ell (m+1)$ boolean variables $x_{i,j}$, ${1\leq i\leq \ell}$, ${1\leq j\leq m+1}$. 
For $1\le i\le\ell$ let $x_i=(x_{i,1},\ldots,x_{i,m+1})$ and
$v(x_i)=x_{i,1}+\cdots+x_{i,m}$. 
Let
\[
U_i\ =\ \left\{\begin{array}{ll}
W_{(v(x_1)+\cdots+v(x_i))\bmod b}, & \mbox{if $x_{i,m+1}=1$;} \\
I,                               & \mbox{if $x_{i,m+1}=0$.}
\end{array}
\right.
\]
Then the variable vector $x=(x_1,\ldots,x_\ell)$ encodes the matrix 
\[
W(x)\ =\  U_{\ell}\bdot U_{\ell-1} \bdot \cdots\bdot U_{1}.
\]
Note that the variables $x_{i,m+1}$ only switch between
$W_{(v(x_1)+\cdots+v(x_i))\bmod b}$ and the identity matrix. In
particular, they do not influence the sum $(v(x_1)+\cdots+v(x_i))\bmod b$.
By Lemma~\ref{lem:myhrc}, for each unitary $n\times n$-matrix~$U$
there is a setting to the $x$-variables 
such that $\Vert U-W(x)\Vert\leq\epsilon$. In the following, $\ell$ is much larger than $k$
such that there are many settings to obtain a certain unitary matrix
in the product approximating~$U$.

Now we define the considered function. Let $\k{1},\ldots,\k{n}$ be the
standard basis of $\C^n$. Let $V_0$ and $V_1$ denote the subspaces
spanned by the first and last $n/2$ of these basis vectors. Let
${0<\vartheta<1/\sqrt{2}}$. The input for the function
$R_{\vartheta,\ell,m,n}$ consists of $3\ell (m+1)$ boolean variables
$a_{i,j},b_{i,j},c_{i,j}, 1\leq i\leq \ell, 1\leq j\leq m+1$, which are
interpreted as universal $(\epsilon,\ell,m)$-codes for three unitary
$n\times n$-matrices $A,B,C$, where $\epsilon = 1/(3n)$. The
function takes the value $z\in\{0,1\}$ if the Euclidean distance
between $CBA\k{1}$ and~$V_z$ is at most $\vartheta$. Otherwise the
function is undefined.

We first prove the upper bound on the size of QOBDDs.

\begin{theorem}\label{the:raz_func_ub}\sloppy
Let $0 < \vartheta < 1/\sqrt{2}$.  The function
${R}_{\vartheta,3k,9kb,n}$ with an input size of $N = {81\,k^2 b+9\,k} = {O(n^5\log^2 n)}$
has QOBDDs with error at most~$\vartheta^2$
and size $O(N^{9/5}/\log^{8/5} N)$.
\end{theorem}

\begin{proof}
Set $\ell = 3k$ and $m = 9kb$. 
We choose the variable order that starts with the $a$-variables ordered as
$a_{1,1},\shortldots,a_{1,m+1}, a_{2,1},\shortldots,a_{2,m+1},\ldots,a_{\ell,1},\shortldots,a_{\ell,m+1}$.
Afterwards the $b$-variables and then the $c$-variables are tested in
analogous orders. We first describe a subgraph~$G_A$ of the QOBDD evaluating
the $a$-variables. Analogous subgraphs~$G_B$ and~$G_C$ 
are constructed for the $b$- and $c$-variables, resp.

The nodes of~$G_A$ are arranged in~$bn$ columns, which we
label by $(r,s)$ with ${0\le r\le b-1}$ and ${1\le s\le n}$, and in levels $1,\ldots,\ell(m+1)+1$.
Let $\k{r}\k{s}\k{t}$ be the vector from an orthonormal basis that 
corresponds to the node of the $t$th level in column $(r,s)$.
The nodes in each of the first~$\ell(m+1)$ levels are labeled by the same
$a$-variable according to the variable order. The last level consists of sinks.
Let $p\in\{1,\ldots,\ell\}$. For $j\in\{1,\ldots,m\}$, 
the node labeled by~$a_{p,j}$ in column $(r,s)$ is left by a single $0$-edge with amplitude~$1$ 
leading to  the node of the next level of the same column and a single $1$-edge with amplitude~$1$
leading to the node of the next level in column $((r+1)\bmod b,s)$. 
For a node labeled by~$a_{p,m+1}$ in column~$(r,s)$,
a single $0$-edge with amplitude~$1$ leaving this node leads to the node in column
$(r,s)$ of the subsequent level. There are $1$-edges connecting this node to the
nodes of the subsequent level such that the mapping 
$\k{r}\k{s}\k{t} \mapsto \k{r}(W_{r}\k{s})\k{t+1}$ is performed,
where $t = (p-1)(m+1)+m+1$.

{\sloppy
It is easy to verify that the graph~$G_A$ constructed in this way is well-formed
and unidirectional. We evaluate $G_A$ according to the semantics 
of QBPs starting from a node on the first level in column~$(r,s)$, i.\,e.,
with the superposition~$\k{r}\k{s}\k{1}$.
Then after reading the variable vectors $a_1,\ldots,a_p$, where $a_i=(a_{i,1},\ldots,a_{i,m+1})$, 
we reach the superposition $\k{r'}(U_p\bdot\cdots\bdot U_1\k{s})\k{t}$
with $r' = {(r+v(a_1)+\cdots+v(a_p))\bmod b}$ and $t = (p-1)(m+1)+m+2$.

}

The QOBDD for ${R}_{\vartheta,\ell,m,n}$ starts with $G_A$, where
the node on the first level in column~$(0,1)$ is chosen as the start node.
Then the amplitude for reaching a node of the $(\ell(m+1)+1)$-st
level in column~$(r,s)$ of~$G_A$ is exactly the $s$th coordinate of $A\k{1}$,
if $r$ is the sum modulo $b$ of all $a$-variables, and $0$ otherwise. 
After reading the $a$-variables, the value of $r$ is no longer needed; however, it 
cannot be erased in a QOBDD. Hence, for each possible value $r$ we add a copy of a 
subgraph~$G_B$ processing the variables encoding $B$ in the same way as described
before for $A$. The sink in column~$(r,s)$ of the $(\ell(m+1)+1)$-st level
of the subgraph~$G_A$ for~$A$ is identified with the node $(0,s)$ of the $r$th copy
of the subgraph~$G_B$ for $B$.  Altogether $b$ copies of the subgraph~$G_B$
are sufficient. In the same way $b^2$ copies of a subgraph~$G_C$ for
processing~$C$ are sufficient. In each copy of $G_C$, the sink in column~$(r,s)$ 
of the last level is a $0$-sink if $s\leq n/2$, and a $1$-sink otherwise. 
For each input, there is exactly one copy of~$G_C$ and exactly one~$r$ such that for all~$s$ 
the amplitude of the node in column~$(r,s)$ of the last level equals the $s$th coordinate of
$CBA\k{1}$. For all other copies of~$G_C$ and for all other~$r$ the amplitudes are~$0$. 

 Let $E_z$
denote the projection to the subspace~$V_z$. If $\k{y} = CBA\k{1}$ has
distance at most~$\vartheta$ from the subspace $V_z$, we have
$\vartheta^2\geq \Vert\k{y}-E_z\k{y}\Vert^2 = 1-\Vert
E_z\k{y}\Vert^2$. The equality follows by an easy calculation. Hence,
the measurement on the level of the sinks leads to the result~$z$ with
probability $\Vert E_z\k{y}\Vert^2 \geq 1-\vartheta^2$.  The size of
the QOBDD is dominated by the $b^2$ copies of~$G_C$.
Each of these copies has size $O(bnN)$. Hence, the size can be estimated by
$O(b^3nN)=O(n^9\log^2 n)=  O(N^{9/5}/\log^{8/5} N)$.
\end{proof}

In order to prove the lower bound, we apply arguments from
communication complexity (see, e.\,g., \cite{Hro97,Kus97} for an introduction).
We first state a result of Raz~\cite{Raz99}, who has proved a lower bound on the
communication complexity for a different function
$R^0_{\vartheta,n}$. Using two rectangular reductions, which
are defined below, we transfer this lower bound to a lower bound on the
communication complexity of $R_{\vartheta,\ell,m,n}$ for any $\ell\ge k$ and $m\ge b-1$.
Finally, by a standard lower bound technique for randomized OBDDs, the lower 
bound on the communication complexity implies a lower bound on the size of
randomized OBDDs.

We define the function~$R^0_{\vartheta,n}$ due to Raz by
describing the corresponding communication problem.
Let $0<\vartheta<1/\sqrt{2}$.  The input of
Alice consists of a unit vector $x\in\IR^n$ and two orthogonal
subspaces $S_0$ and $S_1$ of $\IR^n$ of dimension $n/2$ each. Bob
gets an orthogonal real-valued $n\times n$-matrix $T$ as input. The
output is $c\in\{0,1\}$ if $Tx$ has distance at most $\vartheta$
from~$S_c$, and arbitrary otherwise.  We remark that the usual
definition of communication complexity can easily be extended to the
case of infinite input sets which is considered here. Raz
has proved the following result.

\begin{theorem}[\cite{Raz99}]\label{the:raz}
Let $0 < \vartheta < 1/\sqrt{2}$.
Each randomized communication protocol with bounded error for
 $R^0_{\vartheta,n}$ requires $\Omega(n^{1/2})$ bits of communication.
\end{theorem}

We note that the
considered communication problems are partially defined. 
On inputs for which such a problem is not defined, both outputs $0$ and $1$ are allowed. 
A partially defined communication problem on input sets $X$ and $Y$ can also be
described by a relation $R\subseteq X\times Y\times\{0,1\}$, where
$(x,y,z)\in R$ iff $z$ is a valid output for $(x,y)$.  In particular,
if the problem is undefined for $(x,y)$, we have $(x,y,0),(x,y,1)\in
R$. A \emph{rectangular reduction} from {$R'\subseteq X'\times
Y'\times\{0,1\}$} to {$R\subseteq X\times Y\times \{0,1\}$} consists of
two mappings $f\colon X'\to X$ and $g\colon Y'\to Y$ such that
{$(f(x),g(y),z)\in R \Rightarrow (x,y,z)\in R'$}. It is easy to see that
a lower bound on the communication complexity for $R'$ implies the
same lower bound for $R$ if there is a rectangular reduction from $R'$
to~$R$.

We  observe that the problem $R^0_{\vartheta,n}$ can easily be
reduced to the following infinite precision variant $R'_{\vartheta,n}$
of the considered problem $R_{\vartheta,\ell,m,n}$.
The input of $R'_{\vartheta,n}$ consists of
unitary $n\times n$-matrices $A$, $B$ and $C$, where Alice gets $A$
and $C$, and Bob gets $B$. Their task is to compute $z\in\{0,1\}$ if the
distance between $CBA\k{1}$ and $V_z$ is bounded by
$\vartheta$. (Again, $V_0=\sp\{\k{1},\ldots,\k{n/2}\}$ and
$V_1=\sp\{\k{n/2+1},\ldots,\k{n}\}$.) Obviously,
$R^0_{\vartheta,n}$ is a special case of $R'_{\vartheta,n}$. Instead
of an orthogonal matrix $T$, a unitary matrix $B$ is allowed. The
vector $x$ and the subspaces $V_0$ and $V_1$ are now encoded by the
unitary matrices $A$ and $C$. Hence, the lower bound from
Theorem~\ref{the:raz} also holds for $R'_{\vartheta,n}$.  The second
rectangular reduction is given in the following lemma.

\begin{lemma}\label{lem:red}
For all constants $\vartheta,\vartheta'$ with
$0\le\vartheta'<\vartheta<1/\sqrt{2}$, for all $\ell \ge k$ and $m\ge b-1$, 
and for sufficiently large $n$, $R'_{\vartheta',n}$ is reducible to~${R}_{\vartheta,\ell,m,n}$.
\end{lemma}

\begin{proof}
Let $(A',B',C')$ be an arbitrary input for $R'_{\vartheta',n}$. We map this
input to an input for ${R}_{\vartheta,\ell,m,n}$ consisting of the universal
$(\epsilon,\ell,m)$-codes of
unitary $n\times n$-matrices $A,B,C$ with
\[
  \Vert A-A'\Vert \le \epsilon,\quad
  \Vert B-B'\Vert \le \epsilon,\quad\text{and}\quad
  \Vert C-C'\Vert \le \epsilon,
\]
where $\epsilon = 1/(3n)$. By Lemma~\ref{lem:myhrc}, we can find such
an input $(A,B,C)$ for ${R}_{\vartheta,\ell,m,n}$.  We show that this mapping
is even a rectangular reduction.  Let $E_0$ and $E_1$ be the
projections on the subspaces $V_0$ and $V_1$, resp. Let $\k{y} =
CBA\k{1}$ and $\k{y'} = C'B'A'\k{1}$.

Let w.\,l.\,o.\,g.\ $0$ be a solution of ${R}_{\vartheta,\ell,m,n}$ for the input
$(A,B,C)$. Then either $\Vert \k{y} - E_0 \k{y}\Vert \le \vartheta$,
i.\,e., the only valid output is $0$, or $\Vert \k{y} - E_0 \k{y}\Vert >
\vartheta \wedge \Vert \k{y} - E_1 \k{y}\Vert > \vartheta$, i.\,e., the
outputs $0$ and $1$ are allowed. This is equivalent to 
$\Vert \k{y} - E_1 \k{y}\Vert > \vartheta$. We prove that
$0$ is also a solution of the problem $R'_{\vartheta',n}$ for the
input $(A',B',C')$ by showing that
$\Vert\k{y'}-E_1\k{y'}\Vert>\vartheta'$.
  
By the choice of $A$, $B$ and $C$ and by
Proposition~\ref{prop:product_approx}, we obtain $\Vert \k{y'} - \k{y}\Vert \le 3\epsilon = 1/n$. 
By the assumption, $\Vert E_0\k{y}\Vert= \Vert \k{y} - E_1 \k{y}\Vert > \vartheta$. Hence,
$
  \Vert\k{y}-E_0\k{y}\Vert 
    \ =\  \bigl(1 - \Vert E_0\k{y}\Vert^2\bigr)^{1/2}
    \ <\  \bigl(1-\vartheta^2\bigr)^{1/2}
$
and thus
\[
  \Vert E_1 \k{y'}\Vert
    \ \leq\  \Vert E_1(\k{y'}-\k{y})\Vert + \Vert E_1\k{y}\Vert
    \ \leq\  \Vert\k{y'}-\k{y}\Vert + \Vert\k{y}-E_0\k{y}\Vert
    \ <\  \frac{1}{n} + \bigl(1-\vartheta^2\bigr)^{1/2}.
\]
This implies
$\Vert \k{y'} - E_1 \k{y'}\Vert^2 = 1 - \Vert E_1 \k{y'}\Vert^2 > \vartheta^2 - o(1)$.
Since $\vartheta' < \vartheta$ and both $\vartheta,\vartheta'$ are
constants, it follows that $\Vert\k{y'}-E_1\k{y'}\Vert>\vartheta'$ for
sufficiently large $n$. Hence, $0$ is a solution of
$R'_{\vartheta',n}$ for the input $(A',B',C')$.
\end{proof}

Altogether we obtain a lower bound on the communication complexity of
$R_{\vartheta,\ell,m,n}$ for $\ell \ge k$ and $m\ge b-1$.

\begin{corollary}\label{cor:raz}
\sloppy
Let $0 < \vartheta < 1/\sqrt{2}$, $\ell \ge k$, and $m\ge b-1$.  Each randomized communication
protocol with bounded error for ${R}_{\vartheta,\ell,m,n}$ where Alice has
the matrices $A$ and $C$ and Bob the matrix $B$ requires
$\Omega(n^{1/2})$ bits of communication.
\end{corollary}

Now we can prove the second part of the main result of this section,
the lower bound on the size of randomized OBDDs with bounded error.

\begin{theorem}
\label{the:raz_func_lb}
Let $0<\vartheta<1/\sqrt{2}$. Each randomized OBDD with bounded error for
the function $R_{\vartheta,3k,9kb,n}$ on $N=81k^2b+9k=O(n^5\log^2n)$
variables has size
$2^{\Omega(N^{1/10}/\log^{1/5} N)}$.
\end{theorem}

It remains open to find an example of a \emph{total} function with polynomial
size QOBDDs but only exponential size randomized OBDDs. Using the
currently available techniques, this seems to be difficult
since the known lower bound techniques for randomized
OBDDs, which are based on randomized communication complexity, also
work in the quantum case (see Klauck~\cite{Kla00}).

\begin{proof}[Proof of Theorem~\ref{the:raz_func_lb}]
  Let~$G$ be a given randomized OBDD for ${R}_{\vartheta,\ell,m,n}$
  with $\ell = 3k$ and $m = 9kb$ and with an arbitrary variable
  order.  In general, the variables encoding the matrices $A$,
  $B$, and~$C$ do not occur as contiguous groups in the variable order. Because of
  the redundancy of the encoding of the matrices we can construct a
  suborder where the variables of each of the encodings of $A$, $B$, and~$C$ 
  are grouped together such that the corresponding subproblem of
  $R_{\vartheta,\ell,m,n}$ is still hard.  Then we can apply the above
  communication complexity lower bound. Let $\pi$ denote the order of
  the variables $a_{i,j},b_{i,j},c_{i,j}$, $1\leq i\leq \ell$, $1\leq
  j\leq m$, in $G$.   For $A$ (and similarly $B$
  and~$C$) call each set of variables $a_{i,1},\ldots,a_{i,m}$ in its
  encoding a \emph{block}. The variables $a_{i,m+1}$, $b_{i,m+1}$, and
  $c_{i,m+1}$ do not occur in any block or in $\pi$.

\begin{claim*} There is a suborder~$\pi'$
of~$\pi$ such that for each  matrix of $A$, $B$ and $C$ there are exactly~$k$
consecutive blocks in $\pi'$  that each contain exactly~$b$ 
variables. 
\end{claim*}

\begin{proof}[Proof of the claim]
Think of~$\pi$ as a list of all variables (except $a_{i,m+1}$,
$b_{i,m+1}$, and $c_{i,m+1}$) in the prescribed order.
Observe that there are $9k$ blocks of $m = 9kb$ variables each encoding
some matrix from the set~${\cal G}_n$.

We divide~$\pi$ into $9k$ contiguous parts such that for each
block there is a part that contains at least~$b$ of its variables
and such that for different blocks there are different parts with this
property.  The first of these parts is chosen by searching for the
first position in the variable order $\pi$ where for some block $b$
variables have been tested (and hence for all other blocks less than $b$
variables have been tested). Then this block is chosen and the other
variables up to the chosen position are eliminated. Furthermore, all
other variables of the chosen block are eliminated.  An easy induction
shows that this procedure can be iterated until $9k$ parts are chosen.
Thus we are left with $9k$ smaller blocks with exactly~$b$ variables
each and such that for each original block there is a smaller block in
the list.  

We now use the same idea to partition the list of variable blocks
obtained in the first step into three parts such that for each of the
three matrices there is a part containing at least~$k$ of its blocks
and such that for different matrices there are different parts with
this property. Again we eliminate variables in order to ensure that
for each matrix exactly $k$ consecutive blocks remain in the variable
order. In this way, we obtain a variable order~$\pi'$ with the desired
properties. 
\end{proof}

We replace all eliminated variables with~$0$ and remove the
nodes labeled by these variables in the randomized OBDD and
redirect incoming edges to the $0$-successor. Furthermore, 
if all variables $a_{i,1},\ldots,a_{i,m}$ of a block are eliminated, we also replace
$a_{i,m+1}$ with~$0$ and modify the randomized OBDD accordingly. 
The same is done for the eliminated blocks of $b$- and $c$-variables.  
This yields a randomized OBDD~$G'$ for $R_{\vartheta,k,b,n}$ that is at most as
large as~$G$.

We prove the desired lower bound for~$G'$ using the standard
lower bound technique for randomized OBDDs (see, e.\,g., \cite{Weg00}). 
Observe that the variable order~$\pi'$ consists of three parts 
belonging to the different matrices $A,B,C$ in some arbitrary 
order. Let~$C_1$ be the set of nodes which are reached
by some path on which exactly the variables for the first matrix according to~$\pi'$
have been tested, and let~$C_2$ be the set of nodes which are reached by some path on
which exactly the variables in the first two matrices have been tested. The OBDD
can be used to build a randomized one- or two-round communication protocol for
${R}_{\vartheta,k,b,n}$ where \mbox{Alice} has the variables for~$A$ 
and~$C$ and Bob the variables for~$B$.  The players jointly follow a
computation path in the OBDD from the start node to a sink, using random
bits for decisions at random nodes of the OBDD and communicating the
numbers of nodes in the sets~$C_1$ and~$C_2$.  The communication
complexity of this protocol is bounded by
$\ceil{\log|C_1|}+\ceil{\log|C_2|} \le 2(\log|G'|+1)$. Together with
Corollary~\ref{cor:raz}, this yields the claimed lower bound.
\end{proof}

\subsection{Las Vegas QOBDDs Versus Reversible OBDDs}
 \label{subsec_lvqobdd}

 The main result of this section is that
 $\mbox{ZQP-OBDD}\subseteq\mbox{Rev-OBDD}$.  This means that even the
 zero-error QOBDD model with some failure probability is no more powerful with
 respect to polynomial size than reversible OBDDs.  

The essence of the proof is as follows.
Given a reversible OBDD~$G$ and a Las Vegas QOBDD~$G'$ for the same function and with the
same variable order, we show that $G'$ induces collections of measurements, called measurement
schemes here, that allow to distinguish the subfunctions represented at each of the levels of~$G$.
We further prove that for such a measurement scheme, the dimension
of the underlying Hilbert space can be lower bounded in terms of the number of those 
subfunctions. Altogether, we obtain a lower bound on the size of
the Las Vegas QOBDD~$G'$ in terms of the size of the reversible OBDD~$G$.

\begin{definition}\label{def_meas_scheme}
Let $\H$ be a finite-dimensional Hilbert space and
let $\k{v_1},\ldots,\k{v_m}\in\H$ be different pure quantum
states. Let $X = \{1,\ldots,m\}$ and $Y = \{1,\ldots,n\}$.  Call an
$m\times n$-matrix $A = (a_{ij})$ with entries in $\{0,1,*\}$ and
projective measurements $\M_j = (M_{j,0},M_{j,1},M_{j,{\rm ?}})$ with possible results
$\{0,1,{\rm ?}\}$, where $j=1,\ldots,n$, a \emph{measurement scheme for $\k{v_1},\ldots,\k{v_m}$ with
zero error and failure probability $\epsilon$}, $0\le\epsilon < 1$, if
\begin{shortindent}{(iii)}
\item[(i)]  for all different $i,j\in X$ there is a $k\in Y$ 
  such that $a_{ik},a_{jk}\in\{0,1\}$ and $a_{ik} \neq a_{jk}$;
\item[(ii)] for all $i\in X$ and $j\in Y$,
  if $a_{ij} = *$, then $a_{ik} = *$ for all $j\le k\le n$; and
\item[(iii)] for all $i\in X$ and $j\in Y$,
  if $a_{ij}\in\{0,1\}$, then $\Pr\{ \M_j(\k{v_i}) = a_{ij} \} \ge 1-\epsilon$
  and $\mbox{$\Pr\{ \M_j(\k{v_i}) = \neg a_{ij} \}$} = 0$.
\end{shortindent}
\end{definition}

A measurement scheme allows us to distinguish any pair of vectors from
$\k{v_1},\ldots,\k{v_m}\in\H$ by zero error measurements.
Our aim is to prove a lower bound on the dimension of~$\H$ in terms of~$m$.
For this, we use the following lemma due to Klauck~\cite{Kla00}, which is a Las Vegas
variant of Lemma~\ref{lem:nayak}.

\begin{lemma}[\cite{Kla00}]\label{lem:lv_holevo}
\sloppy
Let $\sigma_0,\sigma_1$ be density matrices over~$\H$ and
let $0\leq p\leq 1$.
Suppose that there is a projective measurement
$\M = (M_0,M_1,M_{\rm ?})$ with possible results~$\{0,1,{\rm ?}\}$ such that
$\Pr\{ \M(\sigma_b) = b \} \ge 1-\epsilon$ and
$\Pr\{ \M(\sigma_b) =\neg b\} = 0$ for all~$b\in\{0,1\}$.
Let $\sigma = p \sigma_0+(1-p)\sigma_1$. 
Then $S(\sigma) \ge p S(\sigma_0) + (1-p) S(\sigma_1) + (1-\epsilon)H(p)$.
\end{lemma}

The following lemma extends a result of
Klauck~\cite{Kla00} that gives a lower bound on the Las Vegas one-way
quantum communication complexity in terms of deterministic one-way
communication complexity. The proof of Klauck provides the main idea of the proof of
Lemma~\ref{lem:lv_mm} for measurement schemes without
``$*$''-entries.

\begin{lemma}\label{lem:lv_mm}
Let $\k{v_1},\ldots,\k{v_m}\in\H$ be different pure quantum states.
If there is a measurement scheme for $\k{v_1},\ldots,\k{v_m}$ with
zero error and failure probability~$\epsilon$, then $\dim(\H)\ge
m^{1-\epsilon}$.
\end{lemma}

\begin{proof} 
Let $A$ be the $m\times n$-matrix with entries from $\{0,1,*\}$,
and let $\M_1,\ldots,\M_n$ be the projective measurements 
in the given measurement scheme for $\k{v_1},\ldots,\k{v_m}$. 
Let $X = \{1,\ldots,m\}$ and $Y = \{1,\ldots,n\}$.
Call two rows of a $A$ \emph{distinguishable} if they differ in a column where
both of them have boolean values. Thus the rows of~$A$ are 
pairwise distinguishable according to the hypothesis. 

In the following we inductively define a mixed state over $\H$ with 
large von Neumann entropy in order to obtain the lower bound on the
dimension of $\H$. The mixed states that we consider are convex
combinations of the pure states $\sigma_i = \k{v_i}\b{v_i}$,
$i=1,\ldots,m$.  For any $I\subseteq X$, $j\in Y$, and $b\in\{0,1\}$
let $I_{j,b} = \{ i\in I \mid a_{ij} = b \}$.
\begin{shortindent}{(ii)}
\item[(i)] 
  For $I\subseteq X$ with $|I|\ge 2$ and $j\in Y$ such that all
  rows in the submatrix $I\times\mbox{$\{j,j+1,\ldots,n\}$}$ of $A$ are
  distinguishable, let
  $\sigma(I,j) = (|I_{j,1}|/|I|)\cdot\sigma(I_{j,1},j+1) + 
    (|I_{j,0}|/|I|)\cdot\sigma(I_{j,0},j+1)$.
\item[(ii)] Let $\sigma(\{i\},j) = \sigma_i$ for $i\in X$ and $1\le j\le n+1$.
\end{shortindent}

If the rows in the submatrix $I\times\mbox{$\{j,j+1,\ldots,n\}$}$ of
$A$ are distinguishable, by condition (ii) of
Definition~\ref{def_meas_scheme} the $j$th column of the submatrix only contains the
entries $0$ and $1$: If it contained an entry~``$*$'', the whole row would
consist of~``$*$'' and would thus not be distinguishable from the other
rows. It follows that $\sigma(X,1)$ is well defined by a recursive
application of the above definition, since (by induction), all rows in~$I$ 
are pairwise distinguishable as long as $|I|\ge 2$, in which case part~(i)
is applicable. After some applications of part~(i), finally 
part~(ii) is applicable. 

\begin{claim*}
For each $I\subseteq X$ and $j\in Y$ such that all rows in the submatrix
$I\times\{j,j+1,\ldots,n\}$ of $A$ are distinguishable, 
$S(\sigma(I,j)) \ge (1-\epsilon)\log|I|$.
\end{claim*}

By the claim $S(\sigma(X,1)) \ge (1-\epsilon)\log m$ and $\dim(\H) \ge
2^{S(\sigma(X,1))}\geq m^{1-\epsilon}$, which implies
Lemma~\ref{lem:lv_mm}.  It remains to prove the claim by an induction
on the definition of $\sigma(I,j)$.

\medskip
{\sloppy
\emph{Induction base (Part (ii) of the definition)}: Then 
$S(\sigma(\{i\},j)) = 0$ for all $i\in X$ and $1\le j\le n+1$.

}

\medskip
\sloppy\hbadness=1500
\emph{Induction step (Part (i) of the definition)}: We consider
$\sigma(I,j) = {p\cdot\sigma(I_{j,0},j+1)} + {(1-p)\cdot\sigma(I_{j,1},j+1)}$,
where $p = |I_{j,0}|/|I|$. Observe that $I = I_{j,0} \cup I_{j,1}$ and
that for $b\in\{0,1\}$, $\sigma(I_{j,b},j+1) = \sum_{i\in I_{j,b}} p_i \sigma_i$
for suitable probabilities $p_i$, $i\in I_{j,b}$, with $\sum_{i\in I_{j,b}} p_i = 1$ (the latter
can also be proved by an easy induction on the definition of the $\sigma(I,j)$). Thus, 
applying the measurement $\M_j$ to $\sigma(I_{j,b},j+1)$ yields
\[
  \Pr\{ \M_j(\sigma(I_{j,b},j+1)) = b \} \ge 1-\epsilon\quad\text{and}\quad
  \Pr\{ \M_j(\sigma(I_{j,b},j+1)) = \neg b \} = 0.
\]
By Lemma~\ref{lem:lv_holevo}, this implies
\[
  S(\sigma(I,j)) \ \ge\  p\cdot S(\sigma(I_{j,0},j+1)) + (1-p)\cdot S(\sigma(I_{j,1},j+1)) + (1-\epsilon) H(p).
\]
By the induction hypothesis, 
$S(\sigma(I_{j,b},j+1)) \ge (1-\epsilon)\log|I_{j,b}|$ for $b\in\{0,1\}$. Thus,
\begin{align*}
  S(\sigma(I,j)) &\ \ge\  p (1-\epsilon)\log|I_{j,0}| + (1-p)(1-\epsilon)\log|I_{j,1}| + (1-\epsilon) H(p)\\
                 &\ =\   (1-\epsilon)\bigl(p\log|I_{j,0}| + (1-p)\log|I_{j,1}| + H(p)\bigr).
\end{align*}
Using that $p|I| = |I_{j,0}|$ and $(1-p)|I| = |I_{j,1}|$, we get
\begin{align*}
  S(\sigma(I,j) &\ \ge\   (1-\epsilon)\bigl(p\log(p|I|) + (1-p)\log((1-p)|I|) + H(p)\bigr)\\
                &\ =\  (1-\epsilon)\bigl(p\log p + (1-p)\log(1-p) + H(p) + \log|I|\bigr)
                 \ =\  (1-\epsilon)\log|I|,
\end{align*}
as desired.  This completes the proof of the claim and thus the proof
of Lemma~\ref{lem:lv_mm}.
\end{proof}

Now we can state and prove the main result.

 \begin{theorem}\label{the:lv_obdds}
 Let $G$ be a minimum size, leveled, reversible $\pi$-OBDD for $f$.
 Let $G'$ be a leveled $\pi$-QOBDD that computes $f$ with zero error and
 failure probability~$\epsilon$, $0\le\epsilon < 1$. For $i=1,\ldots,n+1$,
 let $L_i$ and $L_i'$ be the sets of nodes on level~$i$ in $G$ and $G'$, resp.
 Then $|L_i'|\ge |L_i|^{1-\epsilon}$ for $i=1,\ldots,n+1$.
 In particular, $|G'|\ge |G|^{1-\epsilon}$.
 \end{theorem}

 \begin{corollary}
 $\mbox{Rev-OBDD}=\mbox{EQP-OBDD}=\mbox{ZQP-OBDD}$.
 \end{corollary}

\begin{proof}[Proof of Theorem~\ref{the:lv_obdds}]
W.l.o.g.\ let $G = (V,E)$ and $G' = (V',E')$ have the variable order
$x_1,\ldots,x_n$.  From $G$ and $G'$ we construct some set of vectors
which are intermediate states of the computation of $G'$. We exploit
the relation to $G$ in order to construct a measurement
scheme for these vectors such that the lower bound follows from
Lemma~\ref{lem:lv_mm}.

W.\,l.\,o.\,g.\ $f$ depends on all variables. Let 
$\delta\colon V'\times V'\times\{0,1\}\to\C$ denote the transition amplitudes of~$G'$.
Let $\H$ be the Hilbert space spanned by an orthonormal basis
$(\k{v})_{v\in V'}$ whose elements are identified with the nodes
of~$G'$. Let $s\in V$ and $s'\in V'$ be the start nodes of~$G$
and~$G'$, resp., and let $F\subseteq V'$ be the set of sinks of $G'$.
For a partial input assignment~$a$ to $x_1,\ldots,x_i$, let
$\k{\phi(a)}\in\H$ be the superposition reached in~$G'$ by carrying
out its computation on~$a$. Let $\M_{\rm sink} = (M_{{\rm
sink},0},M_{{\rm sink},1},M_{{\rm sink},{\rm ?}})$ be the projective
measurement of the output label at the sinks in~$G'$.  For
$b\in\{0,1\}$, fix a unitary operator~$U_b$ on~$\H$ such that $U_b
\k{v} = \sum_{w\in V'} \delta(v,w,b)\k{w}$ for all $v\in V'-F$. Such
an operator exists due to the well-formedness of $G'$.

By the assumptions of  the theorem,
$L_i$ is the set of all nodes of~$G$ reached
by partial assignments to $x_1,\ldots,x_{i-1}$, for $i=1,\ldots,n+1$.
Observe that $L_1 = \{ s\}$ and, since $G$ is leveled and $f$ depends on all variables, 
all nodes in $L_i$, ${1\le i\le n}$, are labeled by~$x_i$.  For a node $v\in V$,
let $f_v$ denote the subfunction of~$f$ represented at~$v$ according
to the usual semantics of deterministic OBDDs.

We recursively construct mappings $\asn_i$ for $i=1,\ldots,n+1$ such
that $\asn_i$ maps a node $v\in L_i$ to a partial assignment to
$x_1,\ldots,x_{i-1}$ reaching that node from the start node of $G$.
First, we choose $\asn_1(s)$ as the empty assignment.  Next consider a
level $L_i$ with $i>1$. Let $v_1,\ldots,v_{\ell}$ be all nodes
representing one of the subfunctions~$f_{\rm sub}$ represented at
nodes in $L_i$.  Since $G$ is reversible and of minimum size, there are
a constant $b\in\{0,1\}$ and different nodes $u_1,\ldots,u_{\ell}\in
L_{i-1}$ such that $(f_{u_j})\vert_{x_{i-1}=b} = f_{\rm sub}$ and
there is a $b$-edge from $u_j$ to $v_j$ for
$j=1,\ldots,{\ell}$.  Define $\asn_i(v_j) = (\asn_{i-1}(u_j),b)$ for
$j=1,\ldots,{\ell}$.
For $i=1,\ldots,n+1$, let $C_i = \{ \k{\phi(\asn_i(v))} \mid v\in L_i \}$.

\begin{claim*}
For each $i = 1,\ldots,n+1$, there is a measurement scheme for~$C_i$ 
with zero error and failure probability~$\epsilon$.
\end{claim*} 

By Lemma~\ref{lem:lv_mm}, the claim 
implies $|L_i'| \ge \dim(\sp(C_i)) \ge |L_i|^{1-\epsilon}$
and thus the first part of the theorem. Since
$(x_1+\cdots+x_k)^c \ge x_1^{c}+\cdots+x_k^{c}$
for all $c\ge 1$ and $x_1,\ldots,x_k\in\IR^+_0$,
also $|G'|\ge |G|^{1-\epsilon}$ follows. 

We prove the claim by induction on $i$.
For $i=1$ and $C_1 = \{ \k{\phi(\asn_1(s))} \} = \{ \k{s'} \}$
the empty measurement scheme has the required properties.  

Let $i>1$ and
$L_i = \{ v_1,\ldots,v_m\}$.  Let $Y = \{ y_1,\ldots,y_N \}$, $N =
2^{n-i+1}$, be the set of assignments to $x_i,\ldots,x_n$.  Define the
$m\times N$-matrix $A = (a_{jk})$ by setting $a_{jk} = f_{v_j}(y_k)$
for $1\le j\le m$ and $1\le k\le N$.  For $k=1,\ldots,N$ let $\M_k =
(M_{k,0},M_{k,1},M_{k,{\rm ?}})$ be the projective measurement with
$M_{k,x} = M_{{\rm sink},x}\, U_{y_k}$ where $x\in\{0,1,{\rm ?}\}$ and
$U_{y_k}$ is the unitary transformation carried out by~$G'$ for the
partial input~$y_k$ when started on a superposition of the basis
vectors $(\k{v})_{v\in L_i'}$.

Obviously, $A$ is a boolean matrix where two rows
$j,j'\in\{1,\ldots,m\}$ differ iff the corresponding subfunctions
$f_{v_j}$ and $f_{v_{j'}}$ differ on an input from~$Y$.  Hence, for a
each set of pairwise different rows of $A$ chosen as
representatives for the different subfunctions and vectors in $C_i$
chosen accordingly, the above definitions yield a measurement scheme
due to the fact that $G'$ computes~$f$ with zero error and
failure probability~$\epsilon$.  Our goal is to extend the matrix $A$
and the collection of measurements such that we obtain a measurement
scheme for all vectors in $C_i$. We remark that $A$ does not have
entries ``$*$''.

{\sloppy Consider a subset of rows of $A$ belonging to the same
subfunction $f_{\rm sub}$ and thus containing identical vectors.
W.\,l.\,o.\,g., let $v_1,\ldots,v_{\ell}$ be the respective nodes in
$L_i$ representing $f_{\rm sub}$. Let $u_1,\ldots,u_{\ell}\in L_{i-1}$
and $b\in\{0,1\}$ be as in the definition of the assignments
$\asn_i(v_j)$ above. In particular, $b$ is the same constant for
$u_1,\ldots,u_\ell$. Then $U_b\k{\phi(\asn_{i-1}(u_j))} =
\k{\phi(\asn_{i-1}(u_j),b)} = \k{\phi(\asn_i(v_j))}$.  By induction
hypothesis, there is measurement scheme for $C_{i-1}$. Let $D$ be the
matrix of this measurement scheme, which is of size $|C_{i-1}|\times p$
for some $p$. Consider the sub-scheme for the vectors
$\k{\phi(\asn_{i-1}(u_j))}$, $j=1,\ldots,\ell$, which we obtain from
$D$ by deleting the rows corresponding to the other vectors. Let this
measurement scheme be described by the $\ell\times p$-matrix $B =
(b_{jk})$ and the projective measurements $\P_k =
(P_{k,0},P_{k,1},P_{k,{\rm ?}})$, $k=1,\ldots,p$.  Define $\P_k' =
(P_{k,0}',P_{k,1}',P_{k,{\rm ?}}')$, $k=1,\ldots,\ell$, by $P_{k,x}' =
P_{k,x}\, U_b^{\dag}$ for $x\in\{0,1,{\rm ?}\}$.  

Then for $j\in\{1,\ldots,\ell\}$ and $k\in\{1,\ldots,p\}$ such that
$b_{jk}\in\{0,1\}$,
\begin{align*}
  \Pr\{ \P_k'(\k{\phi(\asn_i(v_j))}) = b_{jk} \} 
  &= \Vert P_{k,b_{jk}}'\k{\phi(\asn_i(v_j))}\Vert^2
   = \Vert P_{k,b_{jk}}\, U_b^{\dag} \k{\phi(\asn_i(v_j))}\Vert^2\\
  &= \Vert P_{k,b_{jk}}\, U_b^{\dag} U_b \k{\phi(\asn_{i-1}(u_j))}\Vert^2\\
  &= \Pr\{ \P_k(\k{\phi(\asn_{i-1}(u_j))}) = b_{jk} \}.
\end{align*}
Hence, the measurements $\P_k'$, $k=1,\ldots,p$, satisfy
property~(iii) in the definition of measurement schemes with
respect to the matrix~$B$.

}

\sloppy 
Let $B_1,\ldots,B_R$ be all submatrices of $D$  obtained by the above
construction for the different subfunctions of $f_v$, $v\in
L_i$. Since in the construction of $B_1,\ldots,B_R$ no columns of $D$ are
deleted, the columns of $B_1,\ldots,B_R$ are labeled by the same
measurements. Hence, we can attach the matrices $B_1,\ldots,B_R$ to $A$
as submatrices in the columns $m+1,\ldots,m+p$ and fill up the
remaining entries with~``$*$'' such that the new matrix $A'$ obtained in
this way and the measurements $\M_1,\ldots,\M_N,\P_1',\ldots,\P_p'$
comprise a measurement scheme for $C_i$ with zero error and failure
probability~$\epsilon$. Since $A$ does not have any ``$*$''-entries,
also property (ii) of Definition~\ref{def_meas_scheme} is fulfilled.
\end{proof}

The above lower bound on the size of zero error QOBDDs in terms of the
size of reversible OBDDs is essentially optimal, as the following
example shows. For $n = 2^{\ell}$ define the {\em index function}
$\IND_n\colon\{0,1\}^{n+\ell}\to\{0,1\}$  on variable vectors
$x = (x_0,\ldots,x_{n-1})$ and $y = (y_0,\ldots,y_{\ell-1})$
by $\IND_n(x,y) = x_{|y|}$, where $|y| = \sum_{i=0}^{\ell-1} y_i 2^i$.

\begin{proposition}
For the variable order $\pi$ described by~$(x_0,\ldots,x_{n-1},y_0,\ldots,y_{\ell-1})$,
each deterministic $\pi$-OBDD representing $\IND_n$ requires size~$2^n$, while the same function
can be computed by zero error $\pi$-QOBDDs with failure probability~$\epsilon$
of size $2^{(1-\epsilon)n+O(\log n)}$.
\end{proposition}

Hromkovi\v c and Schnitger~\cite{Hro01a} have used a similar function
to prove an analogous result for classical Las Vegas and deterministic
one-way communication complexity and the special case
of failure probability~$\epsilon = 1/2$. The proof of the proposition
is by a straightforward adaptation of a simple randomized OBDD 
to the quantum case.

\begin{proof}
The lower bound for deterministic OBDDs is well known and follows from
the fact that $\IND_n$ has maximal one-way communication complexity
with respect to the partition of variables where Alice obtains~$x$
and Bob obtains~$y$.
In the following, we briefly sketch the upper bound construction.

For $\epsilon \ge 1/2$, partition~$x$ into $k = \floor{1/(1-\epsilon)}$ 
blocks of size approximately $(1-\epsilon)n$. The QOBDD chooses one of these blocks 
at random by an unlabeled node at the top (which can be removed later on 
similarly to the proof of Theorem~\ref{the:PERM}) with
outgoing edges having amplitudes~$1/\sqrt{k}$. 
These edges
lead to sub-QOBDDs where the complete chosen block is read
and stored, which requires a binary tree 
with $O(2^{(1-\epsilon)n})$ nodes for each block.   
At each leaf of such a tree, append a tree of size~$O(n)$ reading~$y$ 
and computing~$|y|$. Finally, a sink with the correct output value is
reached if $|y|$ lies in the chosen block, which
happens with probability at least~$1/k \ge 1-\epsilon$. 
Otherwise, the ``?''-sink is reached. 

For $\epsilon < 1/2$, we select $k = \ceil{1/\epsilon}$ blocks of
$x$-variables of size approximately $(1-\epsilon)n$ that cover each
single variable exactly~$k-1$ times. The rest of the construction is
the same as above. The failure probability is obviously bounded above
by $1/k \le \epsilon$.
\end{proof}

\subsection{Comparison of QOBDDs and  Read-Once QBPs}

In this section we observe that, similarly to the classical case, QOBDDs
are a more restricted model of QBPs than  read-once
QBPs. A function separating these two models with respect to polynomial
size is the so-called \emph{indirect storage access function}, which is
defined in the following way. Let $n=2^k$. The input of $\ISA_n$
consists of the variables $y_0,\ldots,y_{k-1}$ and
$x_0,\ldots,x_{n-1}$. The $y$-variables are interpreted as a binary
number $s$. The $x$-variables are partitioned into $b=\lfloor
n/k\rfloor$ blocks of size $k=\log n$, which are numbered beginning
with $0$. If $s\geq b$, the output is $0$. Otherwise the $s$th block is
again interpreted as a binary number $t$ and the output is $x_t$.
It is straightforward to construct a decision tree for $\ISA_n$ of size
$O(n^2/\log n)$, which can also be regarded as a read-once QBP.

The lower bound for QOBDDs for all variable orders is a
straightforward combination of two results. Klauck~\cite{Kla00} proved
the lower bound $\Omega(n)$ on the quantum one-way communication
complexity of $\IND_n$, where Alice gets the $x$-variables and Bob the
$y$-variables. This lower bound directly implies the lower bound
$2^{\Omega(n)}$ on the size QOBDDs for $\IND_n$, where the $x$-variables
are tested before the $y$-variables. Using a rectangular reduction,
it has been shown in~\cite{Sa01a} that an OBDD for $\ISA_n$ and an
arbitrary variable order cannot be smaller than an OBDD for
$\IND_{\lfloor n/\log n\rfloor-1}$ and the variable order mentioned
before.  This also holds for QOBDDs such that we obtain the lower bound
$2^{\Omega(n/\log n)}$ on the size for QOBDDs for $\ISA_n$ and an
arbitrary variable order.

\section{QBPs with Generalized Measurements}\label{sec:mqbps}

The usual unitary quantum mode of computation has turned out to be only of
limited use for such restricted models as quantum OBDDs and quantum
finite automata. In this section we consider a generalization of QBPs 
where in each step 
the performed unitary operation is determined by the result of a previous
measurement. 
We first present the definition of QBPs with generalized
measurements and we discuss the relationship to QBPs and to randomized
BPs. Afterwards, we prove a generic lower bound on the size of QOBDDs
with generalized measurements for so-called $k$-stable functions.

\begin{definition}\label{def_gmqbp}
Let $k\in\IN$ with $k \ge 3$.
A \emph{quantum branching program with generalized measurements (gmQBP)}
over the variable set $X=\{x_1,\ldots,x_n\}$ is a 
directed multigraph $G=(V,E)$ with a start node $s\in V$,
a set of sinks $F\subseteq V$, and transition amplitudes~$\delta$.
Nodes and edges are labeled in the same way as in a 
usual QBP (see Definition~\ref{def:qbp}).
Additionally, there is a partition $(V_0,V_1,V_2,\ldots,V_{k-1})$ of~$V$
such that $V_0$ and $V_1$ consist of the $0$- and $1$-sinks of $G$, resp.
The edge labels of the gmQBP~$G$ have to fulfill the following modified
well-formedness constraint. Let $u,v\in V_{\ell}$, $\ell\in\{2,\ldots,k-1\}$, 
be interior nodes with $\var(u)=i$ and $\var(v)=j$, resp. Then for all assignments
$a=(a_1,\ldots,a_n)$ to the variables in $X$,
\begin{equation}\tag{W$^*$}
\sum_{w\in V}\delta^*(u,w,a_i)\delta(v,w,a_j)\ =\ 
\left\{\begin{array}{ll} 1, & \mbox{if $u=v$;} \\ 0, &
    \mbox{otherwise.}
\end{array}\right.
\end{equation}
Furthermore, gmQBPs are unidirectional, i.\,e., for each $w\in V$, all
$v\in V$ for which a $b\in\{0,1\}$ exists such that $\delta(v,w,b)\neq
0$ are labeled by the same variable. 
\end{definition}

We remark that the well-formedness condition for gmQBPs is weaker than the
well-formedness condition for ordinary QBPs, because it has only to
hold for pairs of nodes of the same set~$V_{\ell}$.

We now define the semantics of gmQBPs.
As in the definition of usual QBPs, nodes correspond to vectors in
an orthonormal basis $(|v\rangle)_{v\in V}$ of $\H = {\C}^{|V|}$
and intermediate results of the computation are superpositions of
these vectors.
As for QBPs, a computation step consists of a measurement
and the subsequent transition to successor nodes according
to the transition amplitudes~$\delta$. In a gmQBP,
the measurement generalizes that allowed for QBPs
as follows. The gmQBP performs the projective measurement
$\M = (P_0,P_1,P_2,\ldots,P_{k-1})$ with results $\{0,1,2\ldots,k-1\}$,
where
\[
  P_r \ =\  \sum_{v\in V_r} |v\rangle\langle v|, \quad
  r\in\{0,1,2,\ldots,k-1\}.
\]
The probability of obtaining the result $r$ is $\|P_r|v\rangle\|^{2}$.
If the result~$r$ is $0$ or $1$, the computation stops with output~$r$.
If $r\geq 2$, the computation
continues with the normalized projection
\[
 |\psi'\rangle 
   \ =\  \frac{P_r|\psi\rangle}{\|P_r|\psi\rangle\|}
   \ =\  \sum_{v\in V_r}\alpha_v |v\rangle.
\]
Then for each node ${v\in V_r}$ with $\var(v) = i$
the gmQBP follows the edges with boolean label~$a_i$ according to 
their amplitudes. This yields the new superposition
\[
  |\psi''\rangle \ =\  \sum_{v\in V_r}\alpha_v\sum_{w\in V}\delta(v,w,a_{\var(v)}) |w\rangle.
\]

The above definition does not allow ``?'' outputs
for simplicity, since we do not consider \mbox{Las Vegas} gmQBPs, anyway.
The modified well-formedness constraint implies that for each result
of the measurement the corresponding mapping can
be extended to a unitary transformation.
Computation time and acceptance modes are defined analogously to
QBPs. Also the definition of QOBDDs with generalized measurements
(gmQOBDDs) is straightforward: The variables are required to be tested
according to a fixed variable order. We remark that gmQBPs have a
simple graphic representation. Additionally to the representation of
QBPs there is merely a partition of the nodes.  

The physical realizability of gmQBPs depends on the ability to perform 
measurements during a computation. 
Based on a standard argument using Neumark's theorem (see, e.\,g., \cite{Per95}), 
such measurements can be described by  unitary transformations in an extended 
Hilbert space. Furthermore, intermediate measurements are also possible, e.\,g.,
in the quantum circuit model defined in the textbook of
Nielsen and Chuang~\cite{Nie00} as well as in the model of Aharonov, Kitaev and Nisan~\cite{Aha98}
which allows gates computing general quantum operations (superoperators).

It is obvious that a QBP is a gmQBP with three possible measurement results.
We show that randomized BPs can easily be transformed
into gmQBPs.

\begin{proposition}
For each randomized BP~$G$ computing some function $f$ there is a 
gmQBP~$G'$ computing the same function with the same acceptance mode,
and the size of~$G'$ is bounded above by the size of $G$.
\end{proposition}

\begin{proof}
We remove all randomized nodes from $G$ by allowing each node to have
several outgoing $0$- and $1$-edges labeled by appropriate
probabilities. In the corresponding gmQBP there are the same edges,
where the probability $p$ is replaced with the amplitude $\sqrt{p}$. 
The partition of the node set consists of the set of $0$-sinks, the
set of $1$-sinks and sets each containing exactly one interior
node. An easy induction shows that for each input the acceptance probabilities of $G$
and $G'$ coincide.
\end{proof}

With the currently available techniques we cannot prove
superpolynomial lower bounds for BPs and for QBPs either (cf.~Proposition~\ref{Prop_BPvsQBP}).
Thus we are not able to prove that polynomial size gmQBPs are 
more powerful than polynomial size QBPs. However, for QOBDDs this is easy,
even for $k=4$, i.\,e., the smallest~$k$ where gmQOBDDs are a
generalization of QOBDDs. In Theorem~\ref{the:NO} we have proved
exponential lower bounds on the size of QOBDDs for the function
$\DISJ$ and $\IP$. On the other hand, it is easy to construct linear
size deterministic OBDDs for $\DISJ$ and $\IP$. A careful inspection
shows that each node of these OBDDs has at most two incoming $0$-edges
and at most one incoming $1$-edge. We partition the internal nodes
into two sets $V_2$ and $V_3$ such that each pair of nodes with the
same $0$-successor is not in the same set. Furthermore, by duplicating
the sinks we ensure that each sink has at most one predecessor. The
sets $V_0$ and $V_1$ are the sets of $0$- and $1$-sinks, resp., that
are obtained in this way. We obtain the following result.

\begin{proposition}
There  are gmQOBDDs of linear size with $k=4$ possible measurement 
results that exactly compute $\DISJ_n$ and $\IP_n$.
\end{proposition}

Finally, we prove a generic lower bound on the size of gmQOBDDs for
$k$-stable functions.  A function $f\colon\{0,1\}^n\to\{0,1\}$ is
called \emph{$k$-stable} if for each set $V$ of variables of size~$k$
and each variable $x_i\in V$ there is a setting of the variables
outside~$V$ such that the resulting subfunction is $x_i$ or
$\wbar{x}_i$. It is well known that $k$-stable functions only have
read-once branching programs of size $2^{k-1}$, and it has been shown
in~\cite{Sa01a} that also randomized OBDDs require size
$2^{\Omega(k)}$. Examples for such functions include the determinant
of an $n\times n$-matrix over~$\ZZ_2$, which is $(n-1)$-stable, and
the function checking whether a graph on $n$~vertices has an
$n/2$-clique, which is $(n/4+1)$-stable. For these and other examples,
see Wegener~\cite{Weg00}.

We remark that the state of a gmQOBDD after performing a measurement
during a computation can be described as a mixed state, i.\,e., a
probability distribution over pure states.
Now we can apply a lower bound on the quantum communication complexity for the
index function (defined at the end of Section~\ref{subsec_lvqobdd}) due to Klauck~\cite{Kla00}.

\begin{theorem}
\label{thsqobdd}
Each gmQOBDD with bounded error for a $k$-stable functions has size
$2^{\Omega(k)}$.
\end{theorem}

\begin{proof}\sloppy
  W.\,l.\,o.\,g.\ let $k=2^{\ell}$. 
  Klauck~\cite{Kla00} has observed
  that the quantum one-way communication complexity of the function
  $\IND_k$ is lower bounded by $\Omega(k)$ for the
  partition where the 
  first player Alice gets the input vector $x = (x_0,\ldots,x_{k-1})$ and the second player
  Bob gets $y = (y_0,\ldots,y_{\ell-1})$. This lower bound also holds for the two-sided
  error model and if Alice may send a mixed state to Bob.
  Let a gmQOBDD for a $k$-stable function~$f$ be given.
  Then $\IND_k$ can be computed by a quantum one-way protocol in the
  following way: Alice may choose the first~$k$ variables
  in the variable order and Bob the remaining variables.
  By the property of $k$-stable functions, for each of Alice's variables,
  Bob can fix his variables such that the gmQOBDD outputs the value of the
  variable or its complement. Hence, it suffices for 
  Alice to perform the computation of the gmQOBDDs of the 
  first~$k$ levels for the given setting of her $x$-variables and to send the
  (mixed) state of the gmQOBDD after her computation to Bob. Bob can then
  compute the output as described. The communication complexity is bounded above by 
  the logarithm of the size (or even the width) of the gmQOBDD. Together
  with the lower bound on the quantum communication complexity for
  $\IND_k$, the theorem follows.  
\end{proof}

\section{Open Problems}

In this paper, we have explored the foundations of space-bounded nonuniform
quantum complexity to some extent, but several interesting problems
nevertheless remain open.

\begin{shortindent}{--}
\item[--]
  It is not clear whether algebraic amplitudes for nonuniform QTMs and short amplitudes for QBPs are the most
  general reasonable sets of amplitudes. Is it possible to provide some formal argument 
  that excludes more general sets of amplitudes
  (as done by Adleman, DeMarrais, and Huang~\cite{Adl97} for the uniform case and
  arbitrary complex amplitudes)?
\item[--]
  For space-bounded nonuniform QTMs with algebraic amplitudes we have proved that the general model can be simulated
  by the unidirectional one. It is open so far whether an analogous simulation also exists for the uniform case.
  Furthermore, for QBPs it is straightforward to define a variant without the requirement of unidirectionality. Can this
  generalized model be simulated by the unidirectional model or is it unreasonably powerful?
\item[--]
  It remains open whether there is a space-efficient simulation of QBPs  by nonuniform QTMs for the 
  cases of error-free and exact quantum computation and, if not, to provide
  some evidence showing that such a simulation is 
  unlikely to exist.
\item[--]
  With respect to the comparison of OBDDs and QOBDDs, the relationship between the classes
  $\mbox{BQP-OBDD}$ and $\mbox{BPP-OBDD}$ for total functions is left
  open. 
\item[--] 
  Prove lower bounds for more general variants of QBPs. 
  While lower bounds for QOBDDs can be obtained using tools from quantum communication complexity, 
  already the proof of lower bounds for (possibly unordered) read-once QBPs seems to require new arguments.
\item[--]
  The model of gmQBPs remains largely open to investigation. 
  In particular, the relationship between the standard model of QBPs and gmQBPs needs to
  be further clarified. Show separation results as that for QOBDDs and gmQOBDDs presented here also 
  for more general variants of QBPs or investigate simulations of gmQBPs by usual QBPs.
\end{shortindent}

\end{document}